\documentclass[a4paper,11pt]{article}
\pdfoutput=1
\usepackage{jheppub}

\usepackage{epsf}
\usepackage{epsfig}
\usepackage{subfigure}
\usepackage{mathtools}
\usepackage{hhline}
\usepackage{float}
\restylefloat{table}
\usepackage{multirow}
\usepackage{nicefrac}
\usepackage{epstopdf}
\usepackage{url}
\usepackage{array}
\usepackage[normalem]{ulem}

\allowdisplaybreaks

\newcommand{\be}{\begin{eqnarray}}
\newcommand{\ee}{\end{eqnarray}}

\newcommand{\ba}{\begin{array}}
\newcommand{\ea}{\end{array}}
\newcommand{\bee}{\begin{equation}\ba{c}}
\newcommand{\eee}{\ea\end{equation}}

\newcommand{\bi}{\begin{itemize}}
\newcommand{\ei}{\end{itemize}}

\title{2$b$ or not 2$b$: on the rejection of $g\to b\bar b$ jets}

\author{Beni Pazar$^1$,}
\author{Enrico Lunghi$^1$}
\affiliation{
$^1$Physics Department, Indiana University, Bloomington, IN 47405, USA \\
}
\emailAdd{bpazar@iu.edu}
\emailAdd{elunghi@indiana.edu}

\abstract{
Motivated by new physics models which lead to final states containing a high multiplicity of bottom and top quarks, we develop a tagging strategy to suppress reducible and non-reducible multi-jet backgrounds. The idea takes advantage of the properties of light parton showers and of the gluon fragmentation into heavy quarks to reject jets that do not originate from a bottom quark.}

\begin{document}

\maketitle

\section{Introduction}
\label{sec:introduction}
Many extensions of the Standard Model (SM) involve heavy new particles with couplings to the third generation of SM fermions and yield final states with multiple bottom and top quarks. When feasible, experimental searches require the presence of high transverse momentum leptons and/or missing energy in order to suppress backgrounds or the use of data driven background estimation techniques. In this paper, we focus on techniques to suppress backgrounds to new physics signals with final states involving multiple third generation quarks (i.e. combinations of bottom and hadronic top quarks).

For instance, in models with extra Higgs Bosons and vectorlike quarks, cascade decays involving Higgses and vectorlike quarks can lead to final states in which the two $b$-quarks do not form a resonance (e.g. $pp \to H \to b_4 \bar b \to Z b \bar b \to  \mu^+ \mu^- b \bar b$)~\cite{Dermisek:2016via, Dermisek:2019heo} and to purely hadronic final states involving a large number of $b$ and $t$-quarks (e.g. $pp\to b_4 \bar b_4 \to Hb \; H\bar b \to b\bar b b\bar b b\bar b$) ~\cite{Dermisek:2020gbr}. In both cases the dominant backgrounds are multi-jets in which low-mass light-flavored jets are mistagged as originating from a $b$-quark in the hard interaction.

The ability to isolate jets containing $b$-quarks is important for studies of QCD (e.g. heavy flavor production), of Higgs and top physics, and of new physics models which often couple to the third generation only (to avoid the strong constraints on couplings to the light quarks). More recently, there has been an increased interest in heavy resonances produced at large transverse momentum and whose decay products are, therefore, emitted in a narrow cone (i.e. with low relative $\Delta R$\footnote{$\Delta R$ between two 4-momenta is defined as $\sqrt{(\Delta \eta)^2 + (\Delta \phi)^2}$, where $\eta$ and $\phi$ are pseudorapidity and azimuthal angle of the two 4-momenta, respectively.}). The identification of these fat jets with a quite complicated substructure has spurred the development of dedicated tagging strategies. In particular, ATLAS has developed taggers aimed precisely at the identification of $X\to b\bar b$ resonances~\cite{ATLAS:2012xna, ATLAS:2022qpg}.

Our interest lies not much in resonances decaying to collimated $b\bar b$ pairs or $b+\text{jets}$ (like in decays of high-$p_T$ top quarks), but in distinguishing pure b-jets originating from a $b$-quark produced directly in the partonic interaction (which we refer to as 1$b$ jets) from jets in which $b$-quarks appear in the parton shower of lighter partons (which we refer to as 2$b$ jets). Previous theoretical investigations of this issue have been presented in ref.~\cite{Goncalves:2015prv} (where the focus was on distinguishing $b$ vs $g\to b\bar b$ jets) and ref.~\cite{Bhattacherjee:2016bpy} (in the context of searches for gluinos at the LHC).

In this work we extend the analysis of ref.~\cite{Goncalves:2015prv}, where the focus is on gluon jets which split to a $b\bar b$ pair with both quarks in the jet radius, by including $b$-tagged gluon jets in which only one $b$-hadron is found amongst the jet constituents or which are mistagged (i.e. no $b$-hadron appears in the shower). The  latter are entirely due to detector effects, which we model using Delphes~\cite{deFavereau:2013fsa} (a simplified detector simulator). With a slight abuse of notation we denote all jets that are $b$-tagged but do not originate from a hard $b$-quark as ``2$b$ jets''. The main result of this work is the construction of a 1$b$2$b$-tagger which has to be applied to $b$-tagged jets and which aims at rejecting $b$-tagged gluon jets that either genuinely split to $b\bar b$ or are mistagged but present similar properties.

The paper is organized as follows. In section~\ref{sec:idea} we discuss the present status of $b$-tagging and its implementation in Delphes; in sections~\ref{sec:observables} and~\ref{sec:1b2b} we present an in-depth analysis of the various jet sub-structure observables that we consider and proceed to the construction of the 1$b$2$b$-tagger; in section~\ref{sec:multijet} we show an example of the performance of the 1$b$2$b$-tagger in multi-jet events; finally, in section~\ref{sec:conclusions} we present our conclusions.

\section{Preliminary considerations}
\label{sec:idea}
In this section we discuss the current status of $b$-tagging at ATLAS and CMS. We first review their standard approach and then describe recent developments aimed at identifying boosted resonances that decay to pairs of $b$-quarks.

The ATLAS approach is based on the MV2 and DL1 algorithms~\cite{ATLAS:2019bwq}. MV2 is a boosted decision tree that combines the outputs of multiple lower level tagging algorithms. The lower level algorithms include kinematics, impact parameters, secondary vertex finders, and the topological multi-vertex finder JETFITTER~\cite{ATLAS:2018nnq}. DL1, in contrast, is a deep feed-forward neural network trained on the same lower level algorithms along with the JETFITTER c-tagging variables. The two have relatively similar efficiencies when distinguishing between $b$/$c$-jets and $b$/light-jets. At a working point of $85\%$ signal efficiency they misidentify approximately $38\%$ of $c$-jets and $3\%$ of light-jets. These algorithms define the flavor of jets in a hierarchical manner: first looking to see if there are any b-hadrons among the jet constituents, then $c$-hadrons, otherwise labeling the jet as ``light''. This method does not distinguish between b-hadrons originating from $b$-quarks produced in the hard interaction and those which come from gluon splitting~\cite{ATLAS:2012xna, ATLAS:2015dex, ATLAS:2017bcq}.
This broad classification leaves the $b$-taggers completely blind to the origin of the tagged $b$-quark. JETFITTER uses the variable $f_E$ which is defined as the fraction of the charged jet energy in all secondary vertices. With this definition, $f_E$ peaks near $1$ for both 1$b$ and 2$b$ jets (for the latter, this is only the case as long as the splitting happens early enough).

CMS has multiple approaches to $b$-tagging~\cite{CMS:2012feb, CMS:2017wtu}. In particular jets containing $b$-hadrons are identified at CMS using the Combined Secondary Vertex (CSV) algorithm (presented in ref.~\cite{CMS:2012feb} for Run 1 and updated in ref.~\cite{CMS:2017wtu} for Run 2), which combines flight distance of the secondary vertex with track based lifetime information (if no secondary vertices are found). On top of this, CMS has a version of this tagger based on a deep neural network (DeepCSV) which yields a slight increase in performance.

Early studies of jets containing multiple $b$-hadrons have been performed at Tevatron (see, for instance, the CDF measurement ref.~\cite{CDF:2004mmv}), were based on direct reconstruction of secondary vertices and had limited success due to the difficulty of identifying and reconstructing two $b$-hadrons (double $b$-tag). The ATLAS and CMS collaborations performed studies focused on tagging $b\bar b$-jets both in the context of gluon splitting and for decays of heavy Higgs like resonances to $b\bar b$. In both cases, the focus was to identify $b\bar b$ at relatively low transverse momentum ($p_T \lesssim 200 \; {\rm GeV}$).

A first approach adopted by ATLAS considers jet substructure variables and used the Toolkit for Multivariate Data Analysis (TMVA)~\cite{Hocker:2007ht} to produce a discriminator based on track-jet width, jet width maximum track $\Delta R$, charged track multiplicity, girth (see eq.~(\ref{eq:girth})), and n-subjettiness~\cite{Thaler:2010tr}. The jet $p_T$ analyzed lie in the $[40,480]$ GeV range. At a signal efficiency of $80\%$ this tagger keeps between $20\%-30\%$ of 2$b$ background~\cite{ATLAS:2012xna}. More recently, ATLAS developed DeXTer (Deep set $X\to b\bar b$ Tagger)~\cite{ATLAS:2022qpg} which uses reconstructed tracks and vertices in a Deep Set Neural Network. The range is $p_T \in [20,200]$GeV and, at a working point of $80\%$, keeps approximately $20\%-25\%$ of background. Additionally, ATLAS developed tagging strategies aimed exclusively at identifying heavy color neutral resonances which decay to $b\bar b$, for which gluon splitting to $b\bar b$ provide the dominant background. In ref.~\cite{atlas:2023}, ATLAS proposed a novel algorithm, GN2X, that supersedes previous incarnations of this strategy~\cite{atlas:2020, atlas:2021} and which is based on a transformer neural network.

In ref.~\cite{cms:2015, CMS:2017wtu}, CMS presented an analysis of $b$-tagging in boosted jets based on kinematics and independent tagging of both $b$-quarks and aimed at the identification of resonances decaying to $b\bar b$ (in particular $H$ and $Z$).

\begin{figure}[t]
\begin{center}
\includegraphics[width=0.49 \linewidth]{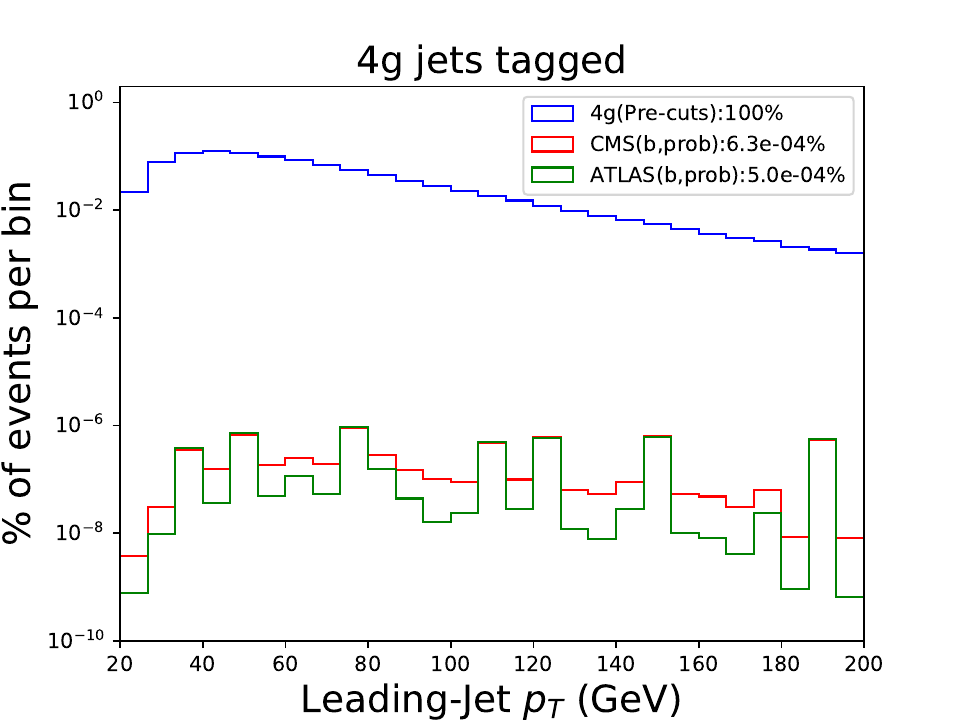}
\end{center}
\caption{Impact of $b$-tagging on 4-jet events originating from an inclusive $pp\to 4g$ hard interaction. The ATLAS vs CMS tagging efficiencies have been modeled within Delphes by rescaling the event weights by the efficiency factors given in Eqs.~(\ref{eq:BTlight})--(\ref{eq:BTbottom}). The statistics of the tagged events is degraded with respect to the pre-$b$-tag distribution because the efficiencies for light jets are much smaller than those for $c$ and $b$ jets (the peaks correspond to the $O(100)$ events in which each of the jets contains an actual $b$ quark originating from gluon splitting).}
\label{fig:AvC4g}
\end{figure}
\begin{table}[t]
\begin{center}
\begin{tabular}{|c||c||c|c|c||c|c|} \hline
\multirow{2}{*}{$p_T$} & Btag & \multirow{2}{*}{f} & \multirow{2}{*}{Btag${}_f$} & \multirow{2}{*}{   $\frac{\text{Btag}_f}{ \text{Btag} }$} & \multirow{2}{*}{prob jet${}_f$} &   \multirow{2}{*}{$\varepsilon_f= \frac{\text{Btag}_f}{\text{prob jet}_f}$}  \\
             & $\scriptstyle =\sum \text{Btag}_f$ &   &  &  &  &  \\
  \hline\hline
  			    &       & b & 0.0062 & 0.27 & 0.012 & 0.52   \\
  20-50 GeV & 0.023 & c & 0.0062 &0.27& 0.048 & 0.13   \\
            &       & l & 0.010 &0.46& 0.94 & 0.011   \\
    \hline
             &       & b & 0.013 & 0.34 & 0.019 & 0.68 \\
  50-100 GeV & 0.037 & c & 0.013 & 0.34 & 0.061 & 0.21 \\
             &       & l & 0.012 & 0.32 & 0.92 & 0.013 \\
  \hline
              &       & b & 0.020 & 0.37 & 0.035 & 0.57 \\
  100-200 GeV & 0.054 & c & 0.018 & 0.34 & 0.085 & 0.21 \\
              &       & l & 0.016 & 0.29 & 0.88 & 0.018 \\
  \hline
              &       & b & 0.025 & 0.40 & 0.04 & 0.63 \\
  200-500 GeV & 0.062 & c & 0.019 & 0.31 & 0.10 & 0.19 \\
              &       & l & 0.018 & 0.29 & 0.86 & 0.021 \\
  \hline
            &       & b & 0.025 & 0.35 & 0.052 & 0.48 \\
  0.5-1 TeV & 0.072 & c & 0.018 & 0.25 & 0.13 & 0.14 \\
            &       & l & 0.029 & 0.40 & 0.82 & 0.035 \\
  \hline
            &       & b & 0.019 & 0.24 & 0.07 & 0.27 \\
  1-1.5 TeV & 0.081 & c & 0.017 & 0.21 & 0.16 & 0.11 \\
            &       & l & 0.045 & 0.55 & 0.77 & 0.058 \\
  \hline
            &       & b & 0.015 & 0.19 & 0.077 & 0.19 \\
  1.5-2 TeV & 0.089 & c & 0.014 & 0.20 & 0.18 & 0.078 \\
            &       & l & 0.059 & 0.61 & 0.75 & 0.079 \\
   \hline
\end{tabular}
\end{center}
\caption{Breakdown of CMS $b$-tagging for gluon initiated jets as a function of transverse momentum. $p_T$ is the transverse momentum range of the jet, Btag is the total $b$-tagging probability, f is the flavor of the jet (b if it contains a bottom, c if it does not contain bottoms but has a charm, l if neither a bottom or charm is found), $\text{Btag}_f$ is the contribution to the total $b$-tagging probability from jets with flavor f (Btag = $\sum \text{Btag}_f$), Btag${}_f$/Btag is the relative contribution of jets with flavor f to the total $b$-tagging probability, prob jet${}_f$ is the probability that a gluon initiated jet has b, c or light flavor, $\varepsilon_f$ is the efficiency for $b$-tagging a jet of a given flavor and originates from Eqs.~(\ref{eq:BTlight})-(\ref{eq:BTbottom}).}
\label{tab:Btag-breakdown}
\end{table}

We now discuss the current status of implementation of the ATLAS and CMS standard $b$-tagging algorithms in the detector simulator Delphes. For both experiments, the tagging efficiencies are modeled following the results presented in refs~\cite{ATLAS:2015dex} and~\cite{CMS:2012feb} by ATLAS and CMS, respectively. The $p_{T}$ dependent efficiency functions depend on the jet flavor (light, charm, bottom) and are:
\begin{align}
\varepsilon_{\rm light}  &=
\begin{cases}
0.01+3.8 \times 10^{-5} \; p_T & \text{CMS} \\
0.002+7.3 \times 10^{-6} \; p_T & \text{ATLAS}
\end{cases}
 \label{eq:BTlight}\\
\varepsilon_{\rm charm}  &=
\begin{cases}
0.25 \; \frac{ \tanh(0.018 \; p_T) }{1+0.0013 \; p_T} & \text{CMS} \\
0.20 \; \frac{ \tanh(0.02 \; p_T) }{1+0.0034\; p_T}  & \text{ATLAS}
\end{cases}
\label{eq:BTcharm}\\
\varepsilon_{\rm bottom} &=
\begin{cases}
0.85 \; \frac{25\; \tanh(0.0025 \; p_T)}{1+0.063 \; p_T}  & \text{CMS} \\
0.80 \; \frac{30\; \tanh(0.003 \; p_T)}{1+0.086 \; p_T} & \text{ATLAS}
\end{cases}
\label{eq:BTbottom}
\end{align}
As an example, in figure~\ref{fig:AvC4g} we show the impact of requiring 4 $b$-tagged jets on a process originating from the inclusive $pp\to 4g$ hard interaction. We used MadGraph5~\cite{Alwall:2011uj} for parton level generation, Pythia8~\cite{Sjostrand:2006za,Sjostrand:2014zea} for parton shower and hadronization, and Delphes for detector simulation. Jets are reconstructed using the anti-$k_t$ algorithm with $\Delta R = 0.5$. For both CMS and ATLAS the fraction of events that survive is about $0.05\%$. In this simulation we do not actually impose the stochastic $b$-tagging but rescale the weight of each event using the efficiencies in Eqs.~(\ref{eq:BTlight})--(\ref{eq:BTbottom}) for CMS and ATLAS.

In table~\ref{tab:Btag-breakdown} we present the $b$-tagging anatomy of a single jet produced with $p_T \in [20\; {\rm GeV},2\; {\rm TeV}]$ and originating from a hard gluon. The flavor ($f$) of the jet is obtained by looking at the event prior to hadronization according to whether a $b$ or $c$ quark is found anywhere in the shower. The $b$-tag probability ($\varepsilon_f$) given in the last column is obtained directly from Eqs.~(\ref{eq:BTlight})--(\ref{eq:BTbottom}) for CMS. The probability of observing a jet with a given flavor depends exclusively on the details of the parton shower as handled by Pythia8. An important feature of these rates is the increased misidentification at high $p_T$, driven by mistagged light jets. Another interesting aspect of these results is that, at moderate $p_T$, $b$-tagged gluon jets are equally likely to originate from jets with showers containing $b$-quarks, $c$-quarks or no heavy quarks at all: the small $b$-tagging probabilities of charm and light jets are compensated for by their significantly larger production rates.

\section{Observables}
\label{sec:observables}
Following ref.~\cite{Goncalves:2015prv} we focus on three observables: the ratio of the energy of the $b$-tagged hadron to the total jet energy ($X_E$), the transverse size of the jet (girth, $g$) and the number of charged tracks in the jet ($N_{\rm ch}$).

$X_E$ is a key discriminator due to the significant difference in the $b$-quark and gluon fragmentation functions. The mass of the $b$-quark greatly suppresses collinear gluon radiation, thus also suppressing secondary partons. This phenomenon is known as the ``dead cone effect''. $b$-quarks tend to retain the majority of their initial energy, leading to an average $X_E \sim 0.8$. In contrast, gluons tend to radiate a considerable fraction of their energy (larger at higher $p_T$) prior to the $b\bar b$ splitting. The latter tend to be produced with similar energy (though asymmetric splitting can and does occur). Both effects tend to lower the energy in the leading $b$-quark found in a gluon shower and the resulting $X_E$ is considerably smaller than typical values observed in $b$-quark initiated jets. Note that, in order to distinguish between 1$b$ and 2$b$ jets, the variable $X_E$ is much more appropriate than $f_E$ (which sums over all displaced vertices) because the latter peaks near 1 for both types of jets (see the discussion of the JETFITTER algorithm discussed at the beginning of section~\ref{sec:idea}). 

\begin{figure}[t]
\begin{center}
  \includegraphics[width=0.65 \linewidth]{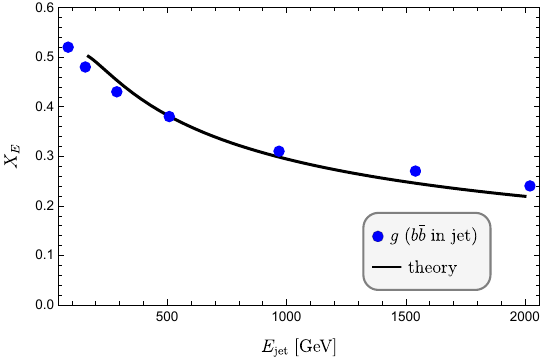}
\end{center}
\caption{Comparison between $X_E$ calculated using Eq.~(\ref{eq:xe_theory}) (solid line) and extracted from Monte Carlo simulations (blue circles). In order to compare directly to the theoretical estimate we require both $b$ and $\bar b$ to be found in the shower and we calculate $X_E$ using the average of the $b$ and $\bar b$ energies.}
\label{fig:XEbreakdown}
\end{figure}

In order to verify the above intuitive arguments, we can start from the theoretically predicted distribution of $b\bar b$ pairs in a jet as a function of the invariant masses of the jet ($m_j^2$) and of the gluon ($k^2$) which splits to $b\bar b$. This is given by~\cite{Mueller:1985zz, Mangano:1992qq,  ellis_stirling_webber_1996}:
\begin{align}
R_{g\to b\bar b} (m_j^2,k^2) &=
\frac{1}{6\pi} \alpha_s(k^2) \left[1+\frac{2 m_b^2}{k^2}\right] \sqrt{1-\frac{4m_b^2}{k^2}} n_g (m_j^2,k^2) \\
n_g (m_j^2,k^2) &= \left[ \frac{\log (m_j^2/\Lambda^2)}{\log (k^2/\Lambda^2)} \right]^a \cosh \left\{
\sqrt{\frac{2 C_A}{\pi b}} \left[ \sqrt{\log(m_j^2/\Lambda^2)}-\sqrt{\log(k^2/\Lambda^2)} \right]
\right\} \label{eq:ng}
\end{align}
where $a = -1/4 -5 n_f/(54 \pi b)$, $b = (33 - 2 n_f)/(12\pi)$ and $\alpha_s (k^2) = 1/ [b \log (k^2/\Lambda^2)]$. This distribution allows us to build a quantity that should resemble $X_E$ by using it to weight the average energy of the  $b$-quarks produced in the $g\to b\bar b$ splitting:
\begin{align}
X_E (m_j^2) &= \int_{4 m_b^2}^{m_j^2} \frac{d k^2}{k^2} R_{g\to b\bar b} (m_j^2, k^2) E_b^{\rm av} (k^2) \; .
\label{eq:xe_theory}
\end{align}
After extracting the relation between gluon virtuality and average $b$-quark energy from Monte Carlo simulations, we plot the resulting $X_E$ in figure~\ref{fig:XEbreakdown}. The blue points in the figure are obtained by selecting jets in which a $b\bar b$ pair is found within the jet radius and calculating the ratio of the average energy of the $b$-quarks and of the jet energy. The agreement between the rough theoretical calculation and the Monte Carlo results is quite satisfactory.

The construction of $X_E$ for a generic jet is done as follows. If a $b$ or a $\bar b$-quark is found anywhere in the shower, we seek the highest energy $b$-hadron in the generator level constituents of the jet and use its energy to construct $X_E$. When a $c$-quark is found instead, we seek the highest energy $c$-hadron, assume that the displaced vertex selected by the $b$-tagger corresponds to this hadron decay and use its energy to calculate $X_E$. If no heavy quarks are found, we simply construct $X_E$ by combining the energies of five randomly chosen charged tracks selected among the jet constituents.

\begin{figure}[t]
  \begin{center}
    \includegraphics[width=0.99 \linewidth]{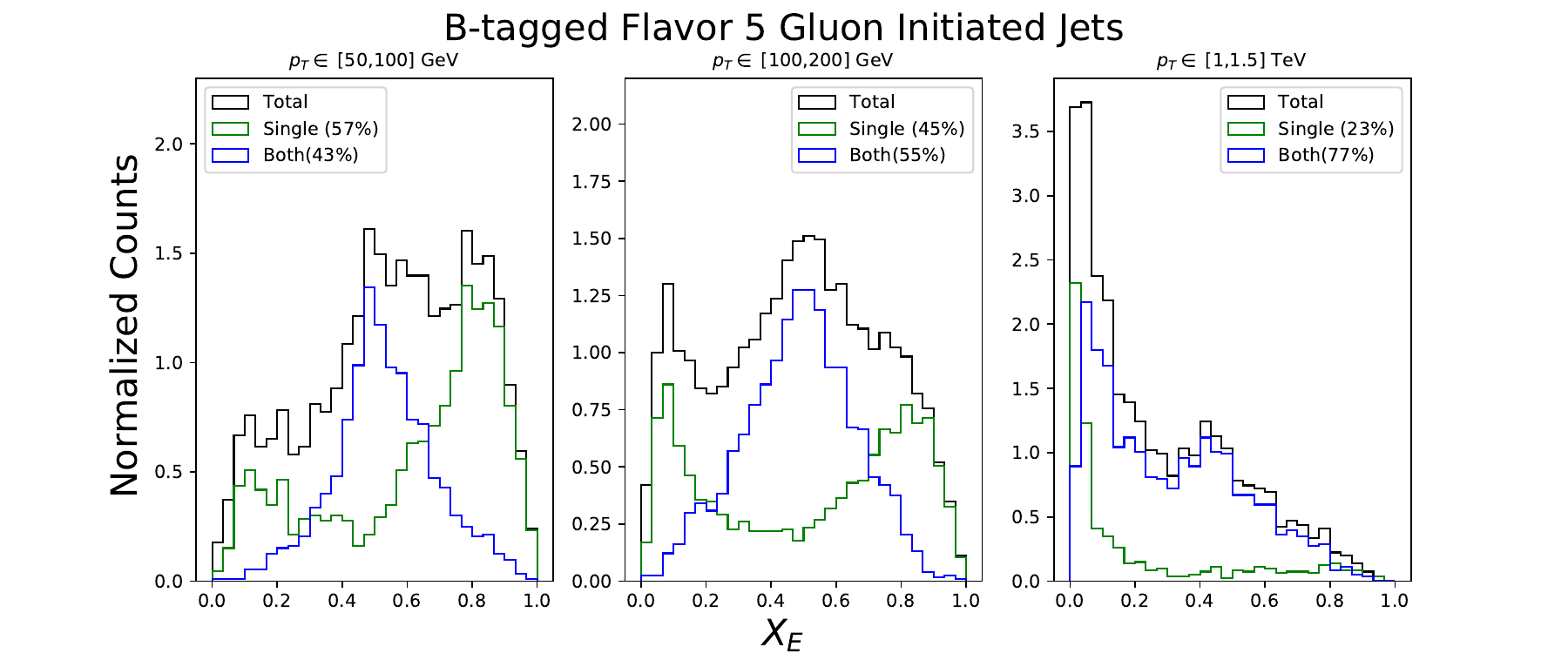}
  \end{center}
  \caption{Normalized $X_E$ distributions for $b$-tagged flavor 5 gluon initiated jets (i.e. jets that originate from a gluon, are $b$-tagged and contain an actual $b$-hadron at some point in the parton shower). Blue and green histograms correspond to events in which only one and both $b$-quarks are found in the shower. In parenthesis we show the fraction of all jets that contain one or both $b$-hadrons.}
  \label{fig:flavor5XE}
\end{figure}

\begin{figure}[t]
  \begin{center}
    \includegraphics[width=0.95 \linewidth]{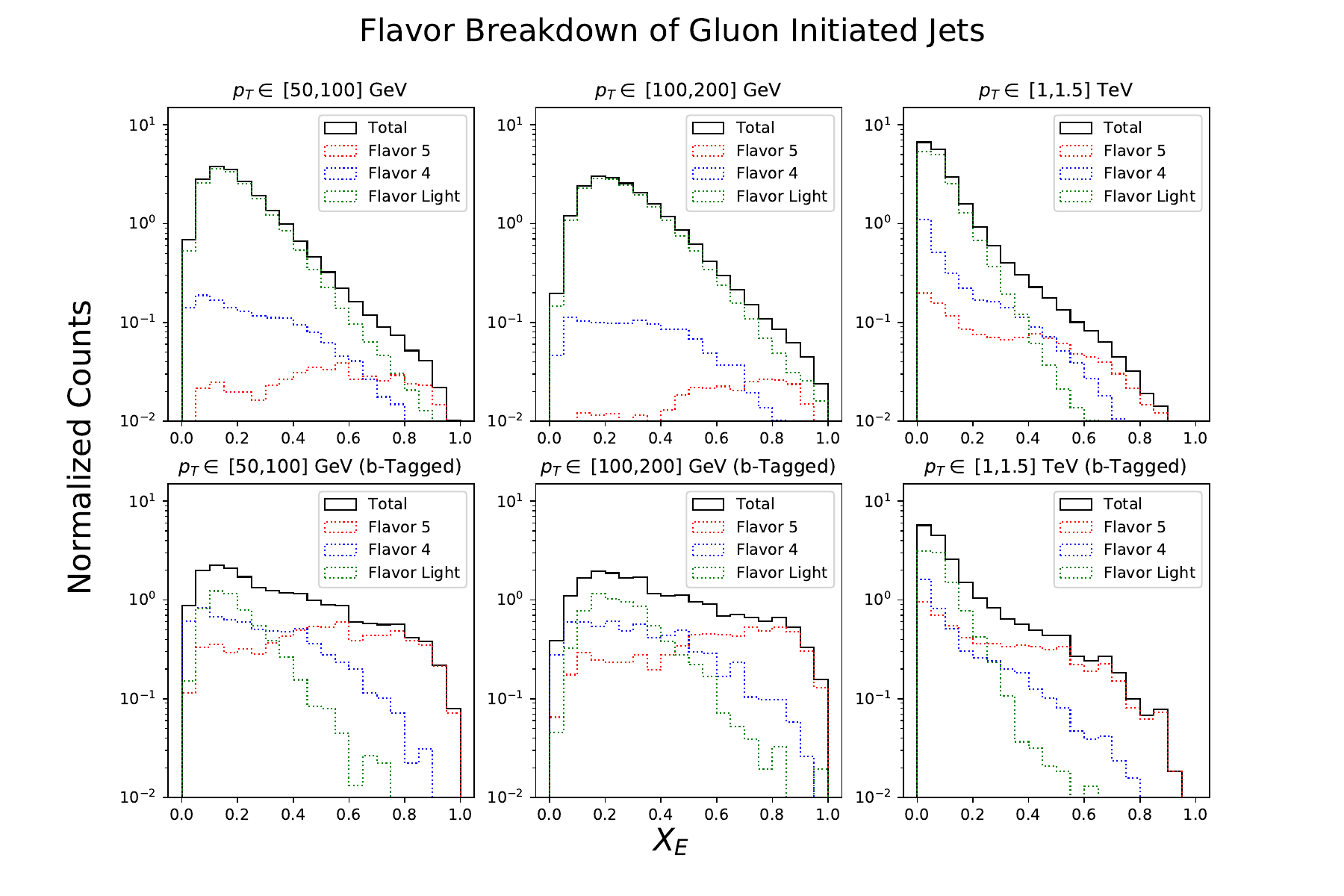}
  \end{center}
  \caption{$X_E$ distribution in gluon initiated jets based on flavor. The distributions are dependent on $p_T$, but for all energies the events which pass $b$-tagging contain proportional amounts of flavors 5,4, and 0. See the text for further details.}
  \label{fig:flavor045}
\end{figure}

\begin{figure}[t]
  \begin{center}
    \includegraphics[width=0.95 \linewidth]{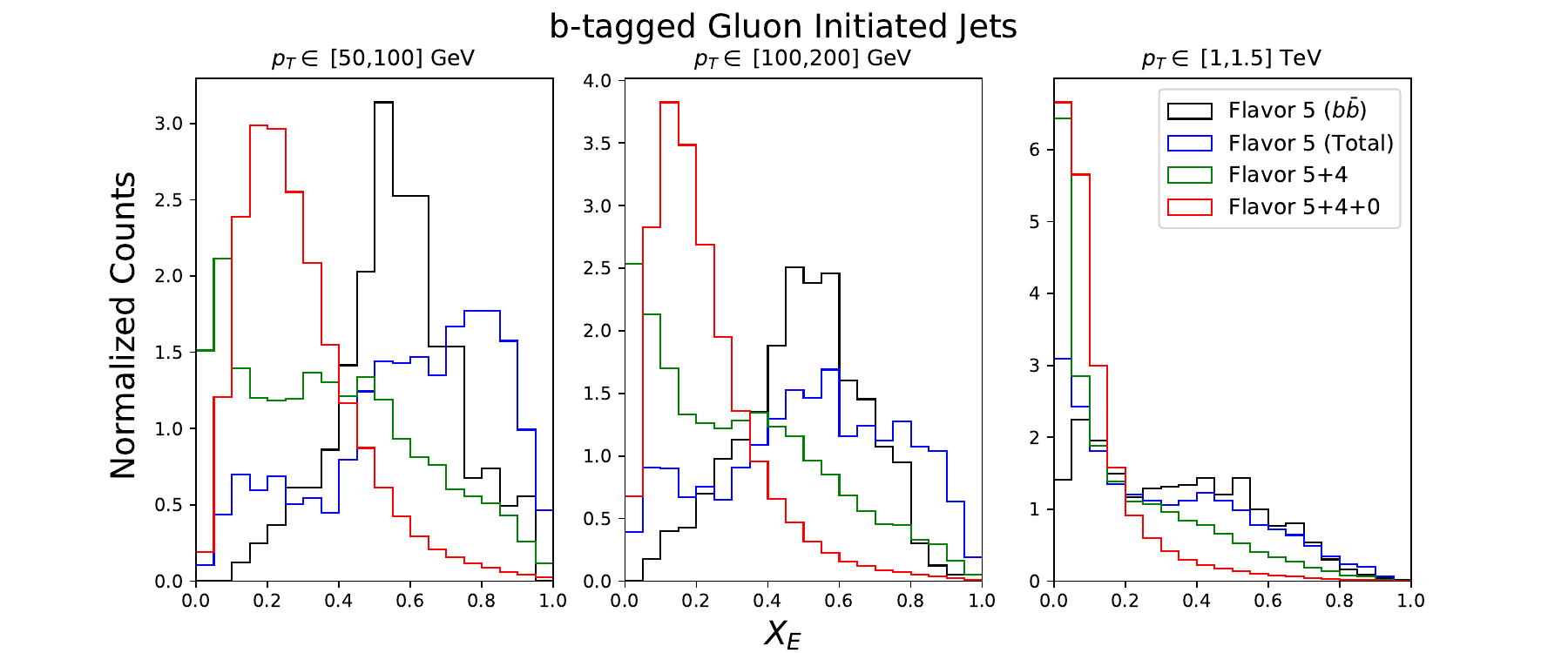}
  \end{center}
  \caption{$X_E$ distributions in $b$-tagged gluon initiated jets. For the black histograms we require both $b\bar b$ to be found in the jet cone; this corresponds to the method used in ref.~\cite{Goncalves:2015prv}. The blue histograms correspond to jets where at least one $b$-quark was found inside the jet. The green histograms correspond to jets where a $b$ or $c$-quark was found inside the jet. The red histograms include jets in which no $b$ or $c$-hadron is found in the jet cone. The $b$-tagging efficiencies for jets containing $b$, $c$ or no heavy quarks are taken from Eqs.~(\ref{eq:BTlight})--(\ref{eq:BTbottom}) for CMS.}
  \label{fig:KraussComp}
\end{figure}

\begin{figure}[t]
  \begin{center}
  \includegraphics[width=0.49 \linewidth]{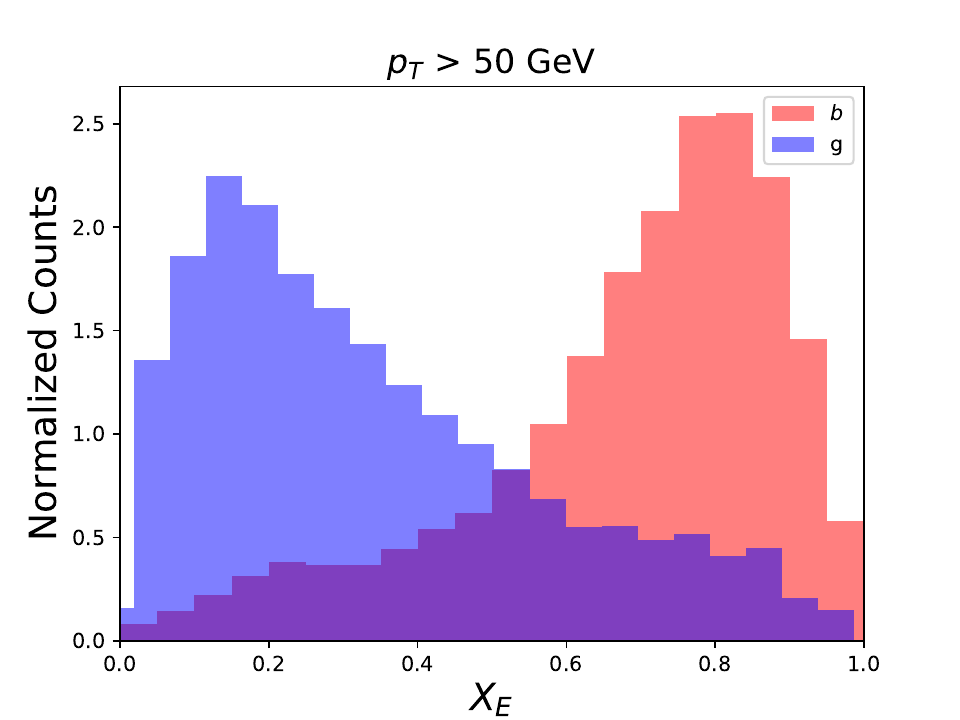}
  \includegraphics[width=0.49 \linewidth]{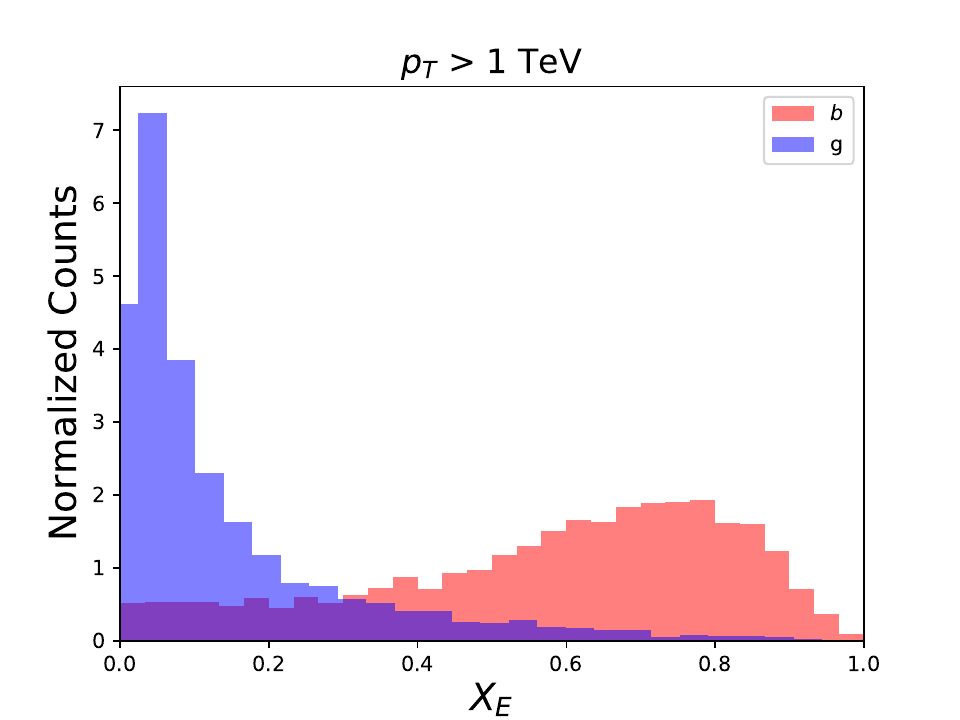}
  \end{center}
  \caption{$X_E$ distributions in gluon and bottom initiated jets for $p_T >50 $ GeV and $p_T >1$ TeV.}
  \label{fig:g_vs_b}
\end{figure}

In figures~\ref{fig:flavor5XE}-\ref{fig:g_vs_b} we present various $X_E$ distributions which we obtain by classifying the jets according to the number of $b$-quarks found in the shower (0, 1, or greater than 2). The purpose of this analysis is to highlight the differences between the observables studied in Ref.~\cite{Goncalves:2015prv} with respect to the setup considered in this work.

In figure~\ref{fig:flavor5XE} we look at gluon initiated jets in which the $b\bar b$ splitting happens at some point in the shower (i.e. at least one bottom quark is found in the cone of the jet). Blue and green histograms correspond to jets with exactly one or two $b$-quarks in a $\Delta R < 0.5$ cone around the jet direction. At relatively low $p_T$ ($p_T\sim (50,100) \; {\rm GeV}$) we find the expected peak at $X_E \sim 0.5$. At large $p_T$ (right panel) the large amount of gluon radiation, described at leading order by the $n_g$ distribution in Eq.~(\ref{eq:ng}), lowers $X_E$ and the peak almost disappears.

In Figure~\ref{fig:flavor045} we illustrate the effects of $b$-tagged gluon jets which do not contain $b$-quarks. We present $X_E$ distributions labeled according to whether $b$'s (flavor 5), $c$'s (flavor 4) or no heavy quarks (flavor 0, light) are found in the shower. Upper panels are prior to $b$-tagging and clearly show the dominant production of flavor 0 jets. The lower panels, in contrast, are after $b$-tagging. For every $p_T$ range, we see that each flavor contributes roughly equal amounts to the final distribution, with flavor 5 jets yielding larger $X_E$ at low $p_T$ (as discussed above). These distributions help explain why the $X_E$ distributions we find yield values of $X_E$ that are much lower than the blue points in figure~\ref{fig:XEbreakdown}.

In figure~\ref{fig:KraussComp} we show multiple $X_E$ distributions. All jets are $b$-tagged according to the CMS algorithm parameterized in Eqs.~(\ref{eq:BTlight})--(\ref{eq:BTbottom}). Black histograms are obtained by requiring the presence of two $b$-hadrons in the jet cone: this corresponds closely to the distributions discusses in Ref.~\cite{Goncalves:2015prv}. Blue histograms include jets for which one of the $b$-hadrons is emitted outside of the jet cone. Green histograms include mistagged jets which contain $c$ but no $b$-hadrons. The red distributions also include mistagged jets which do not contain any heavy hadron and are the distributions upon which we construct the 1$b$2$b$-tagger.

Finally, in figure~\ref{fig:g_vs_b} we show an example of the final distributions that we obtain for bottom and gluon initiated jets. The former peak at $X_E \sim 0.8$ while the latter peak at $X_E \lesssim 0.2$ without any more trace of the peak at $X_E\sim 0.5$.

The two other discriminators we consider, girth ($g$) and charged track multiplicity ($N_{\rm ch}$), are calculated in standard fashion. Girth is the sum of the transverse momentum fractions carried by each jet constituent ($p_T^{(i)}/p_T$) weighted by its angular distance ($\Delta R_{iJ}$) from the jet axis:
\begin{align}
g = \sum_{i\in {\rm jet}}  \frac{p_T^{(i)}}{p_T}  \Delta R_{iJ} \; .
\label{eq:girth}
\end{align}
Gluon initiated jets are expected to have girth distributions that are wider than those for bottom initiated jets. In fact, gluons tend to radiate more than $b$-quarks due to differences in color charges ($N_A=3$ vs $C_F =4/3$) and collinear radiation for the latter is limited by the heavy quark mass. The same argument leads to larger values of $N_{\rm ch}$ for gluon jets.

\section{Construction of the 1$b$2$b$-Tagger}
\label{sec:1b2b}
In this section we present the details of the 1$b$2$b$-tagger which, as already mentioned in the introduction, is designed to provide a further rejection of $b$-tagged gluon initiated jets while retaining most genuine $b$-quark initiated jets.

We consider Monte Carlo samples containing gluon or $b$-quark initiated jets with $p_T$ between 20 GeV and 2 TeV. Events have been generated as discussed at the end of section~\ref{sec:idea}. The three jet sub-structure observables that we consider ($X_E$, $g$ and $N_{\rm ch}$) are calculated as detailed in section~\ref{sec:observables}.

We build the effective tagging variable using a Logistic Regression, a statistical model whose output is the probability of an event occurring~\cite{logistic-regression-hosmer}. We chose this strategy over a straightforward Principal Component Analysis (PCA), because the latter is highly dependent on the variance of the input variables. This requires the data to be standardized to a fixed mean and standard deviation. Standardization is especially important since girth and charged track multiplicity differ dramatically in scale. On the other hand, the logistic regression approach has the advantage of being able to take discrete and continuous inputs which can posses very different scales and variances. This is especially important in our case because $N_{\rm ch}$, in particular, has no typical scale compared to $X_E$ and girth.

Logistic regression algorithm takes as input triplets of $(X_E,g,N_{\rm ch})$ for both signal and background event files. The inputs are combined into a single discriminant $ \beta=\beta_0 + \beta_{X_E} X_E + \beta_g g + \beta_{N_{\rm ch}} N_{\rm ch} $ which is then fed to the sigmoid function $\sigma(\beta) = 1/(1+e^{-\beta})$. An optimization strategy is then used to find a set of coefficients $(\beta_0 , \beta_{X_E}, \beta_g , \beta_{N_{\rm ch}})$ that achieve maximum separation of background and signal events ($\sigma(\beta)$ near 0 and 1, respectively). We used the logistic regression as implemented in the Python package scikit-learn~\cite{scikit-learn} and adopted the Limited-memory Broyden–Fletcher–Goldfarb–Shanno algorithm. The fact that the log-odds (logarithm of the inverse of the sigmoid) is simply a linear combination of the input observables means that the regression is easy to train and implement. And since the coefficients are directly tied to each parameter it's easy to interpret the importance of each parameter. Despite the simplicity of the logistic regression, it is a strong classifier.

Since we expect the dominant variable to be $X_E$ we present the linear combination of the input parameters selected by the logistic regression as:
\begin{align}
X_E^{\rm eff} = X_E + \alpha_{\rm ch}\; N_{\rm ch} +  \alpha_g\; g
\label{eq:Xeff}
\end{align}
The cut-off values for this variable are constructed such that for $X_E^{\rm eff} > X_E^{\rm cut}$ the jet is positively tagged (i.e. is identified as originating from an initial $b$-quark). The efficiency rate for $b$-quark initiated jets and the mistag rate for gluon initiated jets are indicated as $\varepsilon_b$ and $\bar \varepsilon_g$, respectively.

\begin{table}[t]
\begin{center}
\begin{tabular}{|cc|c|c|c|c||c|c|}\hline
  $p_T$ range & $p_T^{\rm avg}$ & $\varepsilon_b = 0.6$ & $\varepsilon_b = 0.7$ & $\varepsilon_b = 0.8$ & $\varepsilon_b = 0.9$ & $\alpha_{\rm ch}$ & $\alpha_g$ \\
  \hline\hline
  20--50 GeV  & 33 GeV & 0.817 & 0.766 & 0.699 & 0.584 & 0.00581 & 0.416 \\
  \hline
  50--100 GeV & 69 GeV & 0.747 & 0.692 & 0.615 & 0.488 & 0.00309 & -0.038 \\
  \hline
  100--200 GeV & 134 GeV & 0.652 & 0.581 & 0.483 & 0.329 & -0.00061 & -0.463\\
  \hline
  200--500 GeV & 263 GeV & 0.551 & 0.469 & 0.356 & 0.182 & -0.00466 & -0.750\\
  \hline
  0.5--1 TeV & 0.61 TeV & 0.482 & 0.391 & 0.269 & 0.114 & -0.0059 & -0.084\\
  \hline
  1--1.5 TeV &  1.15 TeV & 0.490 & 0.404 & 0.287 & 0.126 & -0.00397 & 0.309\\
  \hline
  1.5--2  TeV & 1.66 TeV & 0.481 & 0.389 & 0.275 & 0.121 & -0.00363 & 0.609\\ \hline
\end{tabular}
\end{center}
\caption{Coefficients $\alpha_{\rm ch}$ and $\alpha_g$ of the logistic regression output ($X_E^{\rm eff}$) and cut-off values $X_E^{\rm cut}$ for several signal efficiency working points ($\varepsilon_b \in[ 0.6,0.9]$).}
\label{tab:logreg}
\end{table}

\begin{table}[t]
\begin{center}
\begin{tabular}{|c|c|c|c|c|}\hline
  $p_T$ range & $\varepsilon_b = 0.6$ & $\varepsilon_b = 0.7$ & $\varepsilon_b = 0.8$ & $\varepsilon_b = 0.9$\\
  \hline\hline
  20--50 GeV  & 0.123 & 0.16 & 0.212 & 0.321 \\
  \hline
  50--100 GeV & 0.117 & 0.15 & 0.206 & 0.332 \\
  \hline
  100--200 GeV & 0.104 & 0.142 & 0.216 & 0.365 \\
  \hline
  200--500 GeV & 0.08 & 0.119 & 0.195 & 0.348 \\
  \hline
  0.5--1 TeV & 0.057 & 0.094 & 0.157 & 0.265 \\
  \hline
  1--1.5 TeV & 0.038 & 0.058 & 0.102 & 0.193 \\
  \hline
  1.5--2  TeV & 0.024 & 0.041 & 0.072 & 0.143 \\ \hline
\end{tabular}
\end{center}
\caption{Mistag rate for gluon initiated $b$-tagged jets ($\bar{\varepsilon}_g$), for fixed signal efficiencies ($\varepsilon_b$).}
\label{tab:ebarg}
\end{table}

\begin{figure}[t]
\begin{center}
  \includegraphics[width=0.48 \linewidth]{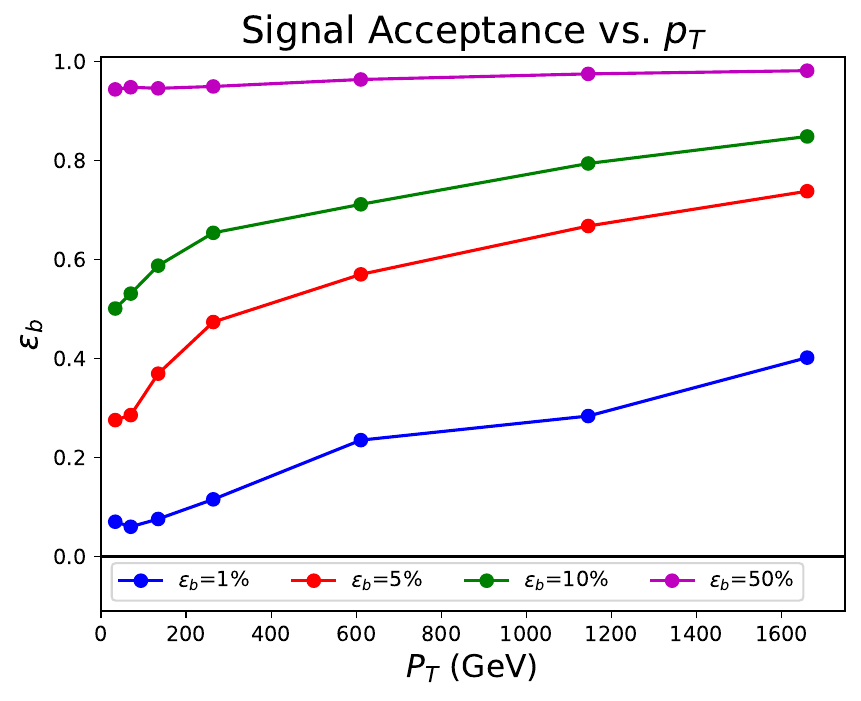}
  \includegraphics[width=0.49 \linewidth]{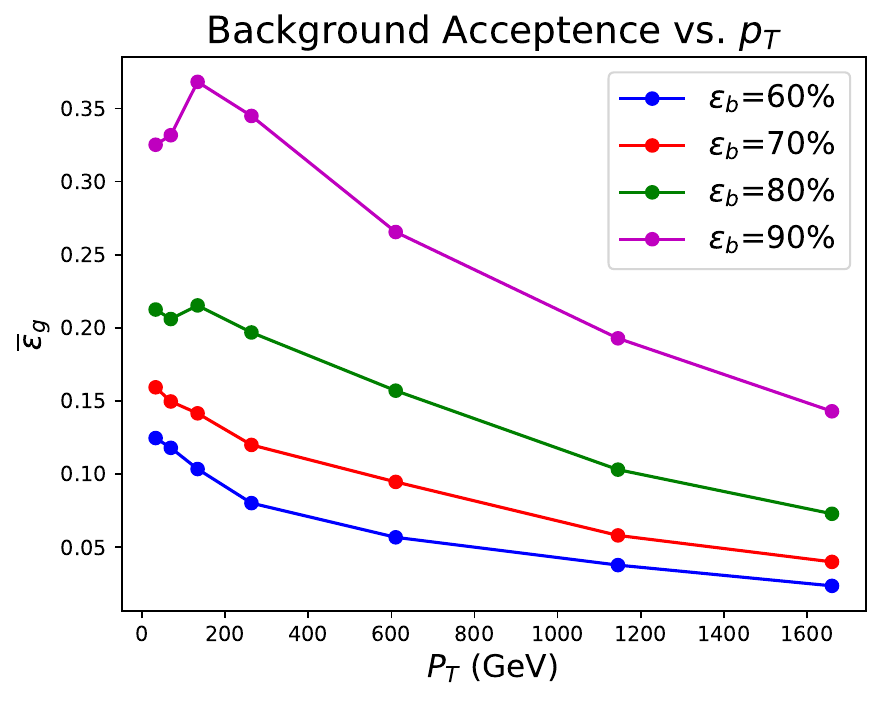}
  \includegraphics[width=0.49 \linewidth]{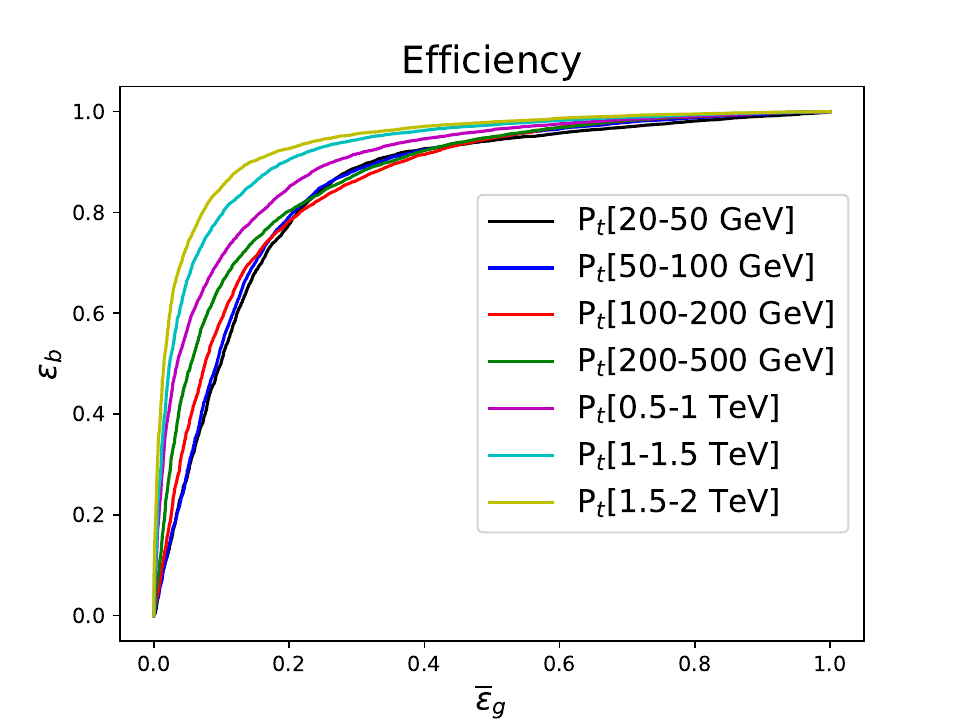}
\end{center}
\caption{Tag and mistag rates as a function of $p_T$. Top left panel: signal tag rate ($\varepsilon_b$) as function of $p_T$ for various values of the background rejection rate ($0.01 \leq \bar\varepsilon_g  \leq 0.5$). Top right panel: background mistag rate ($\bar\varepsilon_g$) as a function of $p_T$ for various values of the background rejection rate ($0.6 \leq \varepsilon_g  \leq 0.9$). Lower panel: signal tag rate as a function of the background mistag rate for various values of $p_T\in [20 ,2000] \; {\rm GeV}$.}
\label{fig:summary}
\end{figure}

In table~\ref{tab:logreg} we present the coefficients of $X_E^{\rm eff}$ and the cut-off values corresponding to various signal efficiencies $\varepsilon_b$ for various jet $p_T$ ranges. In table~\ref{tab:ebarg}, which is one of the main results of this analysis, we present the corresponding gluon mistag rates $\bar\varepsilon_g$. These results are presented graphically in figure~\ref{fig:summary}. The top two panels present the dependence of the signal tag and background mistag rates as a function of $p_T$. The bottom panel shows the Receiver Operating Characteristic (ROC) curves, namely the tag vs mistag rates of $b$ and $g$ initiated jets.

Details about the one and two dimensional distributions of the input variables $X_E$, g, and $N_{\rm ch}$ for different jet transverse momenta are presented in Appendix~\ref{sec:appendix} (figures~\ref{fig:50}-\ref{fig:2000}). From these distributions it is clear that $X_E$ is the dominant discriminator and that the tagger becomes more powerful at larger values of the jet $p_T$.

\section{Application of the 1$b$2$b$  Tagger to multi-jet final states}
\label{sec:multijet}
An important application of the tagger we are proposing is the suppression of SM background to new physics signals with multiple (typically more than four) prompt $b$-quarks. Under these circumstances requesting the presence of multiple 1$b$2$b$-tagged jets (which have all been previously $b$-tagged), yields extremely tiny survival probabilities. For instance, starting with a combination of four jets events originating from a combination of gluons and light quarks with $\Delta R > 0.5$ and $p_T > 20$ GeV, the requirement of four 1$b$2$b$-tagged jets reduces the cross section by a factor of about $10^6$. Simulations under these conditions are clearly unmanageable. In this section we explore how well a simple re-weighting of the events using the $p_T$ dependent efficiencies presented in table~\ref{tab:ebarg} reproduces the actual implementation of the tagger on an event-by-event basis.

\begin{figure}[t]
\begin{center}
  \includegraphics[width=0.49 \linewidth]{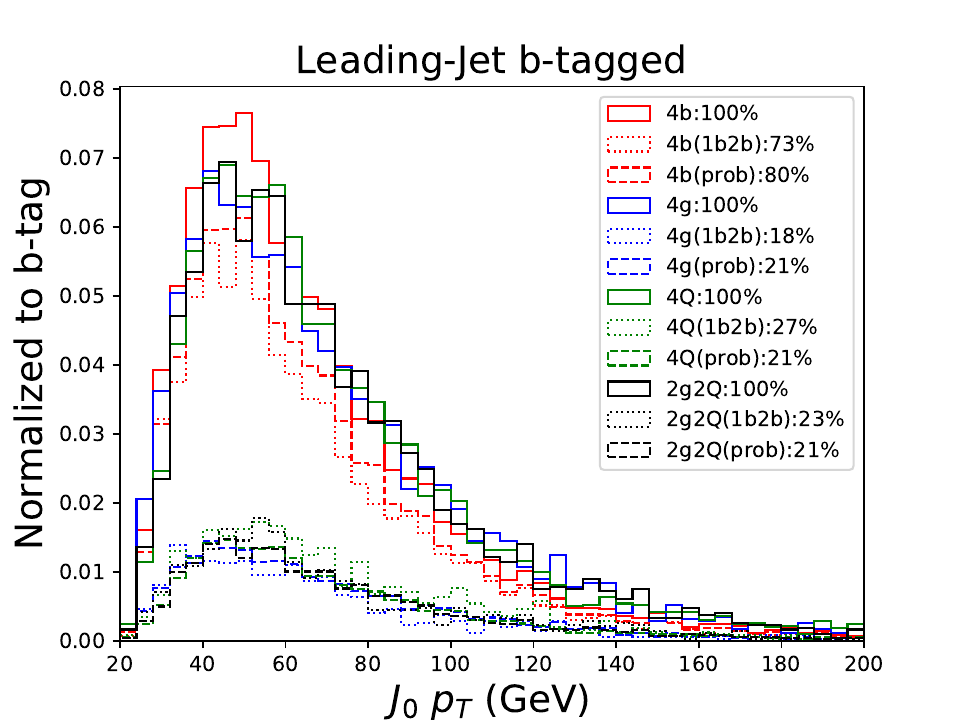}
  \includegraphics[width=0.49 \linewidth]{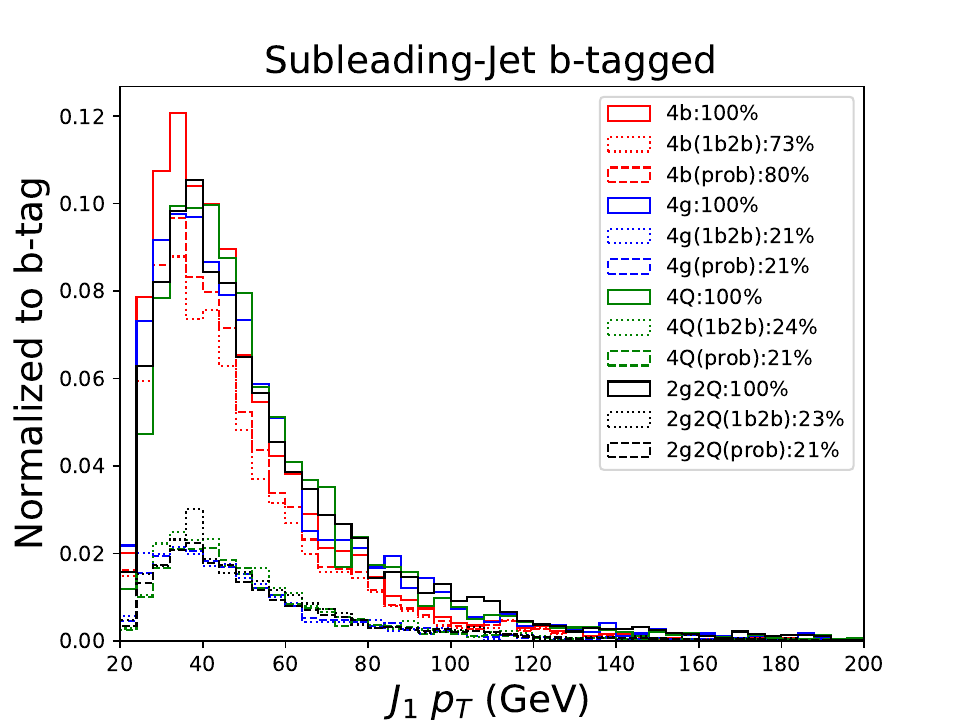}
\end{center}
\caption{Distributions of the leading (left) and subleading (right) jet $p_T$. Initially with only the $b$-tagging from Delphes (solid). Then applying the 1$b$2$b$-tagger directly (dotted) or as a probability (dashed). All distributions are normalized to the $b$-tag distribution to better show the impact of the 1$b$2$b$-tagger}
\label{fig:J01Apl}
\end{figure}

We generated a large sample ($5\times 10^5$ events) of $pp \to (4b, 4g, 2g2q, 4q)$ events at $\Delta R_{gg} > 0.5$ and $p_T > 20$ GeV ($q = u,d,s,c$). As a first step we checked whether the leading or subleading jet 1$b$2$b$-tagging is well reproduced by the efficiencies in table~\ref{tab:ebarg}. The results of this test are presented in the two panels of figure~\ref{fig:J01Apl}, where solid lines are normalized distributions after $b$-tagging. The application of the 1$b$2$b$-tagger is performed either exactly (dotted lines) or using the efficiencies in table~\ref{tab:ebarg} (dashed lines). It is clear that both for the leading and subleading jet 1$b$2$b$-tagger is very accurately reproduced by use of the efficiency table.

\begin{figure}[t]
\begin{center}
  \includegraphics[width=0.49 \linewidth]{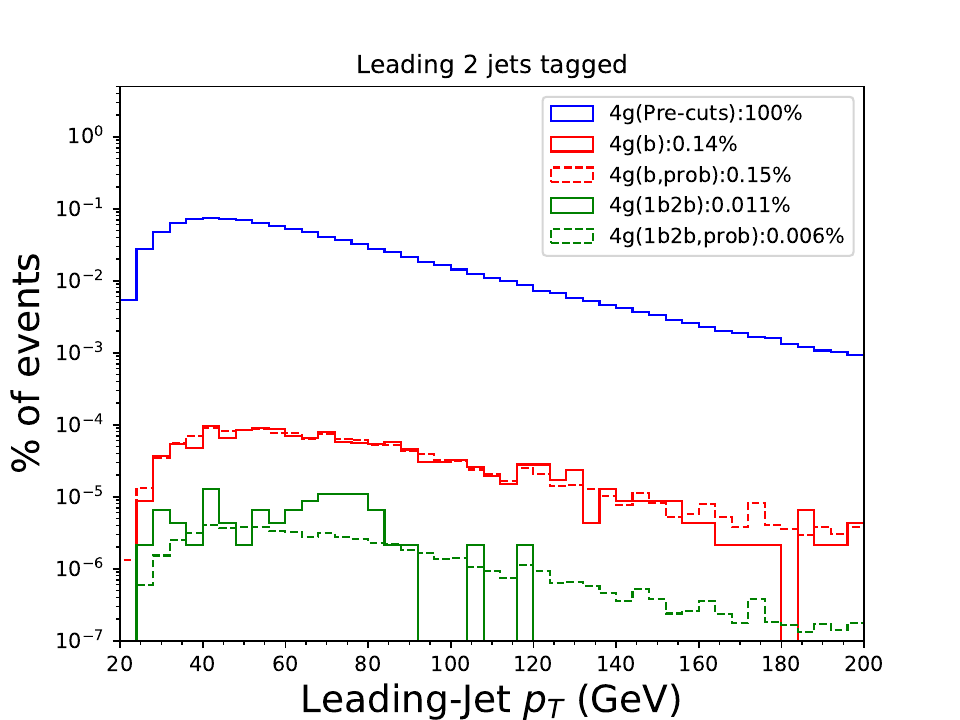}
  \includegraphics[width=0.49 \linewidth]{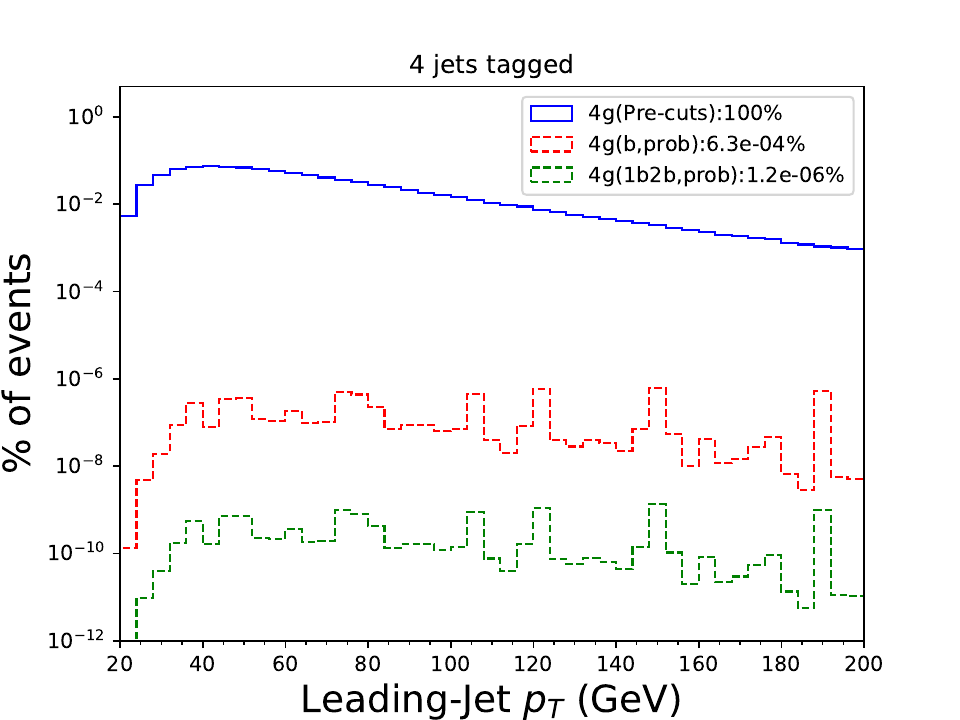}
\end{center}
\caption{In the left panel we present the $p_T$ distribution of the leading jet in $pp\to 4g$ events generated as described in the text and for which the leading two jets are both $b$ and 1$b$2$b$-tagged. The blue, red and green solid lines correspond to the initial selection, the impact of double $b$-tagging and of the additional double 1$b$-2$b$-tagging, respectively. The red and green dashed lines are obtained by reweighting the events using the CMS $b$-tagging efficiencies in Eqs.~(\ref{eq:BTlight})--(\ref{eq:BTbottom}) and the 1$b$2$b$-tagging efficiencies in table~\ref{tab:ebarg}. In the right panel, we request four $b$ and 1$b$2$b$-tagged jets: in this case we can only adopt the reweighting method because of the very low overall efficiencies.}
\label{fig:J24Applied}
\end{figure}

In the left panel of figure~\ref{fig:J24Applied} we restrict the histograms to the $pp\to 4g$ case and impose the 1$b$2$b$ requirement on leading two and four jets in each event (left and right panels, respectively). The solid blue lines are the normalized distributions prior to $b$-tagging, red lines show the impact of $b$-tagging and green lines that of 1$b$2$b$-tagging. When considering the leading two jets only (left panel), we were able to compare the exact implementations (solid red and green lines) with the approximate one (dashed lines). It is clear that the use of the efficiency tables perfectly simulates the 1$b$2$b$-tagger even when imposed simultaneously on two jets. In the right panel, only the approximate tagging is displayed because the low probabilities would have required several million events to obtain reasonable distributions.
The strength of the probabilistic application of the taggers is most evident when requiring multiple jets to be tagged. It not only maintains the same cross section as the exact taggers, but also the overall distribution of the data. Whereas the exact tagger will no longer show the distribution when there are too few events.

\section{Conclusions}
\label{sec:conclusions}
In this paper we studied the properties of $b$-tagged jets originating from prompt $b$-quarks, light quarks as well as gluons and proposed a strategy to build a tagger able to isolate the former and reject the latter two. This work builds on the approach of ref.~\cite{Goncalves:2015prv} and is motivated by the need to reject multi-jet backgrounds to new physics signals consisting, at the parton level, of multiple $b$ and $t$-quarks. When top quarks are presents, search strategies requiring leptons and missing energy are possible but suffer from smaller branching ratios due to leptonic decays of the tops (see, for instance, ref.~\cite{Han:2018hcu}). On the other hand, for purely hadronic multi-$b$ decays (for instance, in ref.~\cite{Dermisek:2020gbr}, models with extra Higgses and vectorlike fermions yield final states with up to 6 $b$-quarks), experimental searches require a rejection of multi-jet QCD backgrounds beyond what conventional $b$-taggers allow.

Both ATLAS~\cite{atlas:2023} and CMS~\cite{cms:2015} have studied tagging strategies focused on jets with two $b$-quarks in connection to isolating heavy color neutral resonances decaying to $b\bar b$ (e.g. heavy Higgses). One of the main discriminants in this case is the fraction of the total energy found in {\it all} secondary vertices: the propensity of gluons to radiate usually leads to $b\bar b$ pairs carrying a small fraction of the total jet energy.

We focus on three jet substructure variables (girth, charged track multiplicity and the fraction of the jet energy in the leading $b$-hadron candidate) and show how it is possible to construct a 1$b$2$b$-tagger that leads to typical gluon rejection efficiencies in the 7--20\% range (with the stronger rejection corresponding to higher $p_T$ jets), while allowing roughly $80\%$ of prompt $b$-jets.

We show that this tagger works in multi-jet final states but when generating backgrounds for which more than two jets are required to be 1$b$2$b$-tagged, it is necessary to simulate the impact of the tagger by rescaling event weights using the efficiencies we present in table~\ref{tab:ebarg}. The impact of this tagger on experimental studies of explicit new physics models with final states involving multiple $b$-quarks will be presented in a forthcoming publication.

\appendix
\section{Additional figures}
\label{sec:appendix}

\begin{figure}[H]
\begin{center}
  \includegraphics[width=0.32 \linewidth]{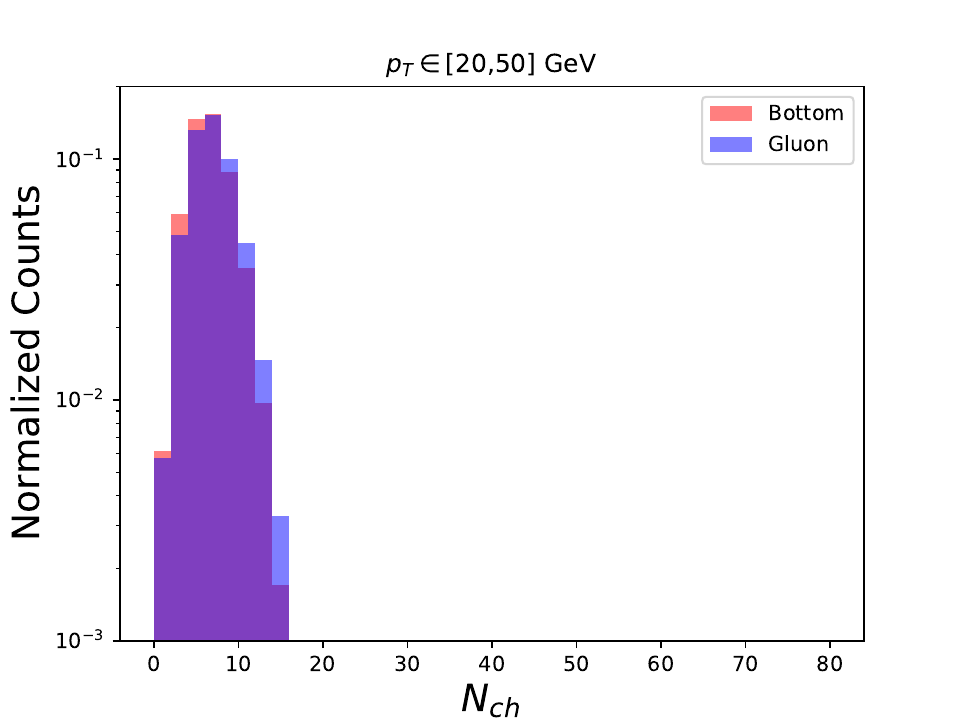}
  \includegraphics[width=0.32 \linewidth]{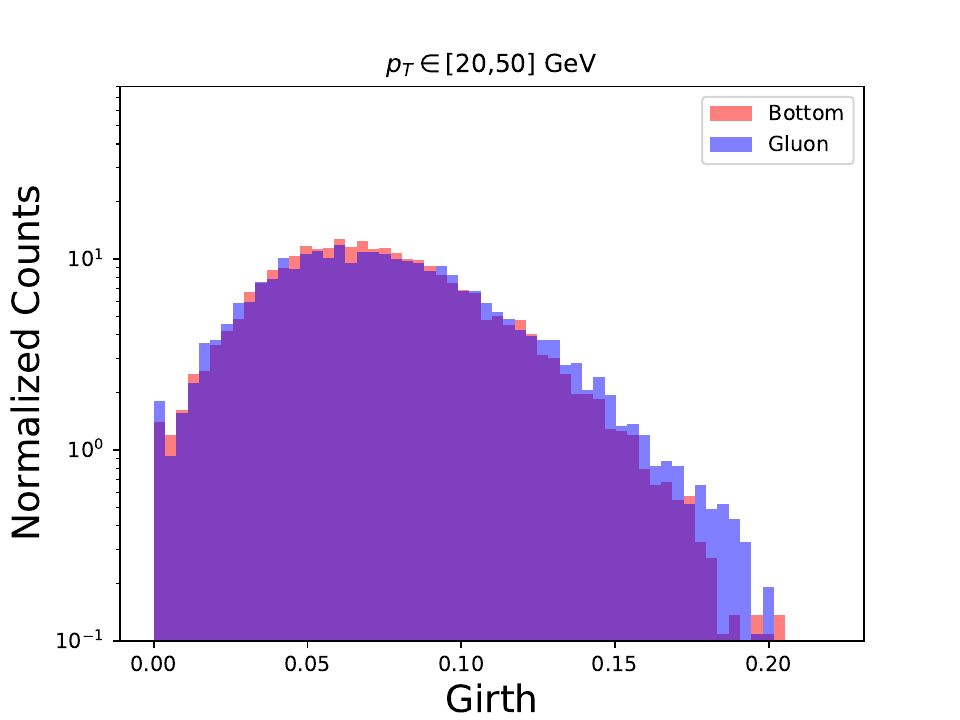}
  \includegraphics[width=0.32 \linewidth]{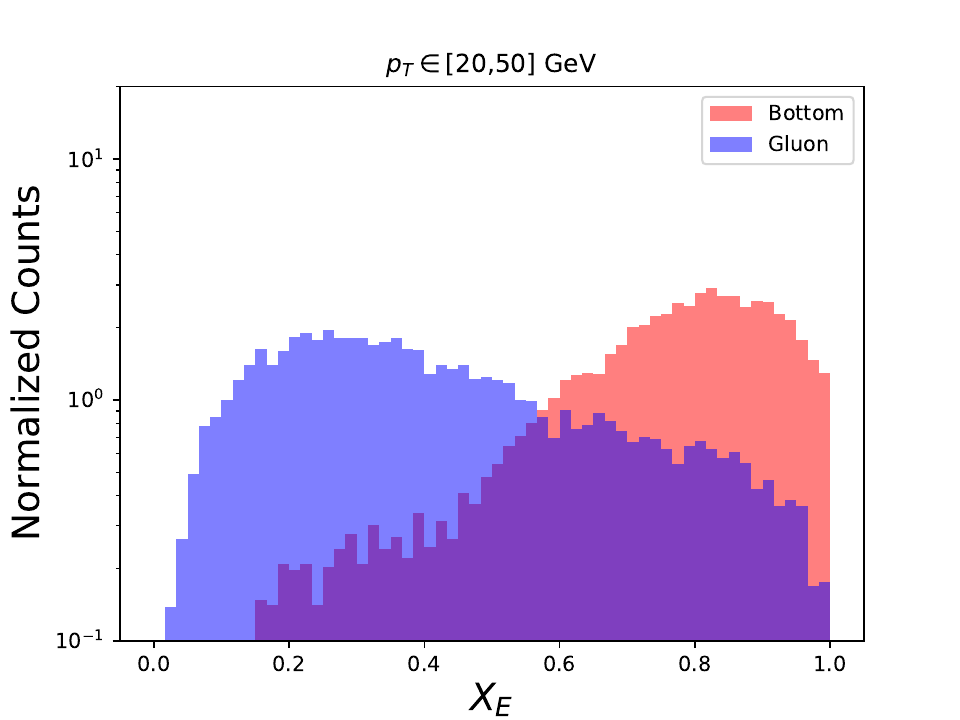}
  \includegraphics[width=0.32 \linewidth]{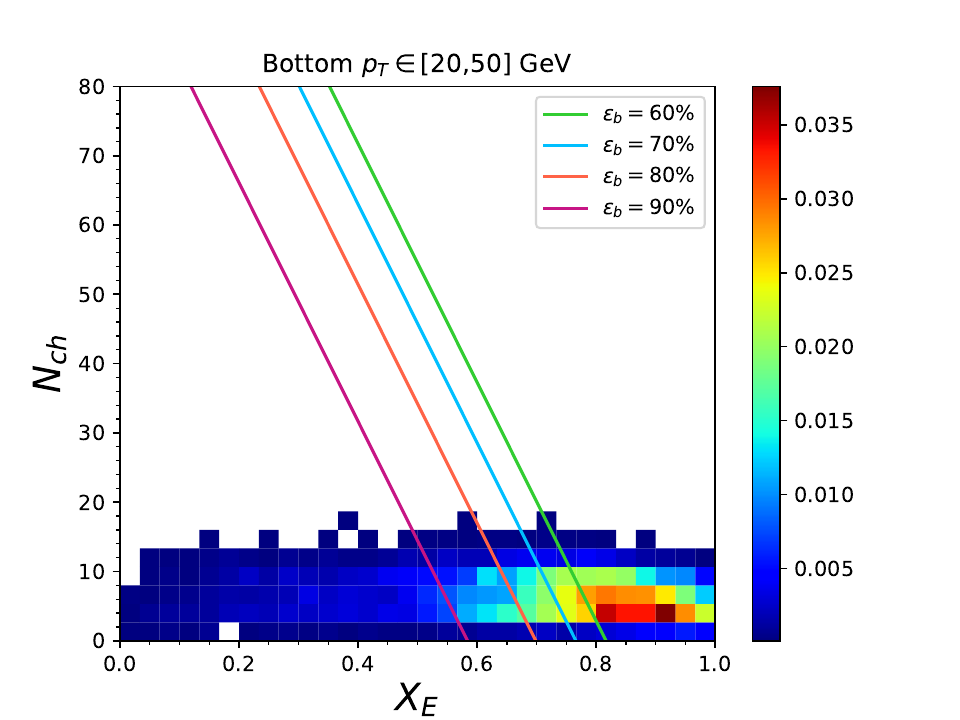}
  \includegraphics[width=0.32 \linewidth]{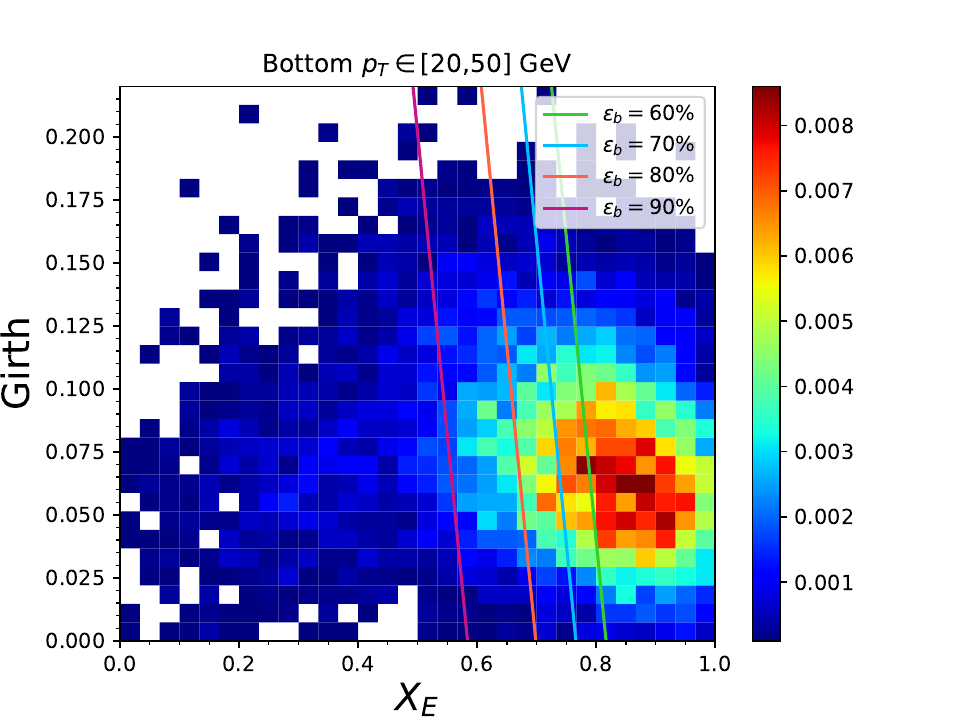}
  \includegraphics[width=0.32 \linewidth]{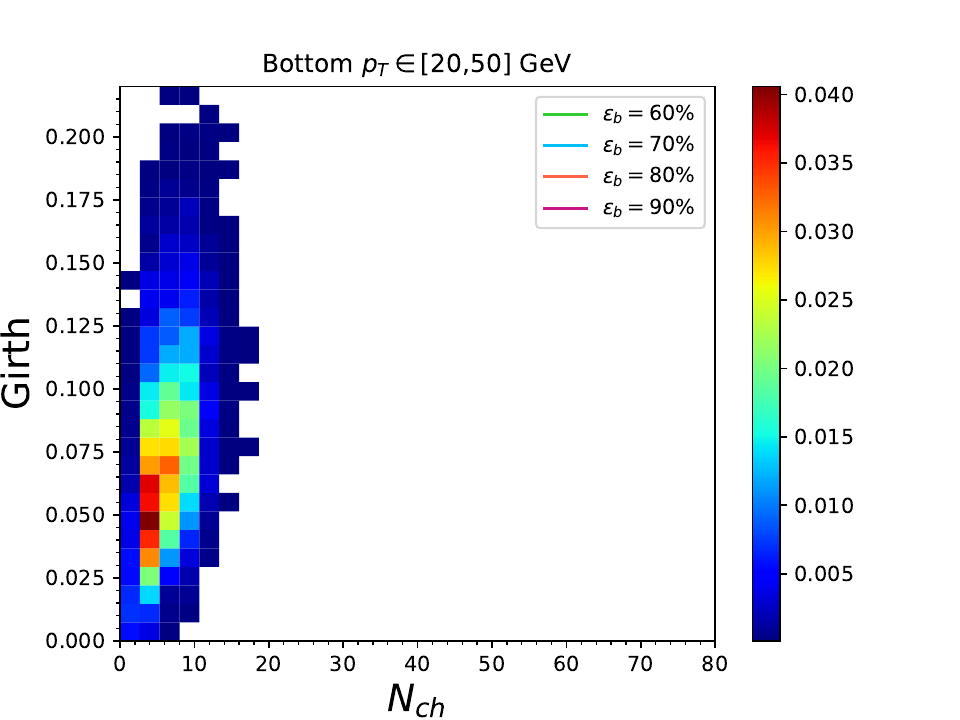}
  \includegraphics[width=0.32 \linewidth]{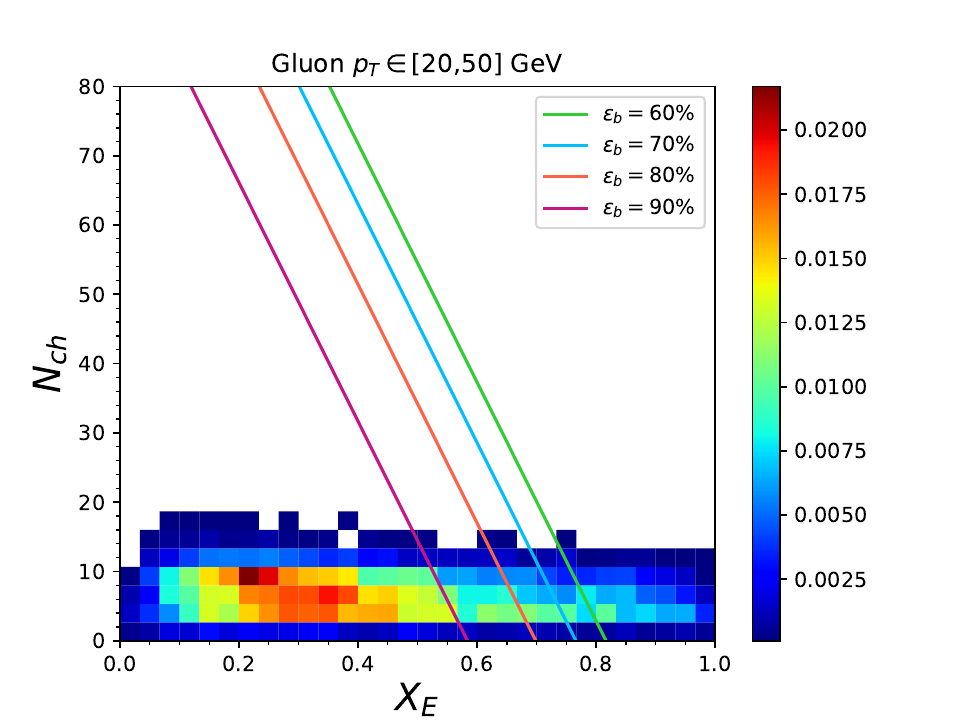}
  \includegraphics[width=0.32 \linewidth]{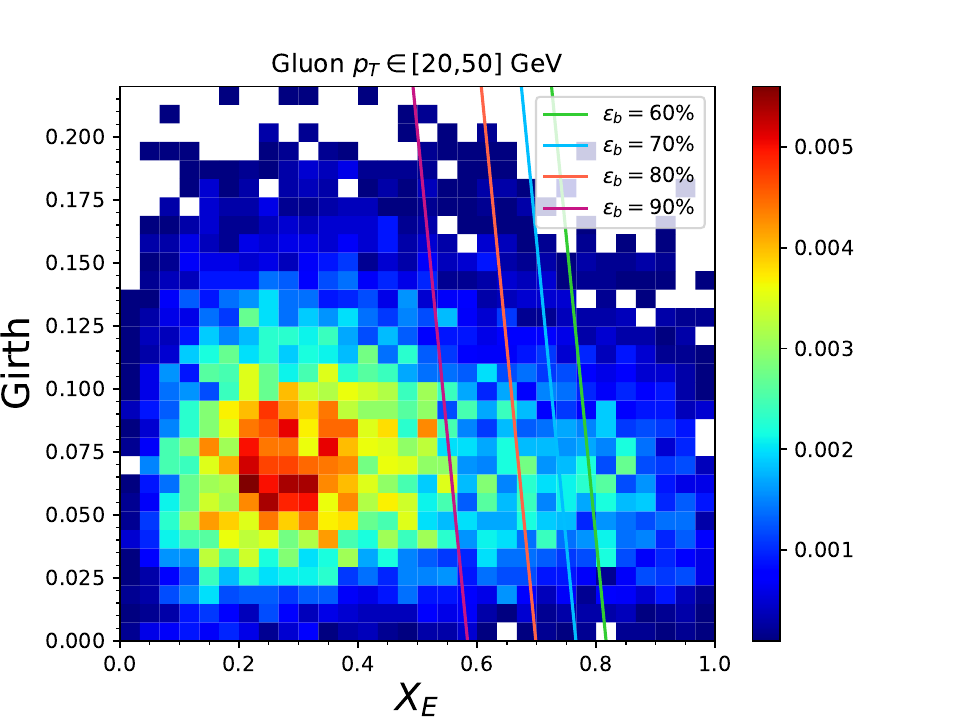}
  \includegraphics[width=0.32 \linewidth]{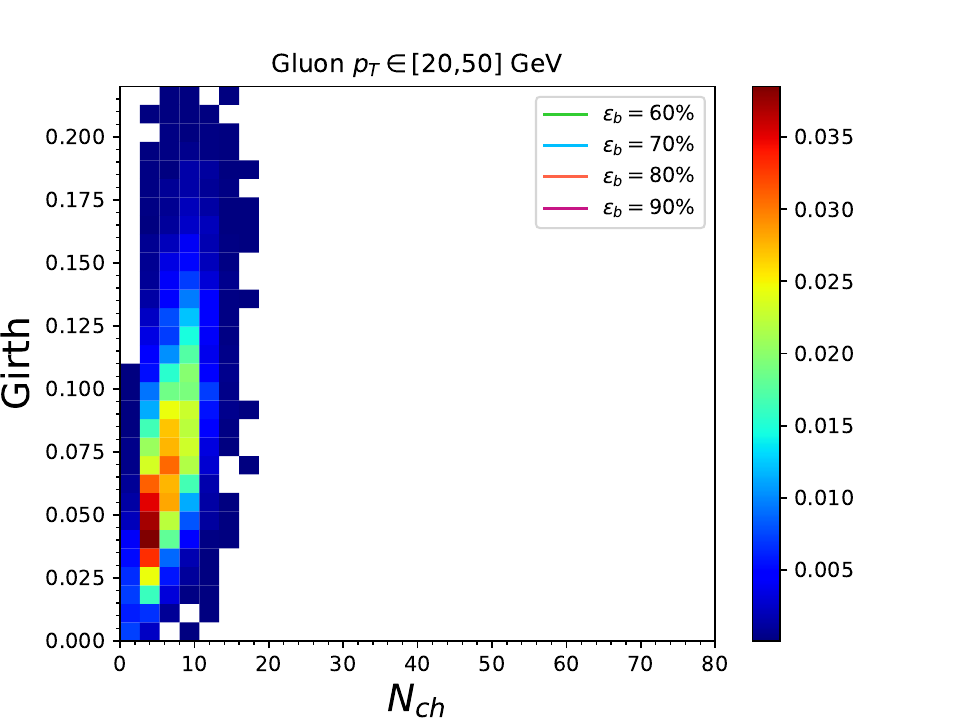}
  \includegraphics[width=0.32 \linewidth]{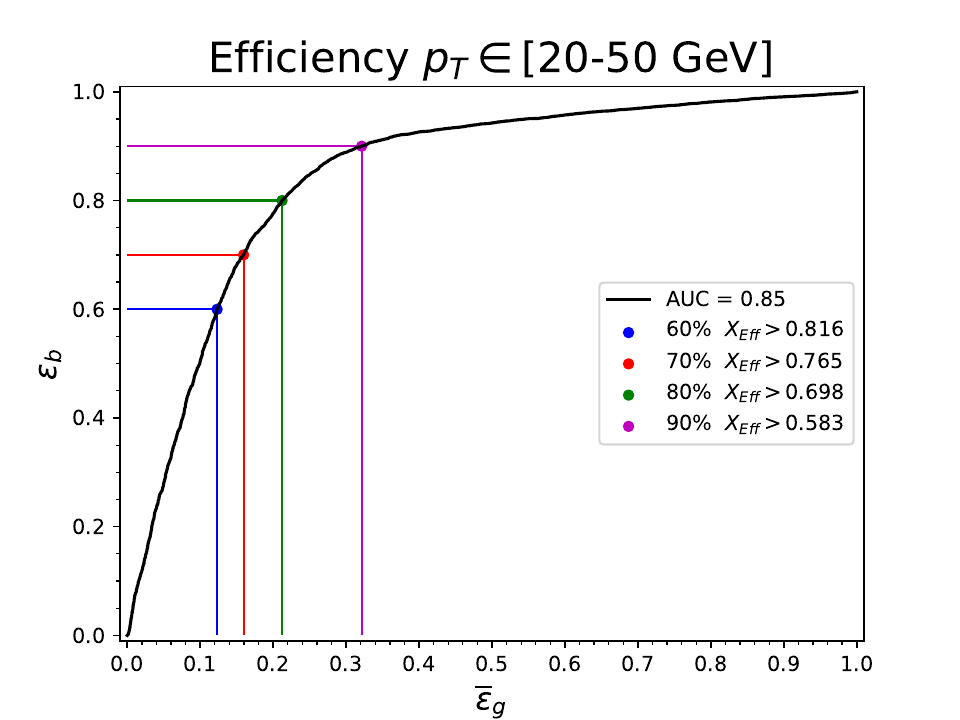}
\end{center}
\caption{Charged track multiplicity, girth and $x_E$ for jets with $p_T\in[20,50] \; {\rm GeV}$.
}
\label{fig:50}
\end{figure}

\begin{figure}[H]
\begin{center}
  \includegraphics[width=0.32 \linewidth]{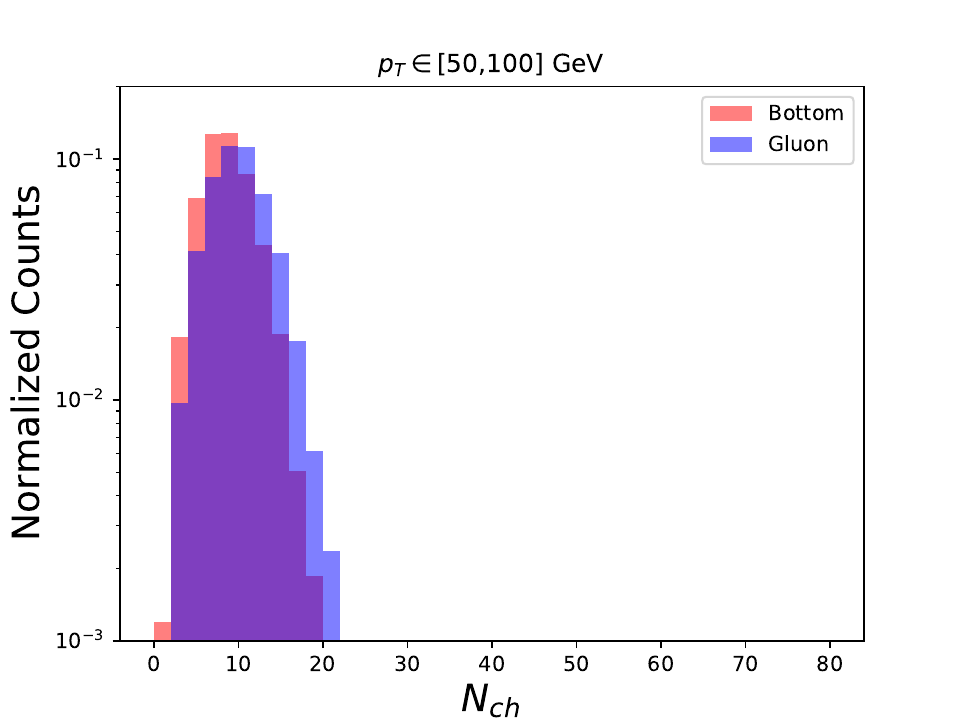}
  \includegraphics[width=0.32 \linewidth]{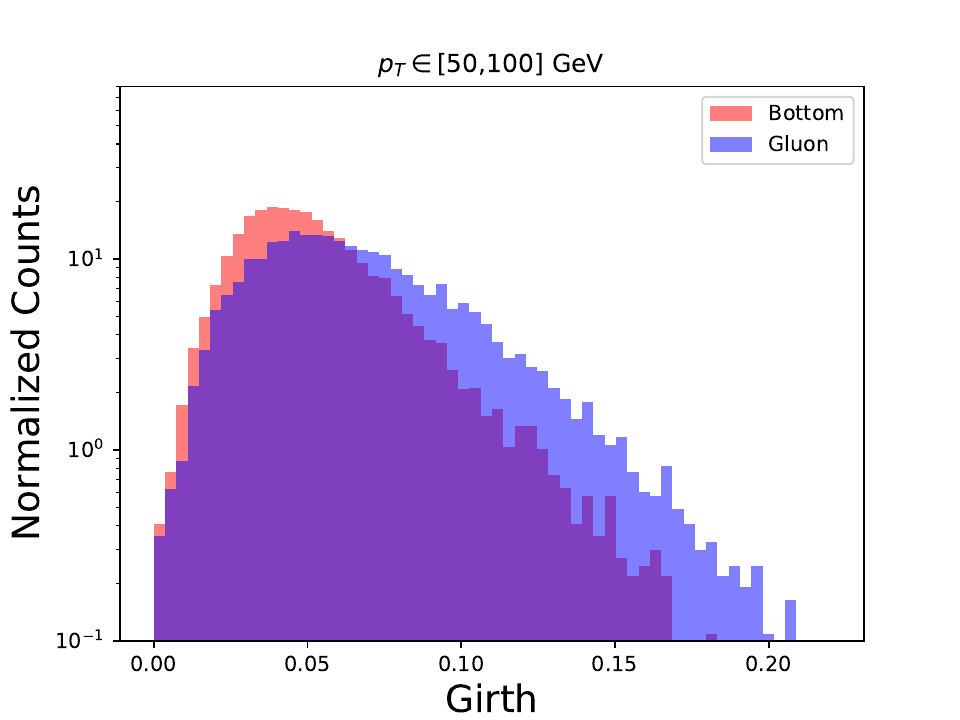}
  \includegraphics[width=0.32 \linewidth]{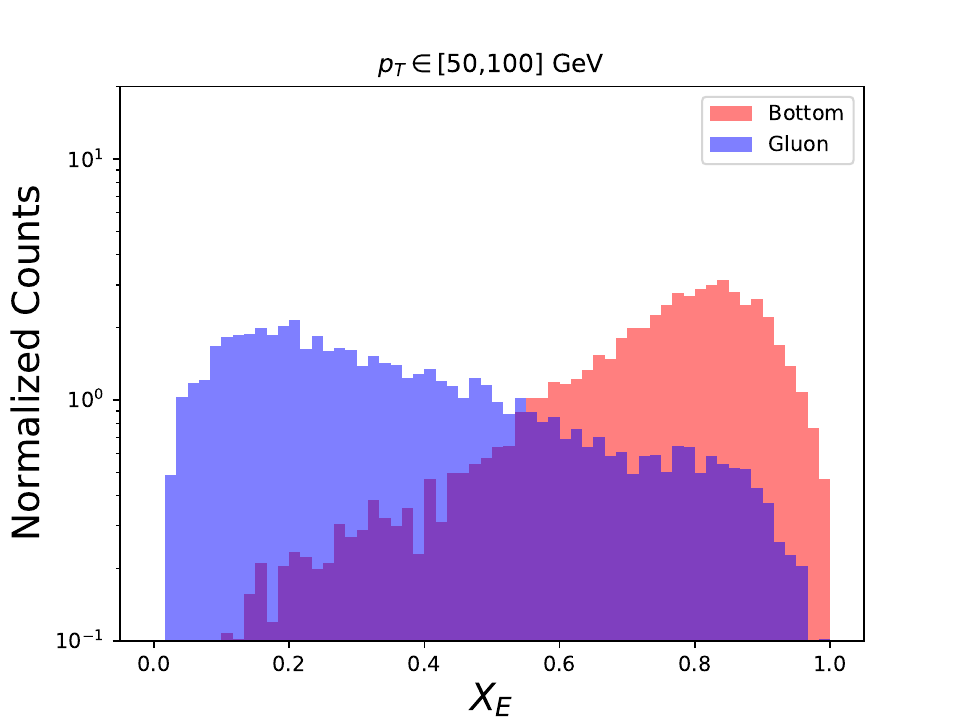}
  \includegraphics[width=0.32 \linewidth]{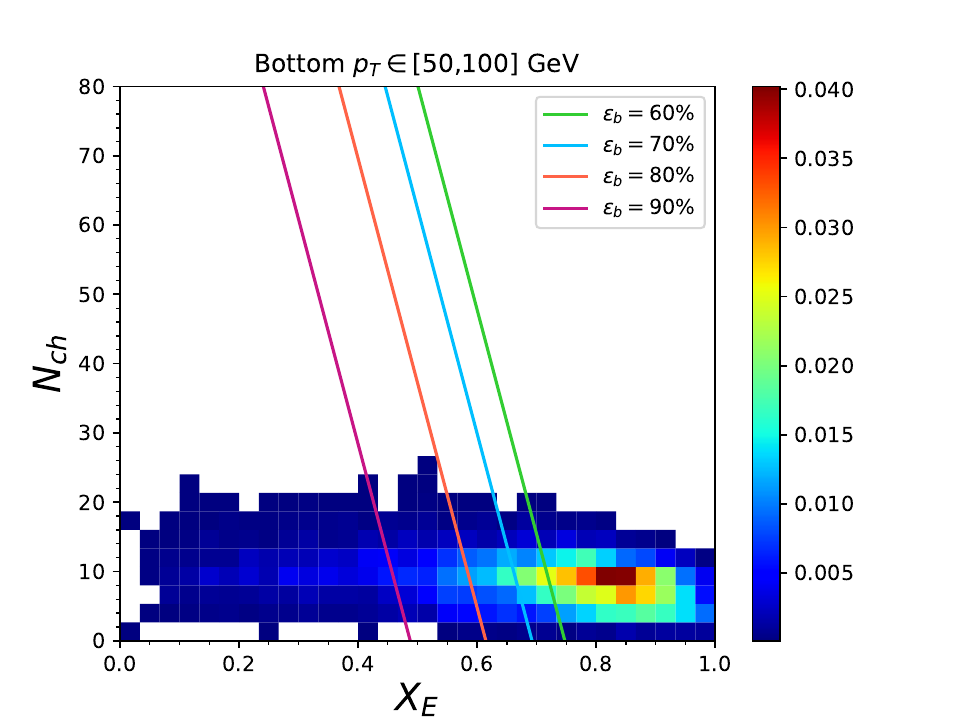}
  \includegraphics[width=0.32 \linewidth]{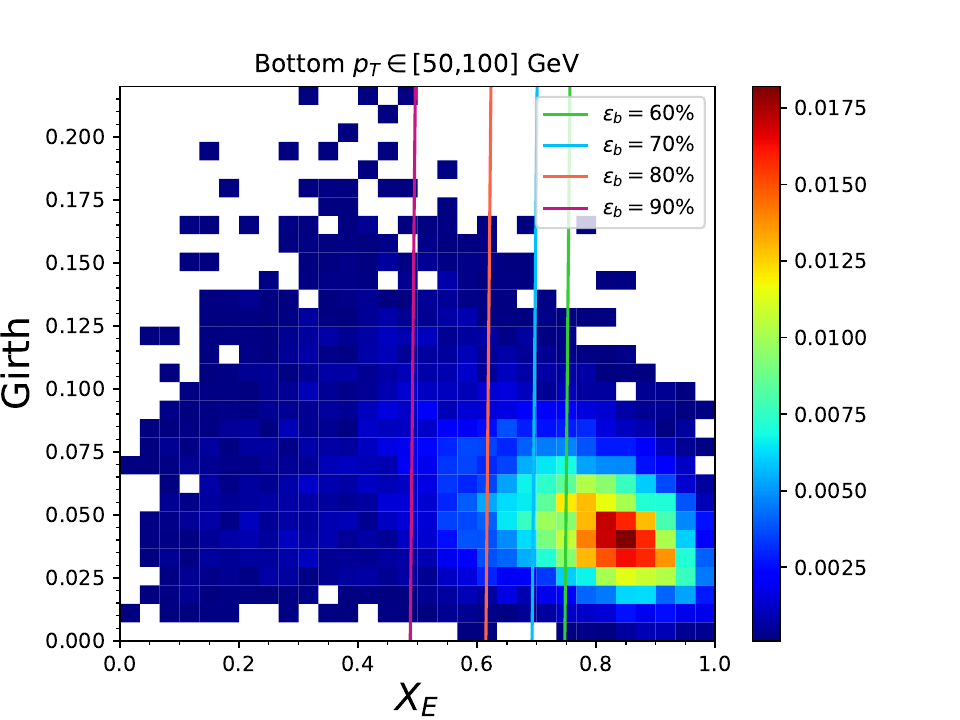}
  \includegraphics[width=0.32 \linewidth]{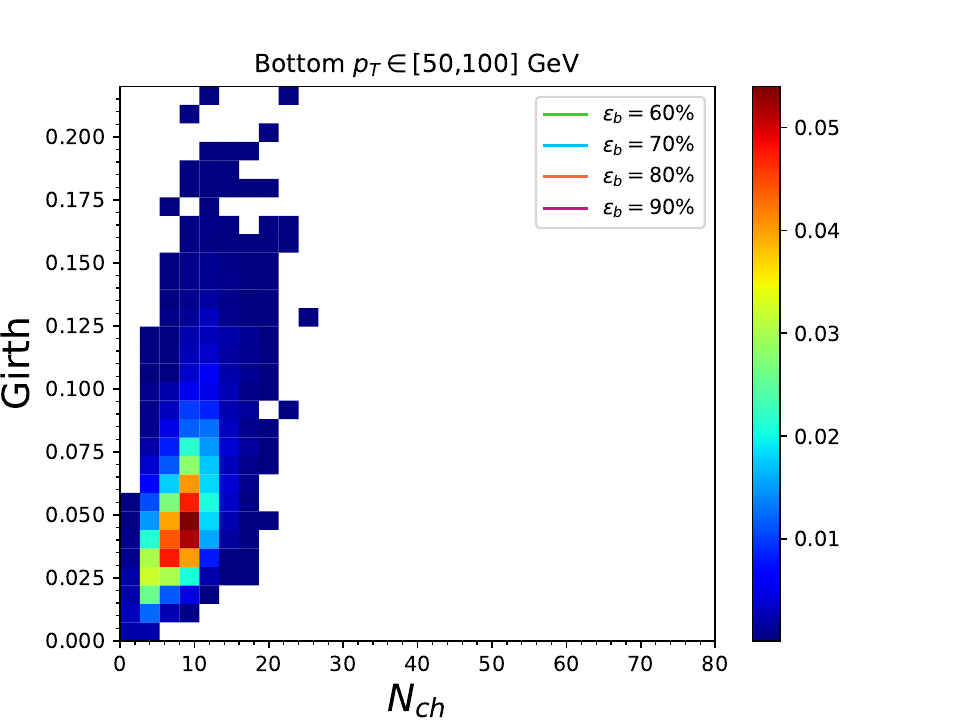}
  \includegraphics[width=0.32 \linewidth]{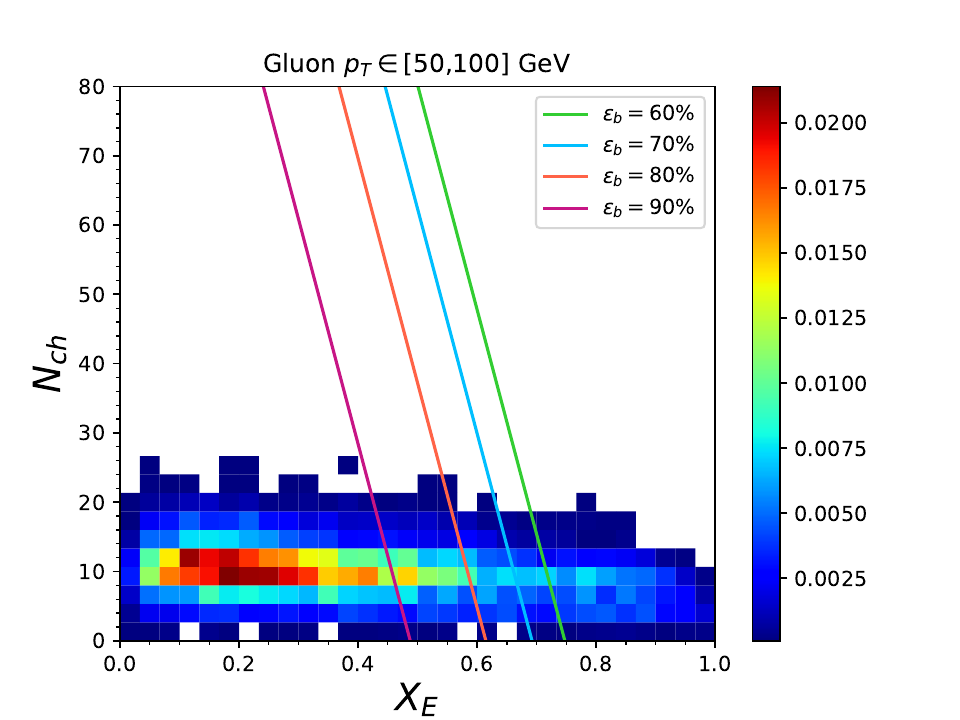}
  \includegraphics[width=0.32 \linewidth]{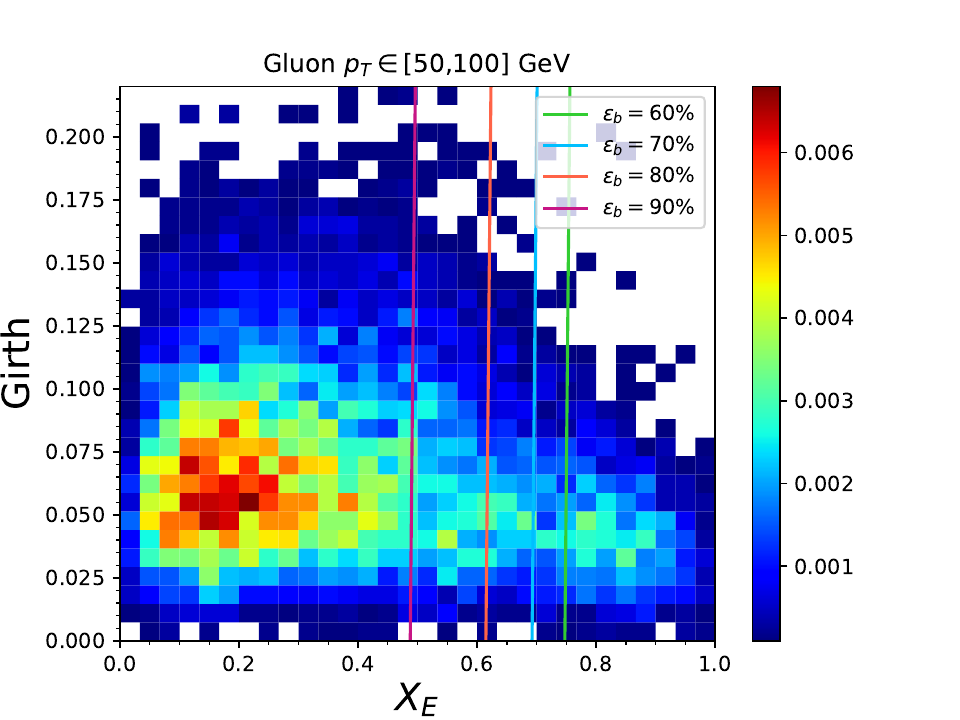}
  \includegraphics[width=0.32 \linewidth]{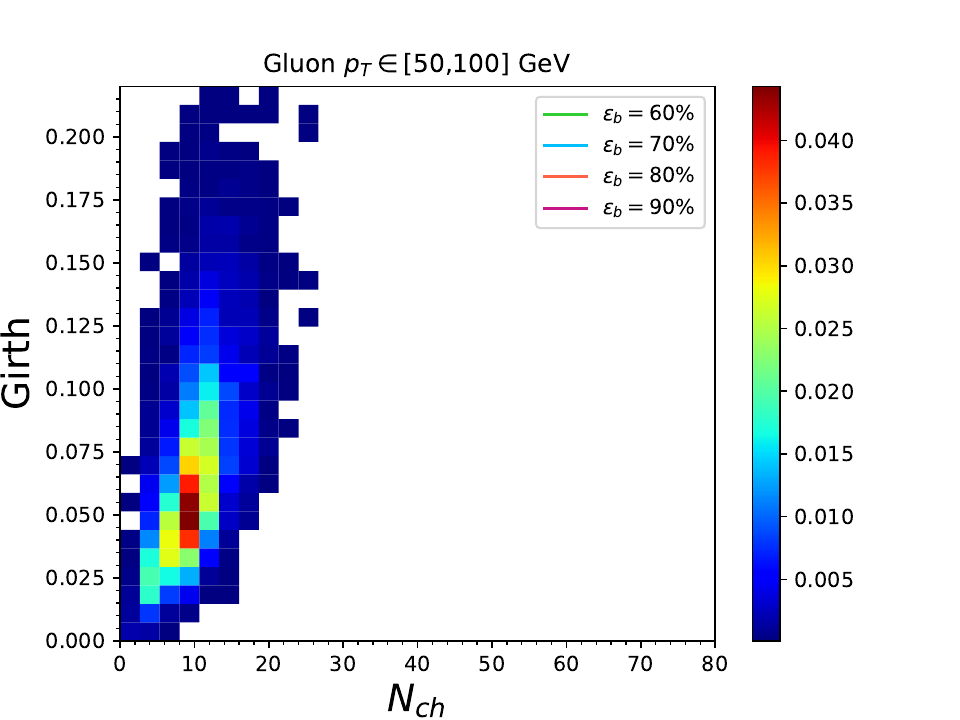}
  \newline
  \includegraphics[width=0.32 \linewidth]{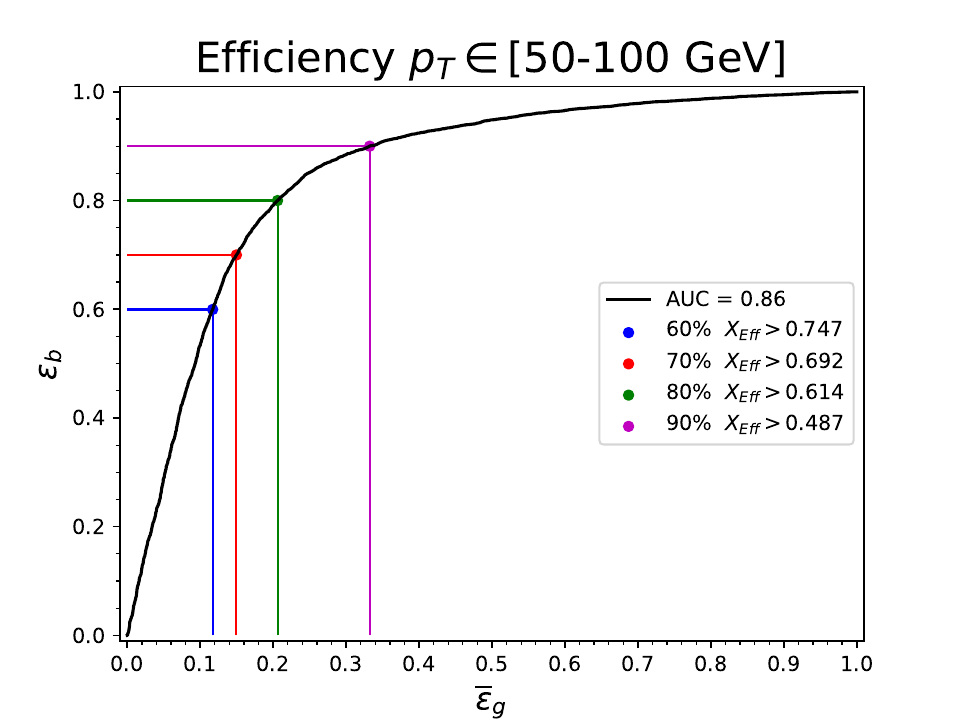}
\end{center}
\caption{Charged track multiplicity, girth and $x_E$ for jets with $p_T\in[50,100] \; {\rm GeV}$.}
\label{fig:100}
\end{figure}

\begin{figure}[H]
\begin{center}
  \includegraphics[width=0.32 \linewidth]{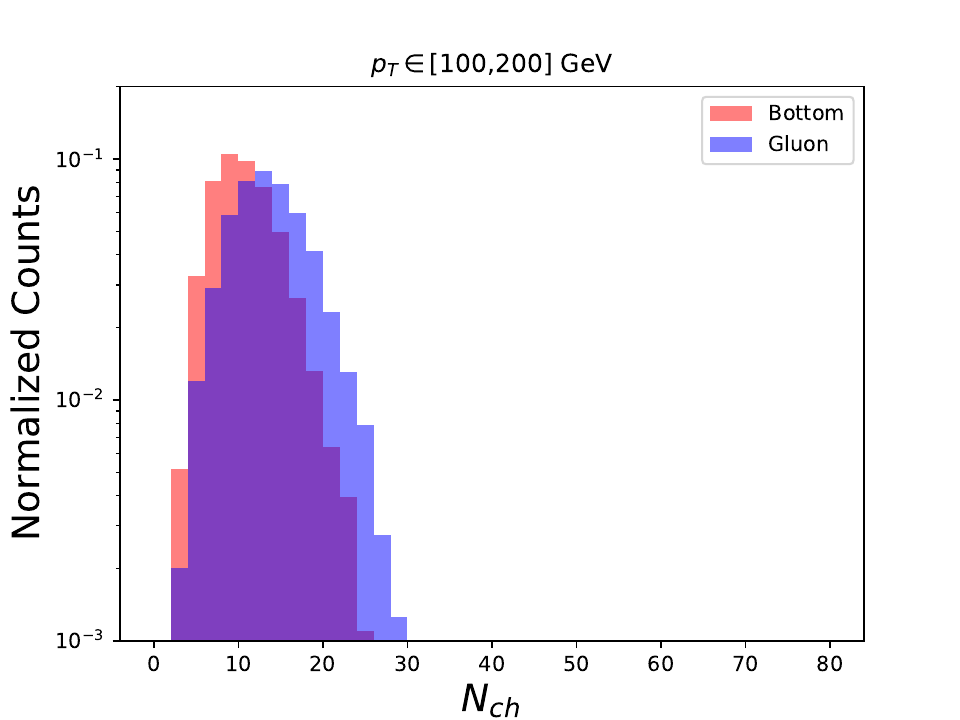}
  \includegraphics[width=0.32 \linewidth]{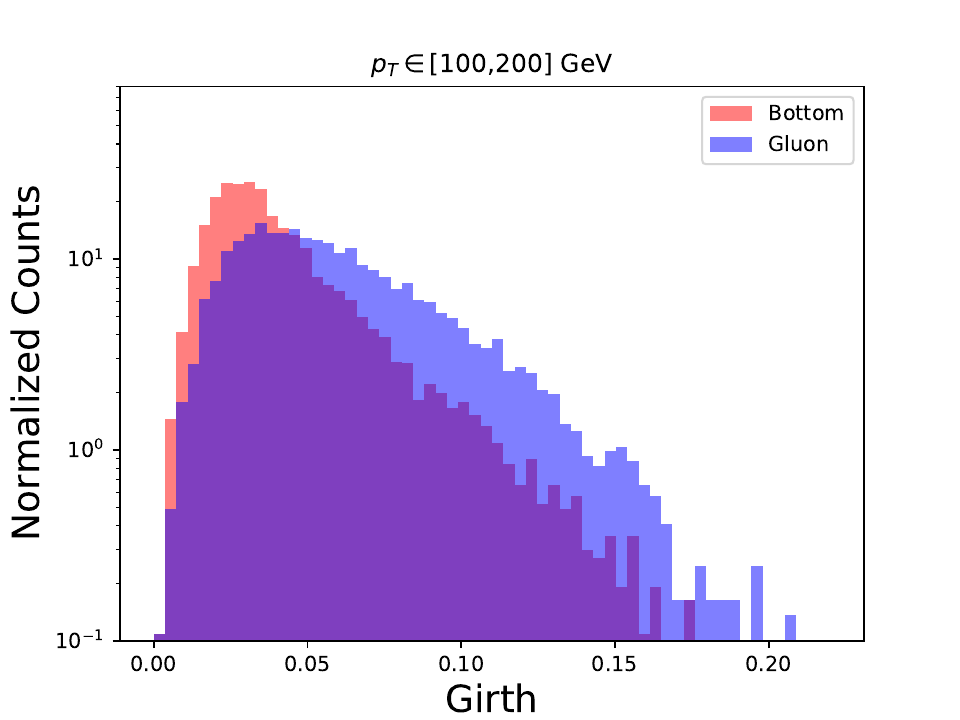}
  \includegraphics[width=0.32 \linewidth]{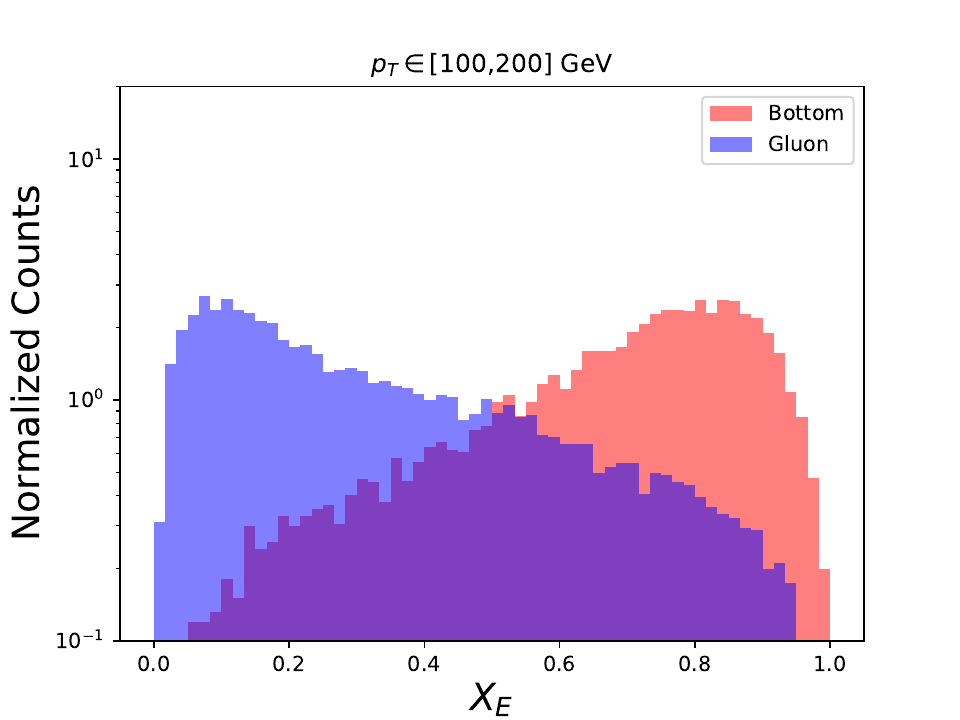}
  \includegraphics[width=0.32 \linewidth]{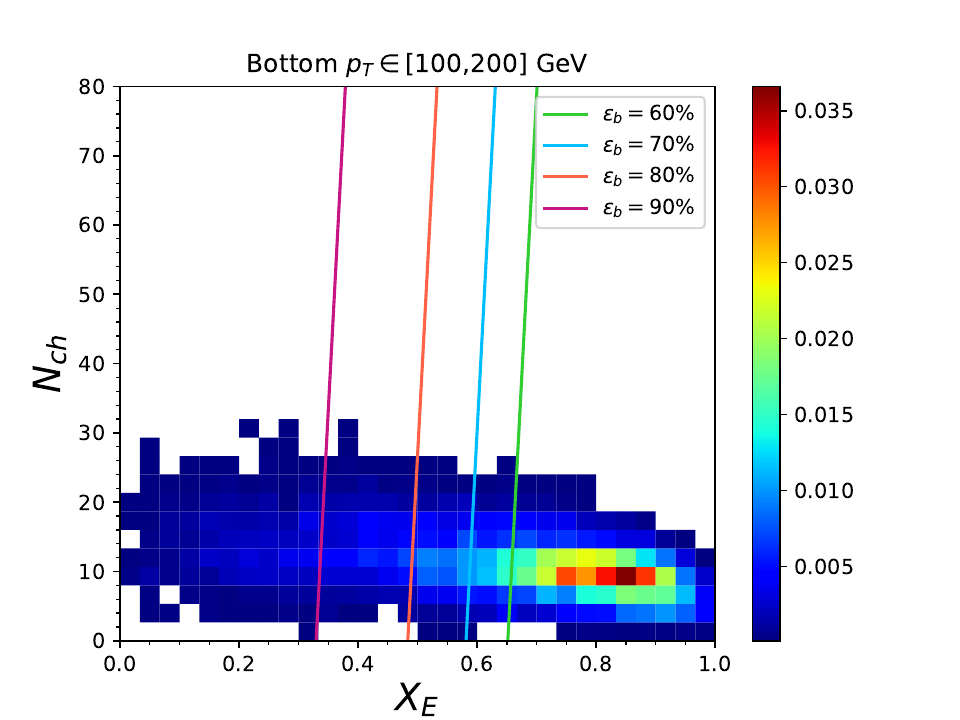}
  \includegraphics[width=0.32 \linewidth]{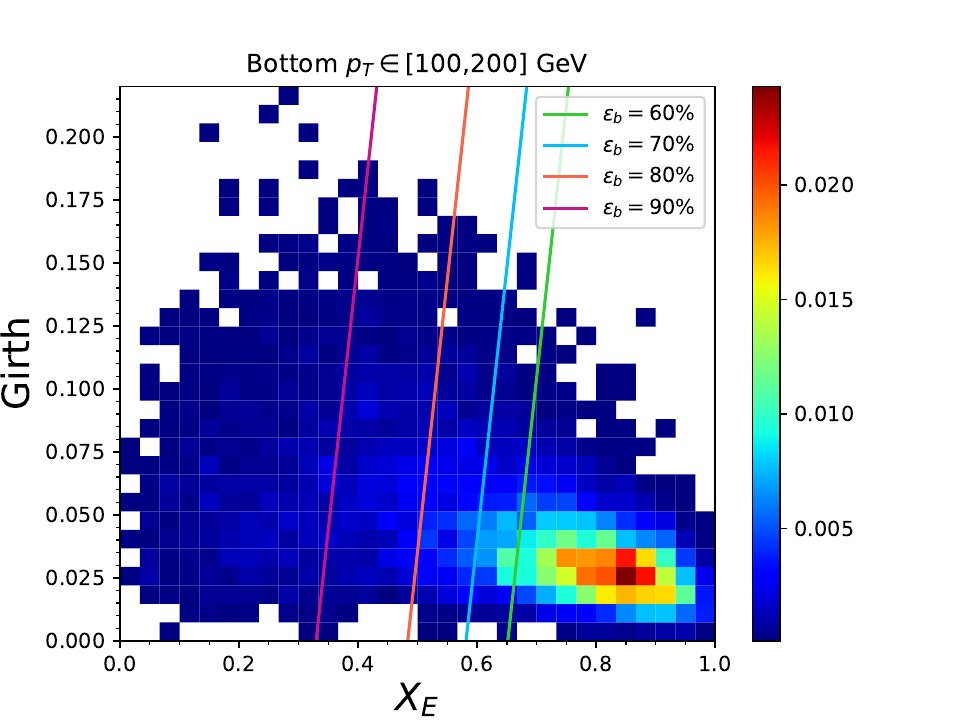}
  \includegraphics[width=0.32 \linewidth]{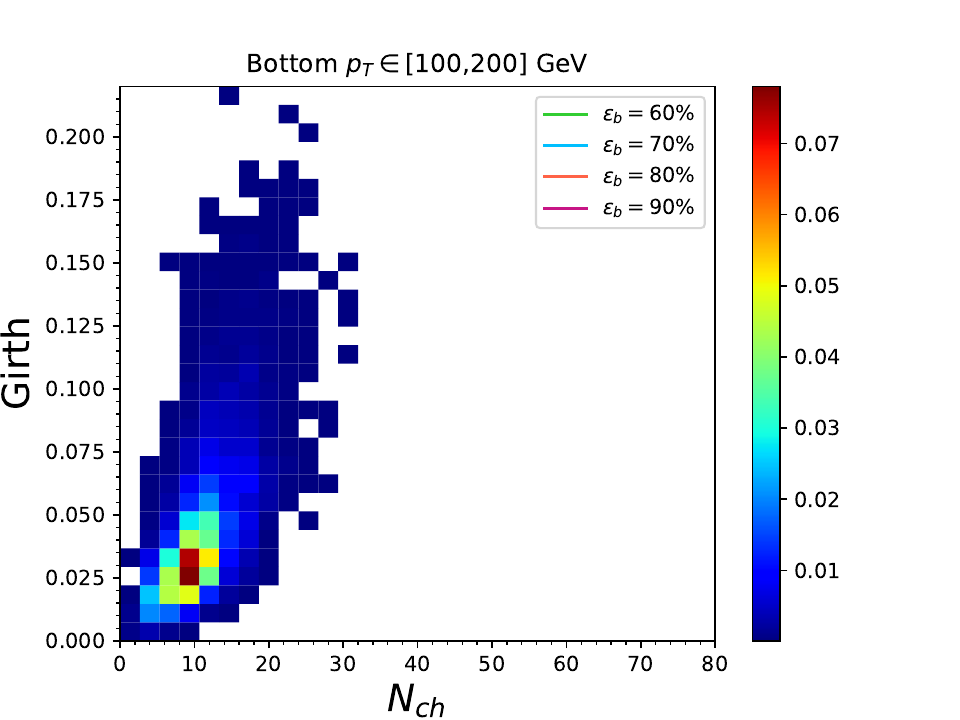}
  \includegraphics[width=0.32 \linewidth]{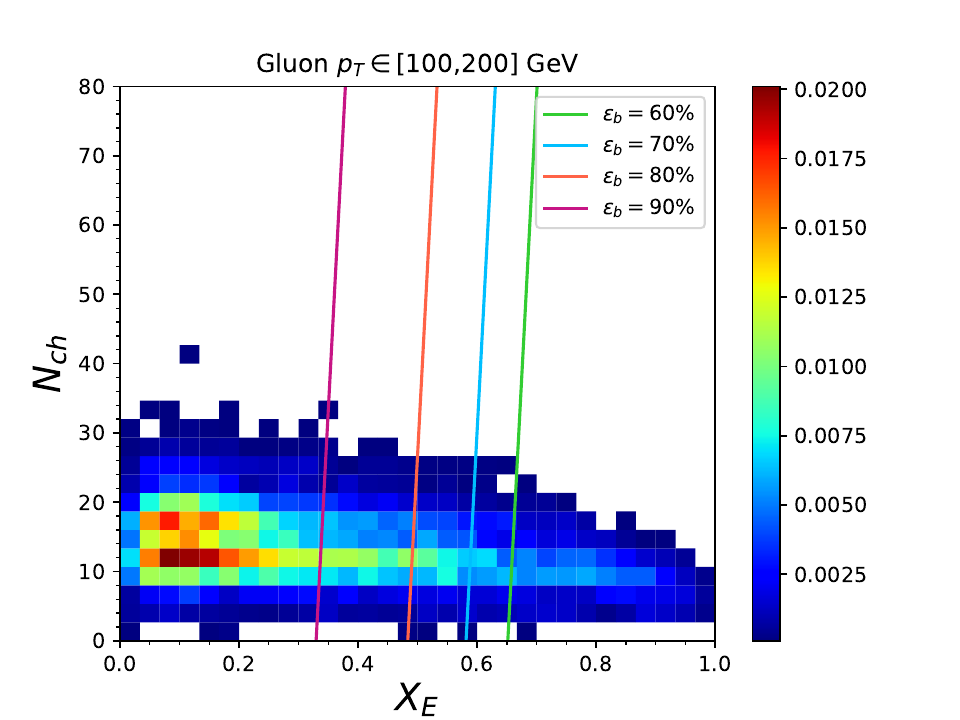}
  \includegraphics[width=0.32 \linewidth]{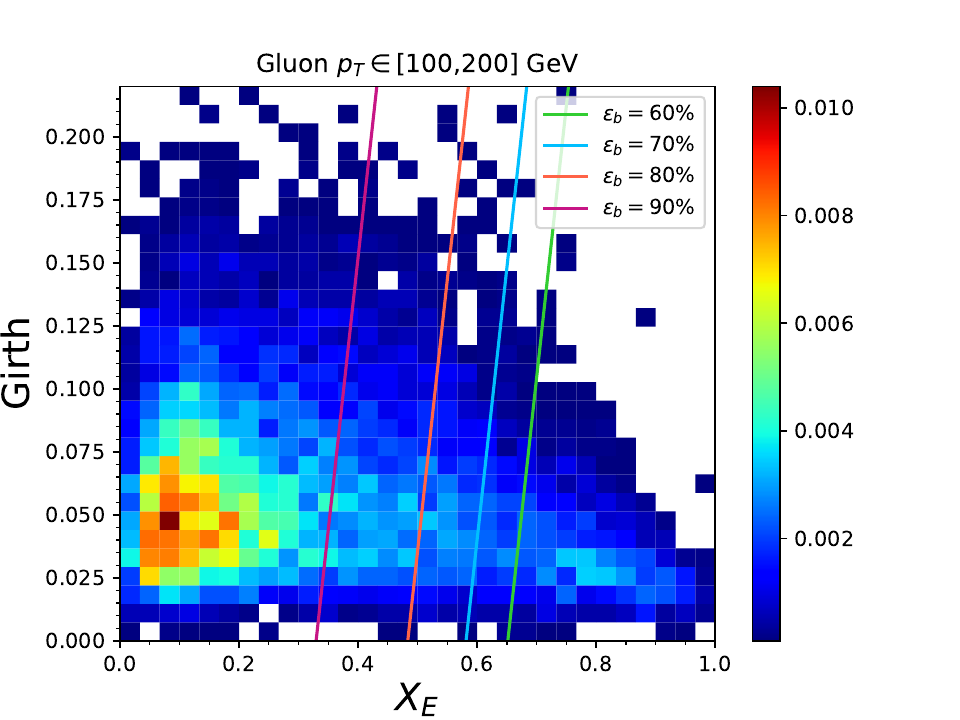}
  \includegraphics[width=0.32 \linewidth]{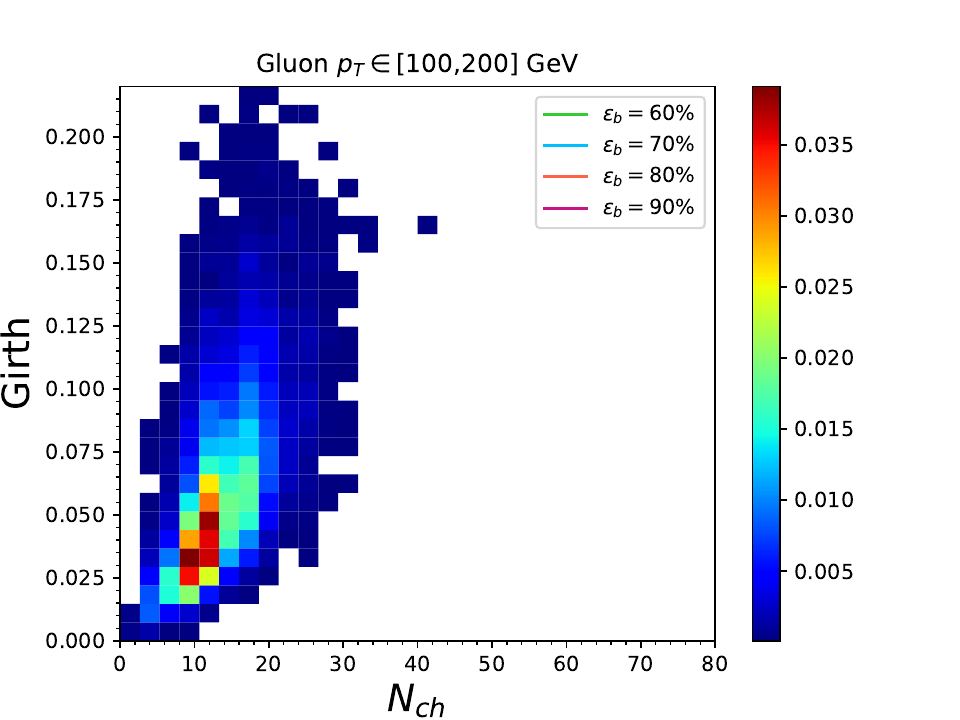}
  \newline
  \includegraphics[width=0.32 \linewidth]{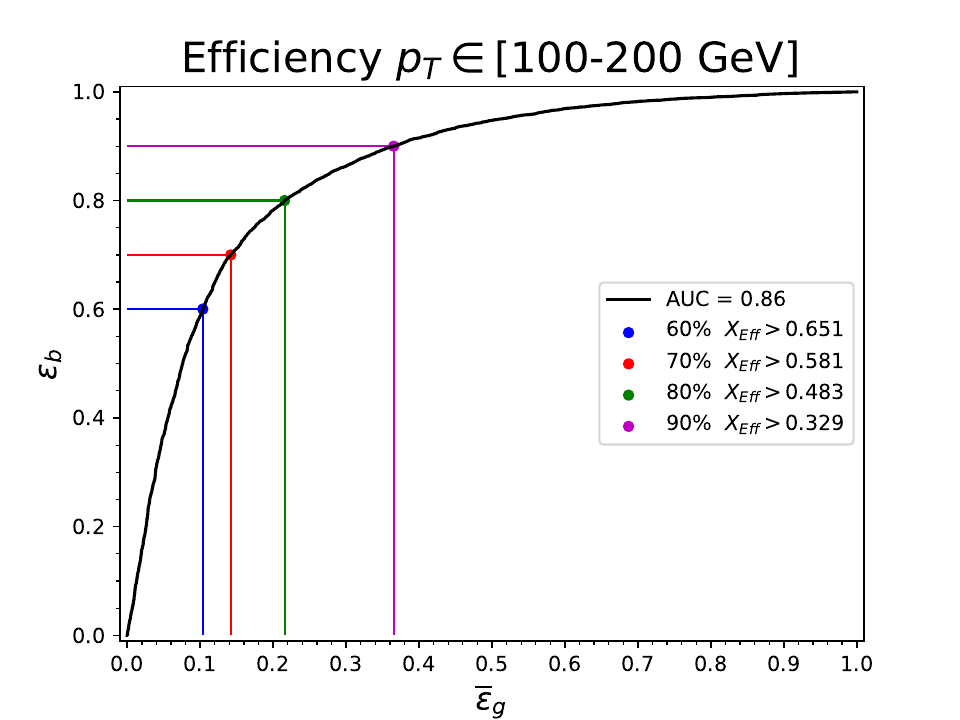}
\end{center}
\caption{Charged track multiplicity, girth and $x_E$ for jets with $p_T\in[100,200] \; {\rm GeV}$.}
\label{fig:200}
\end{figure}

\begin{figure}[H]
\begin{center}
  \includegraphics[width=0.32 \linewidth]{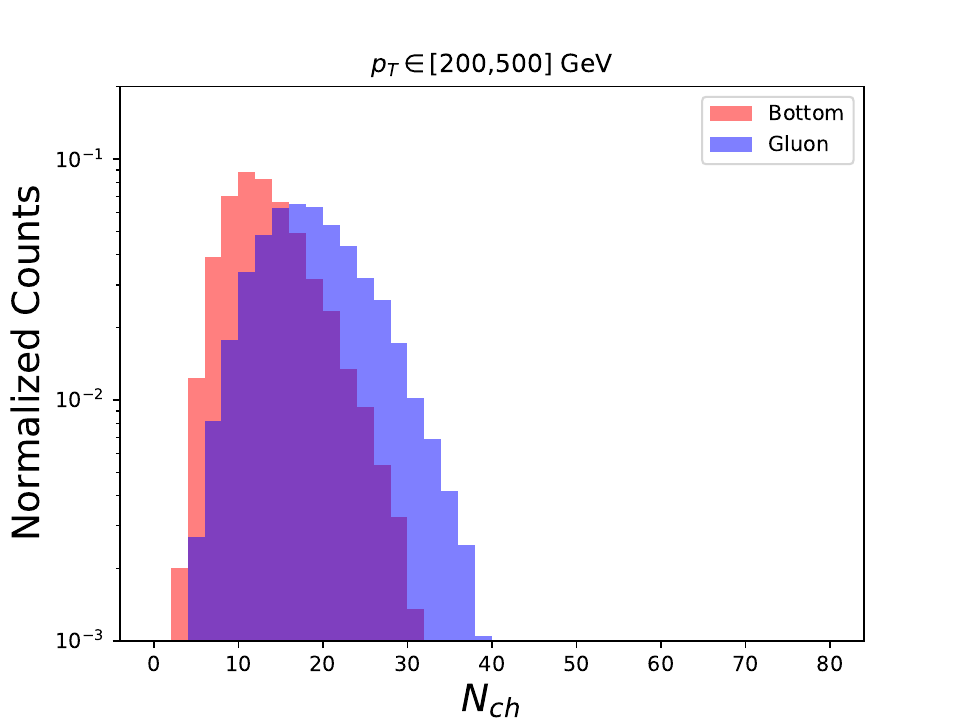}
  \includegraphics[width=0.32 \linewidth]{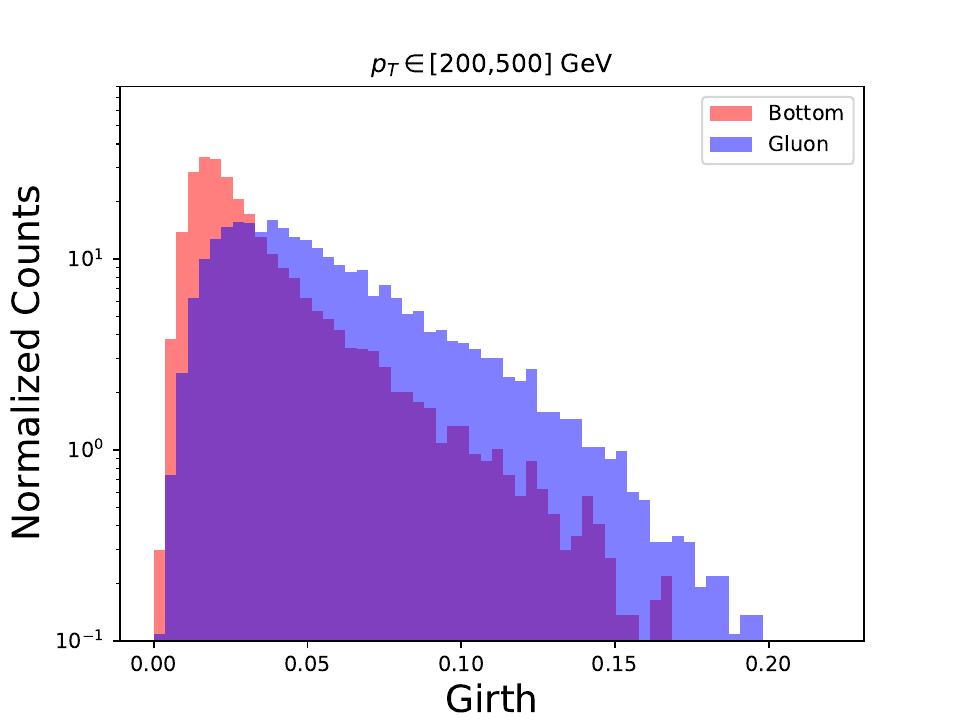}
  \includegraphics[width=0.32 \linewidth]{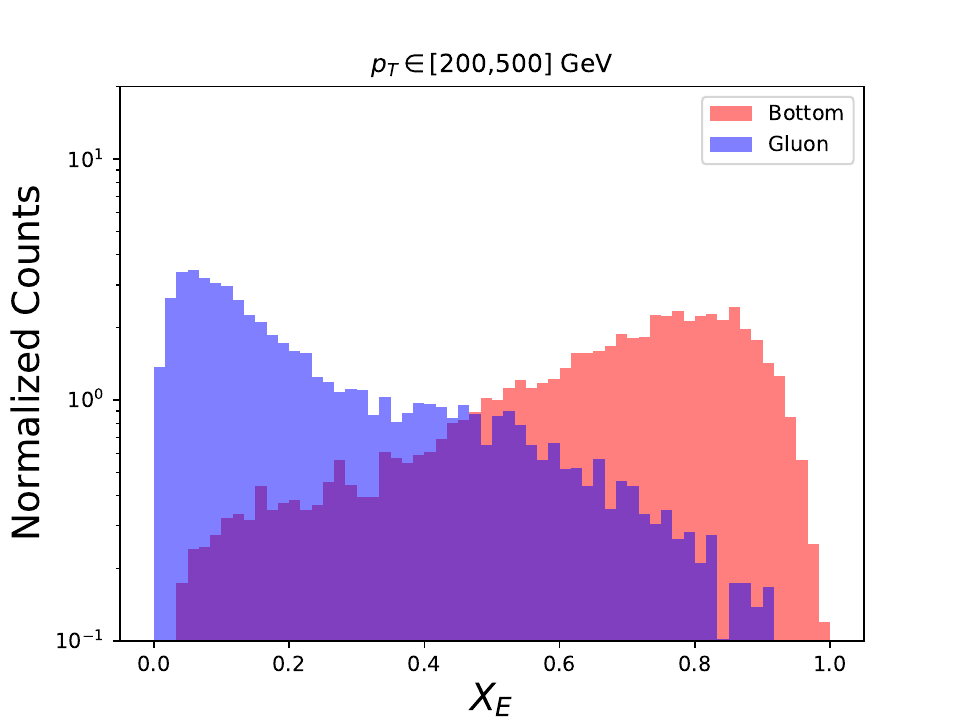}
  \includegraphics[width=0.32 \linewidth]{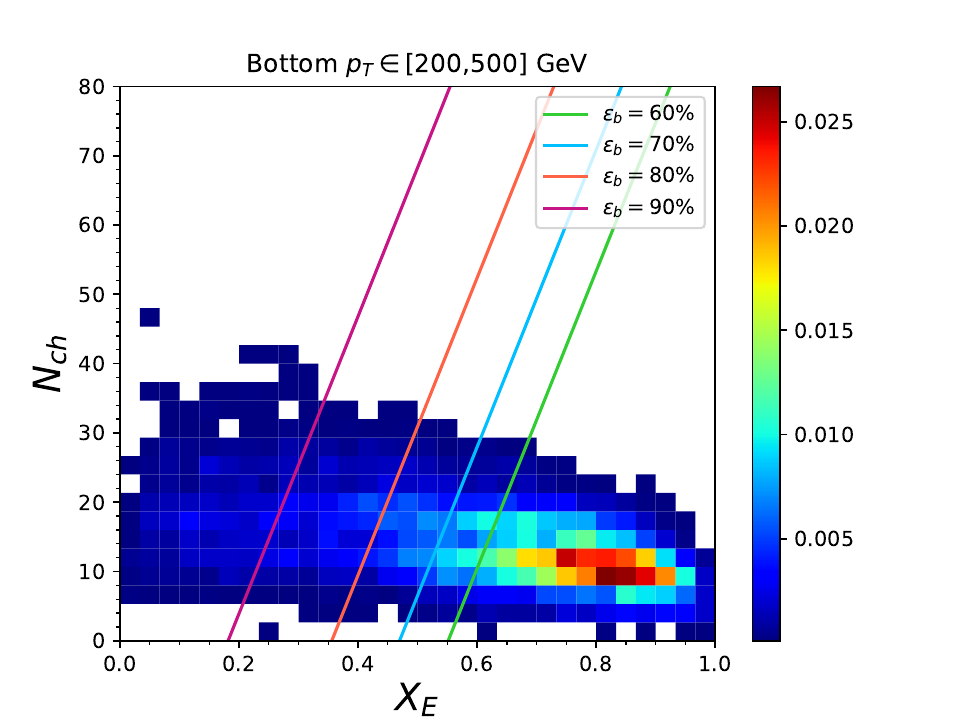}
  \includegraphics[width=0.32 \linewidth]{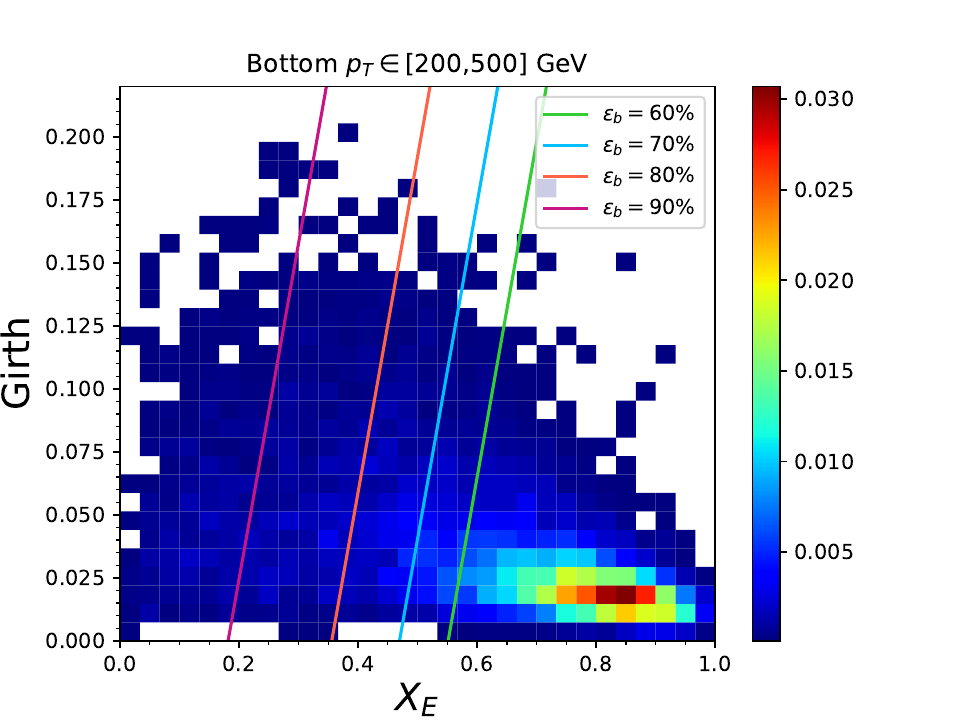}
  \includegraphics[width=0.32 \linewidth]{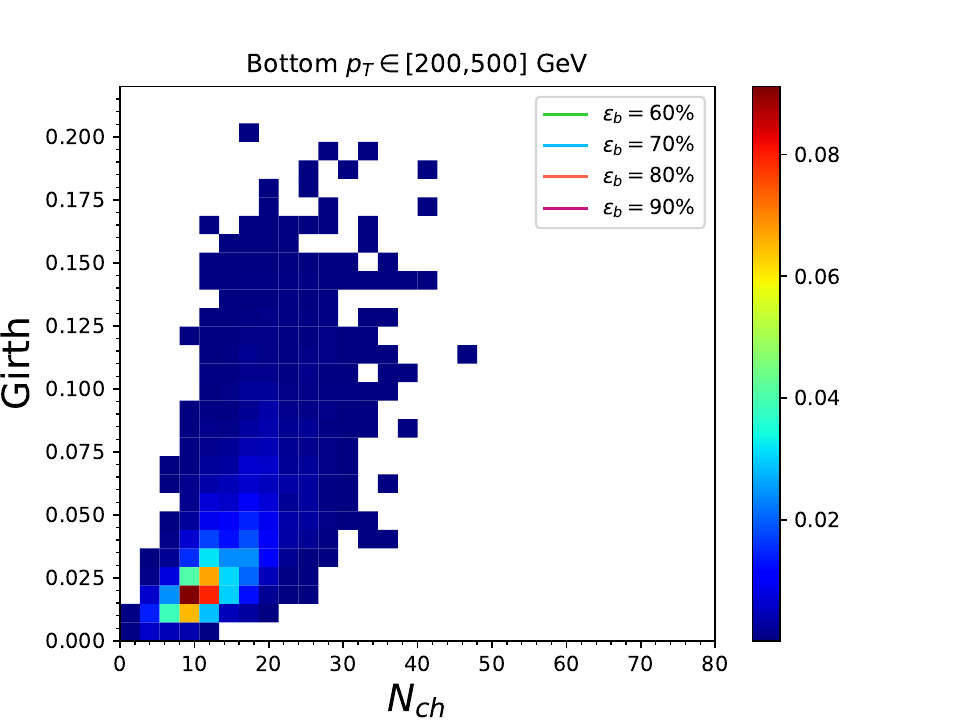}
  \includegraphics[width=0.32 \linewidth]{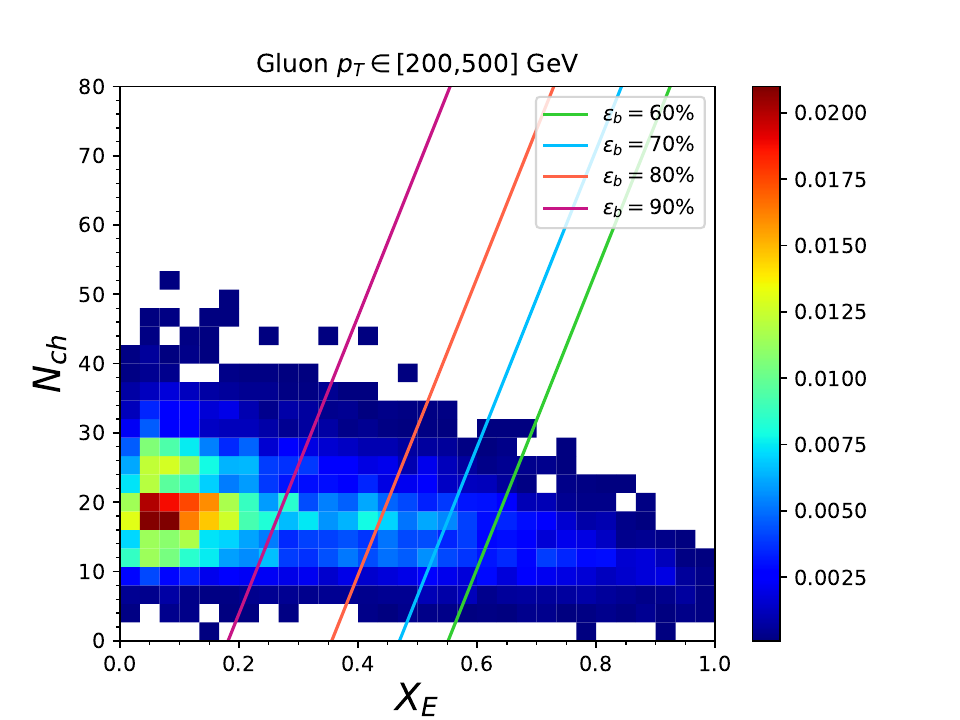}
  \includegraphics[width=0.32 \linewidth]{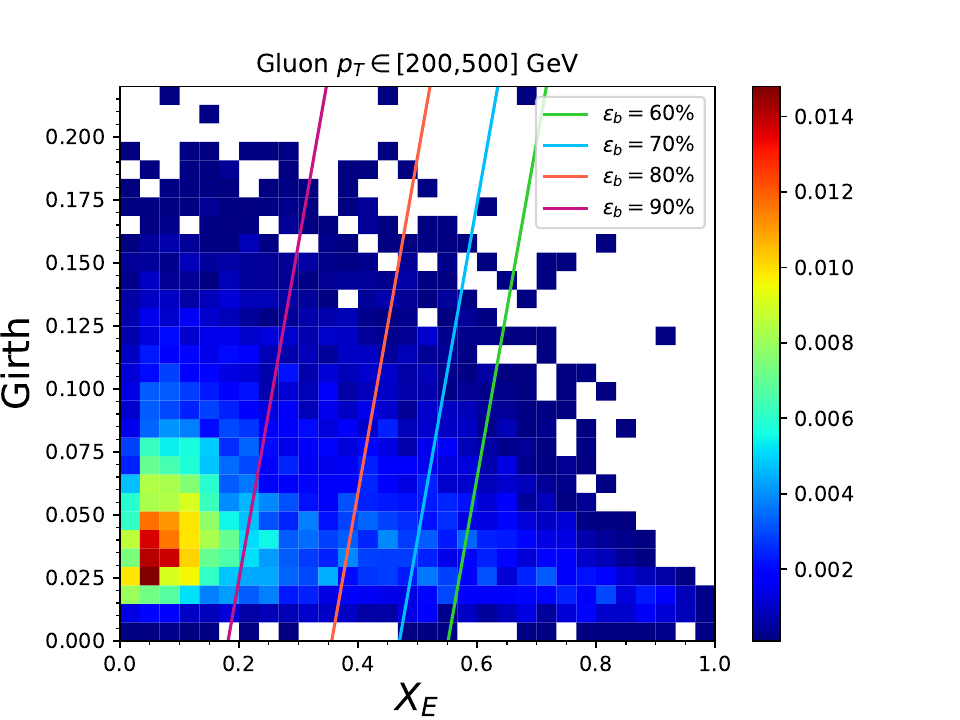}
  \includegraphics[width=0.32 \linewidth]{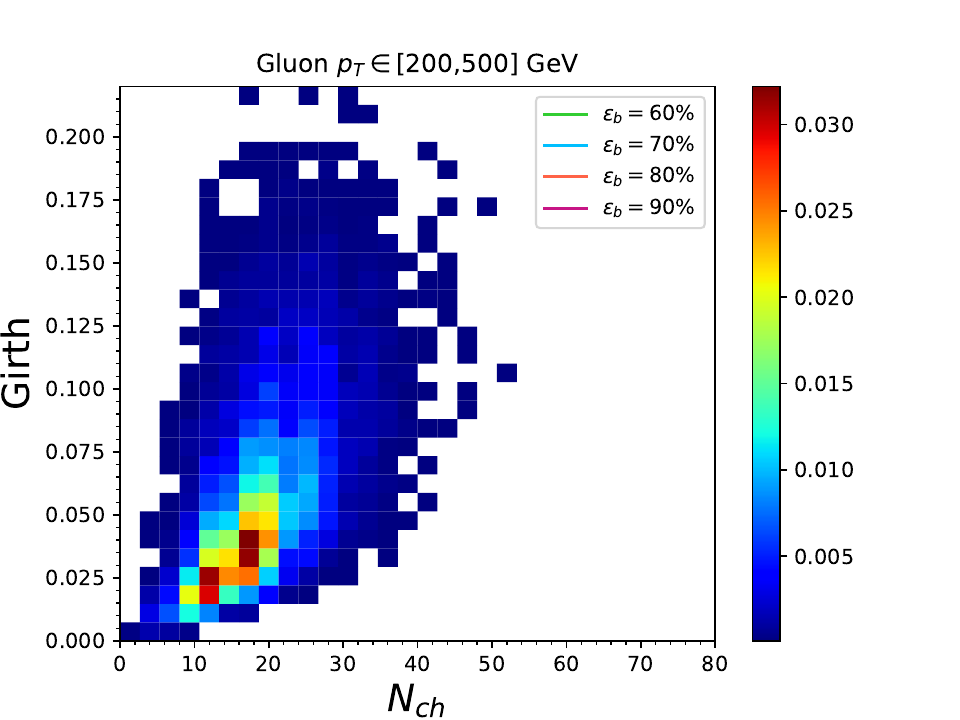}
  \newline
  \includegraphics[width=0.32 \linewidth]{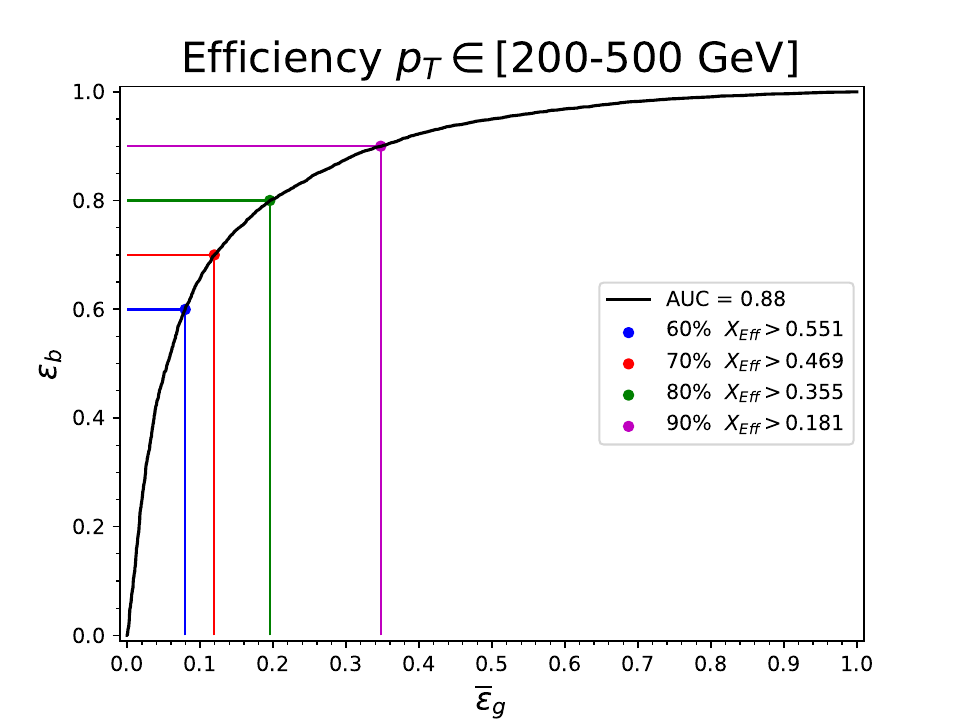}
\end{center}
\caption{Charged track multiplicity, girth and $x_E$ for jets with $p_T\in[200,500] \; {\rm GeV}$.}
\label{fig:500}
\end{figure}

\begin{figure}[H]
\begin{center}
  \includegraphics[width=0.32 \linewidth]{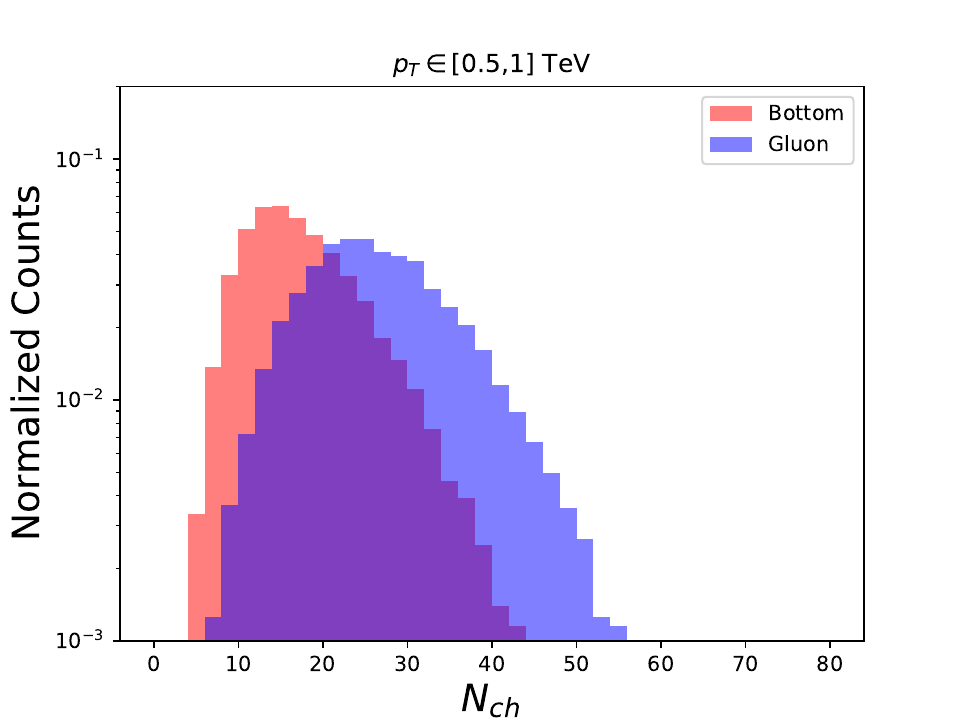}
  \includegraphics[width=0.32 \linewidth]{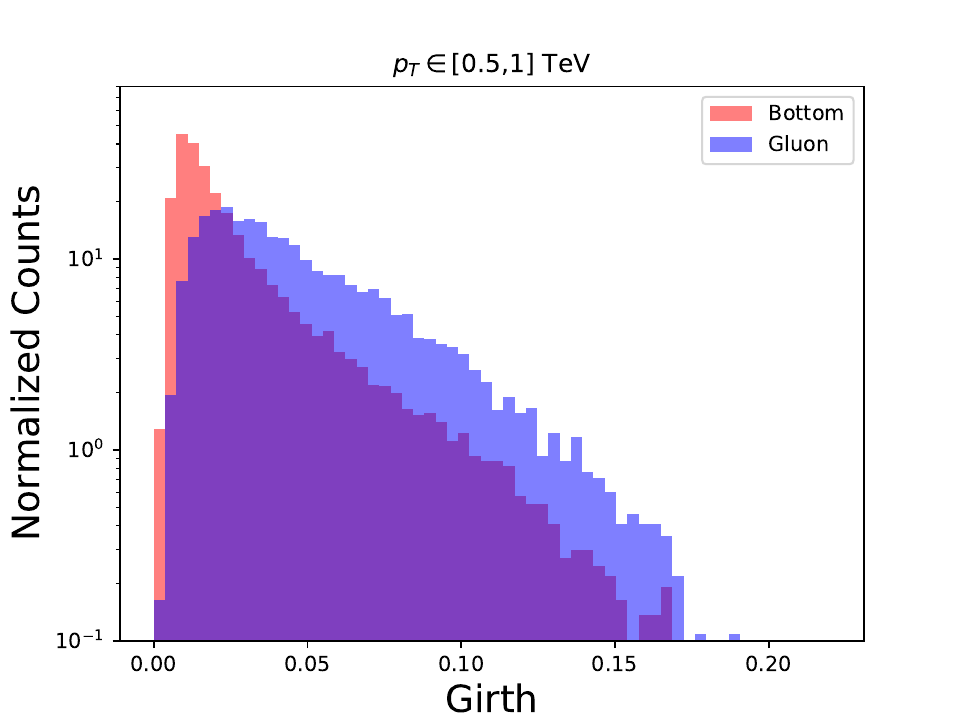}
  \includegraphics[width=0.32 \linewidth]{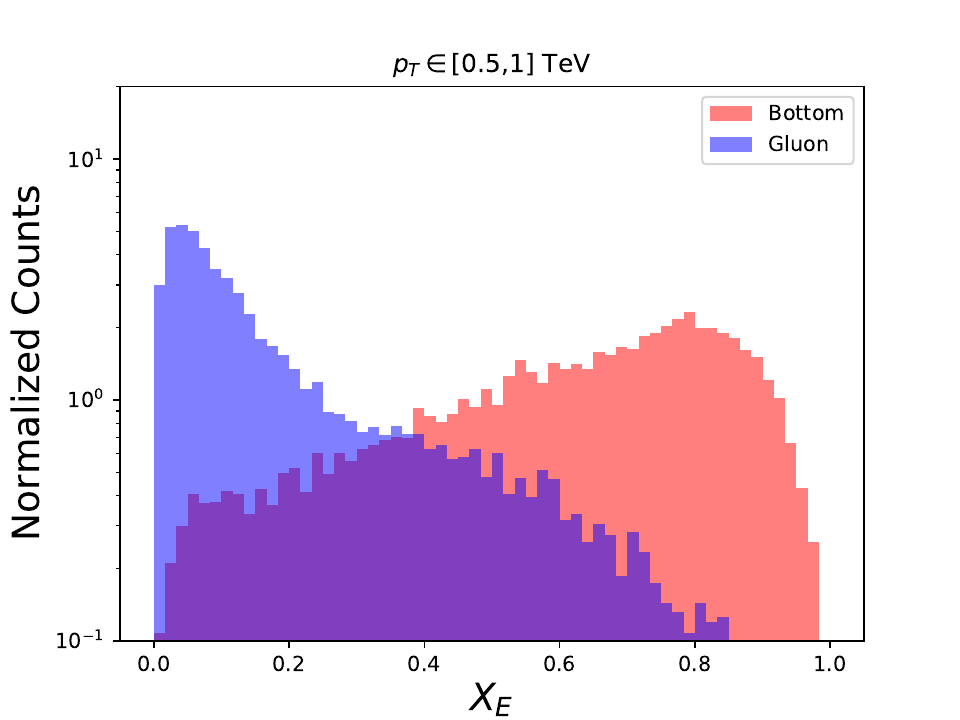}
  \includegraphics[width=0.32 \linewidth]{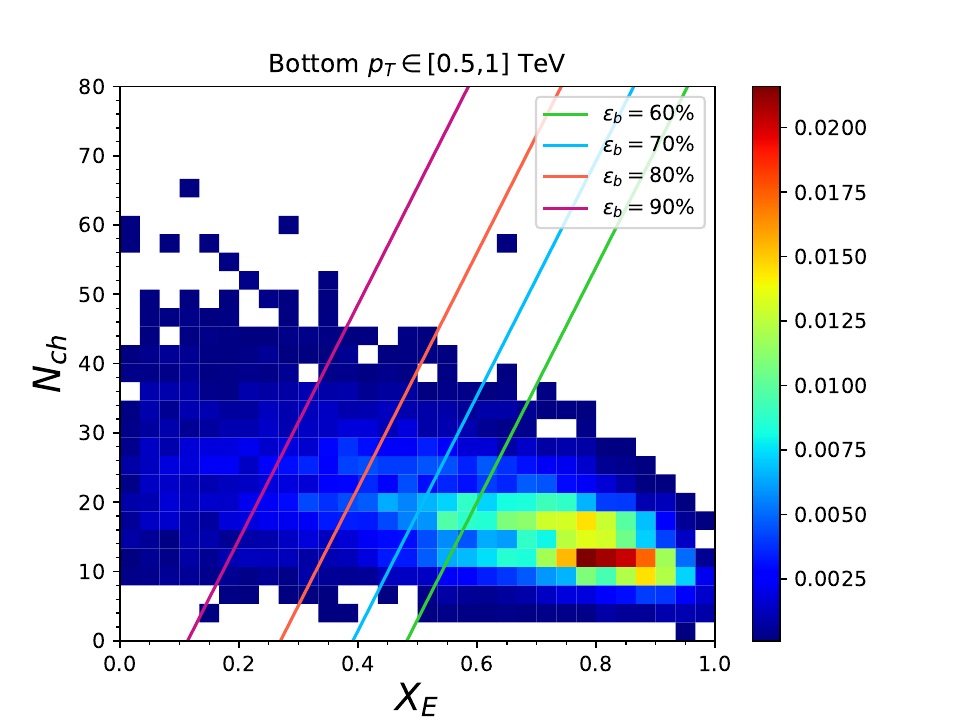}
  \includegraphics[width=0.32 \linewidth]{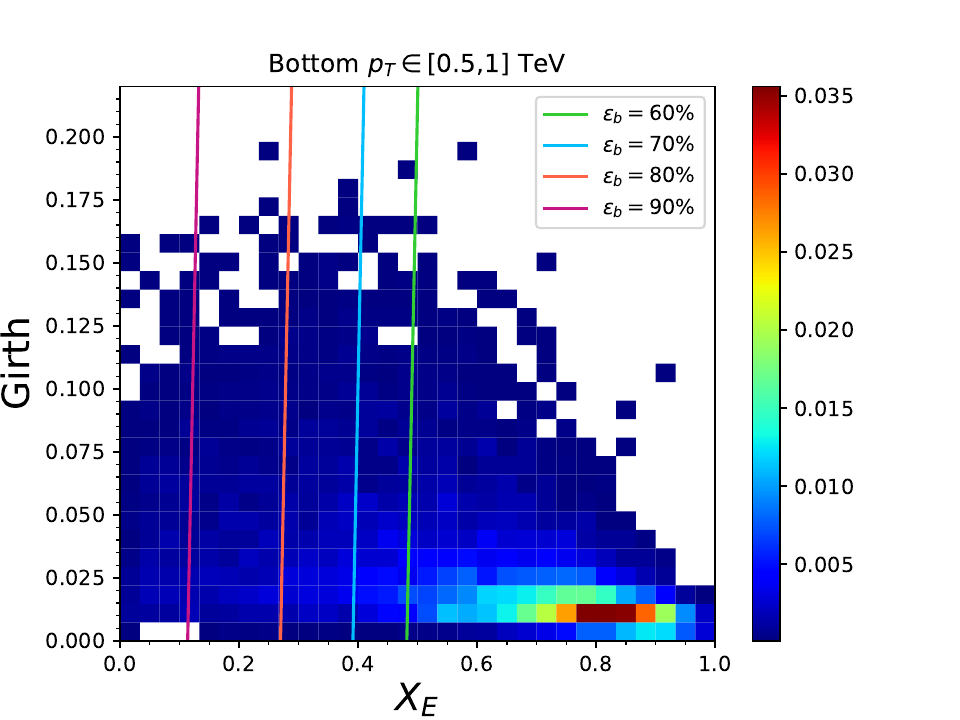}
  \includegraphics[width=0.32 \linewidth]{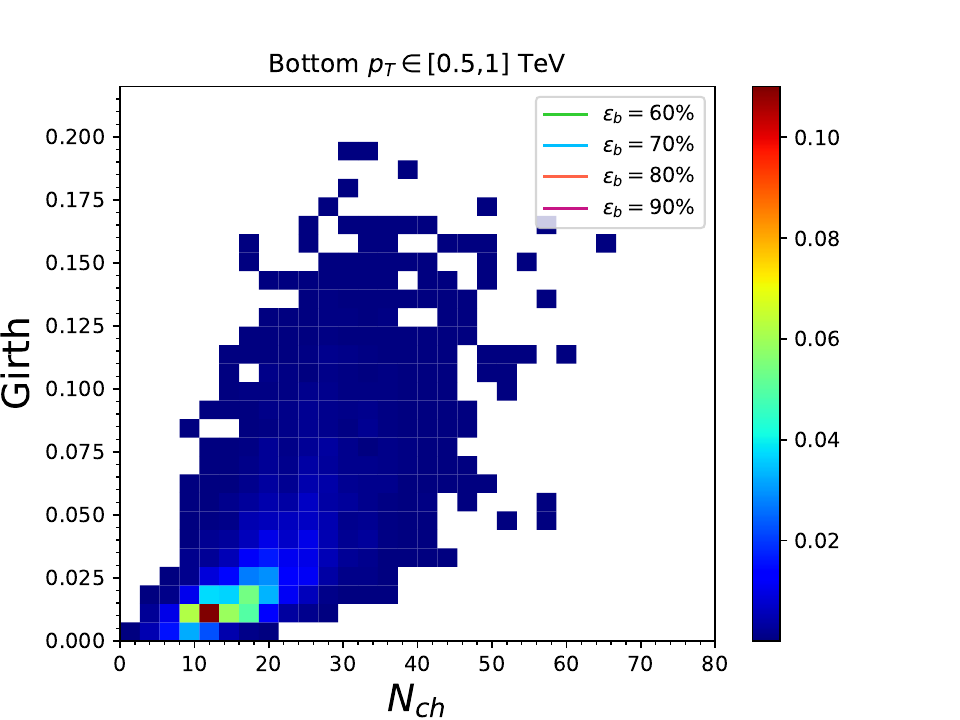}
  \includegraphics[width=0.32 \linewidth]{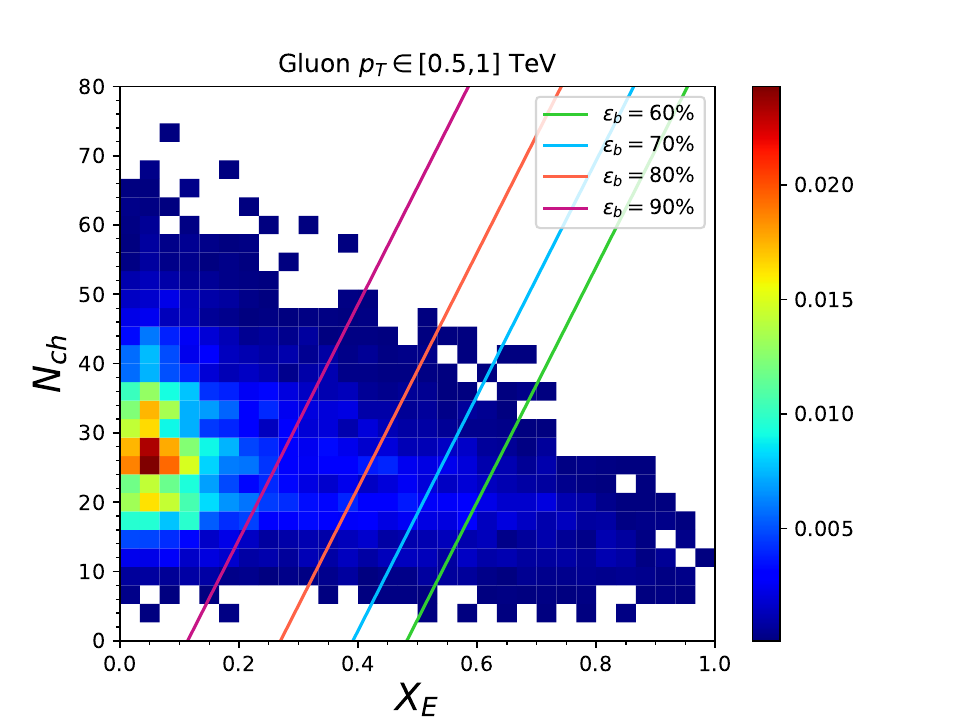}
  \includegraphics[width=0.32 \linewidth]{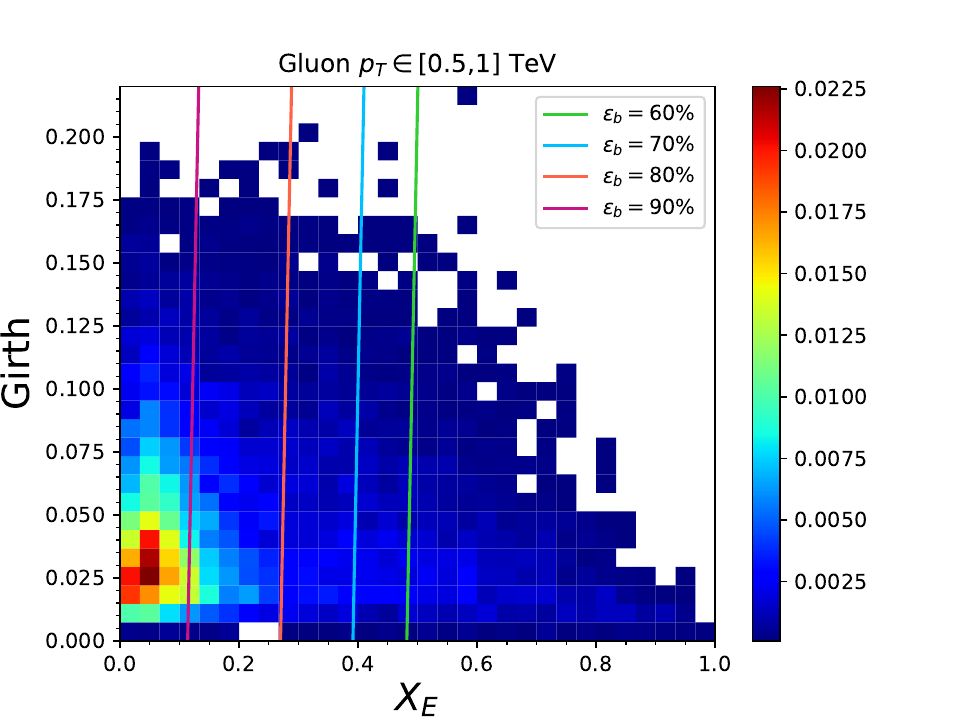}
  \includegraphics[width=0.32 \linewidth]{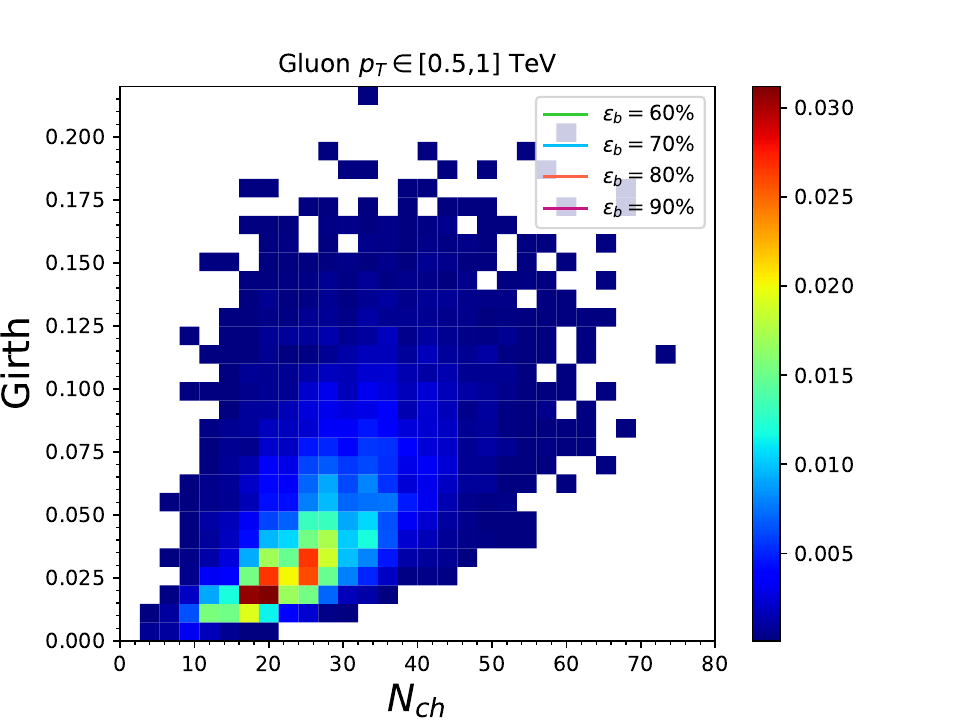}
  \newline
  \includegraphics[width=0.32 \linewidth]{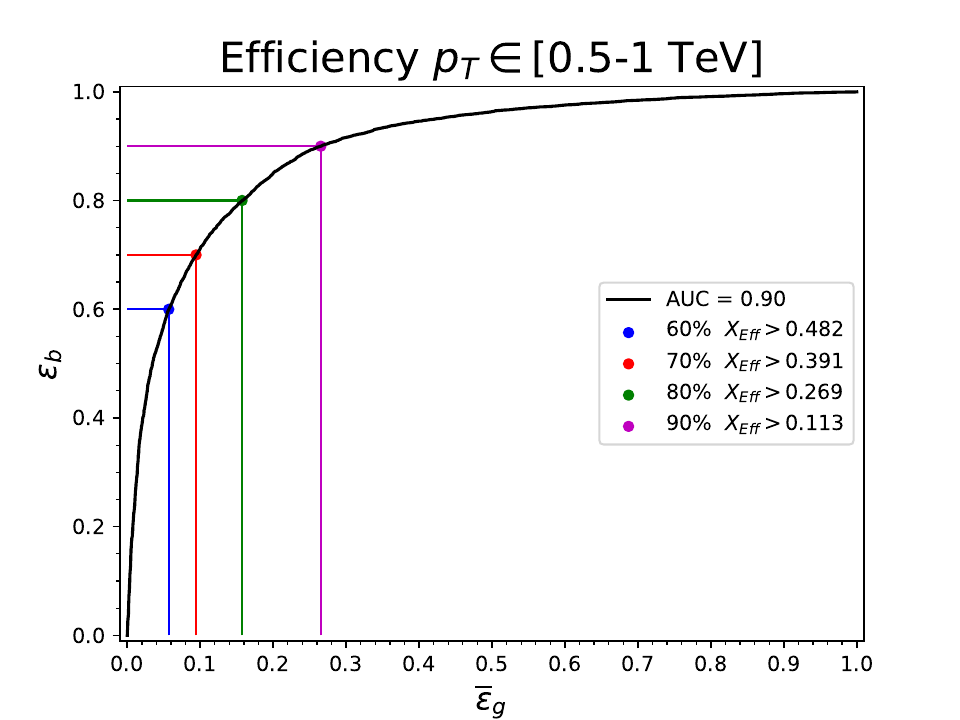}
\end{center}
\caption{Charged track multiplicity, girth and $x_E$ for jets with $p_T\in[500,1000] \; {\rm GeV}$.}
\label{fig:1000}
\end{figure}

\begin{figure}[H]
\begin{center}
  \includegraphics[width=0.32 \linewidth]{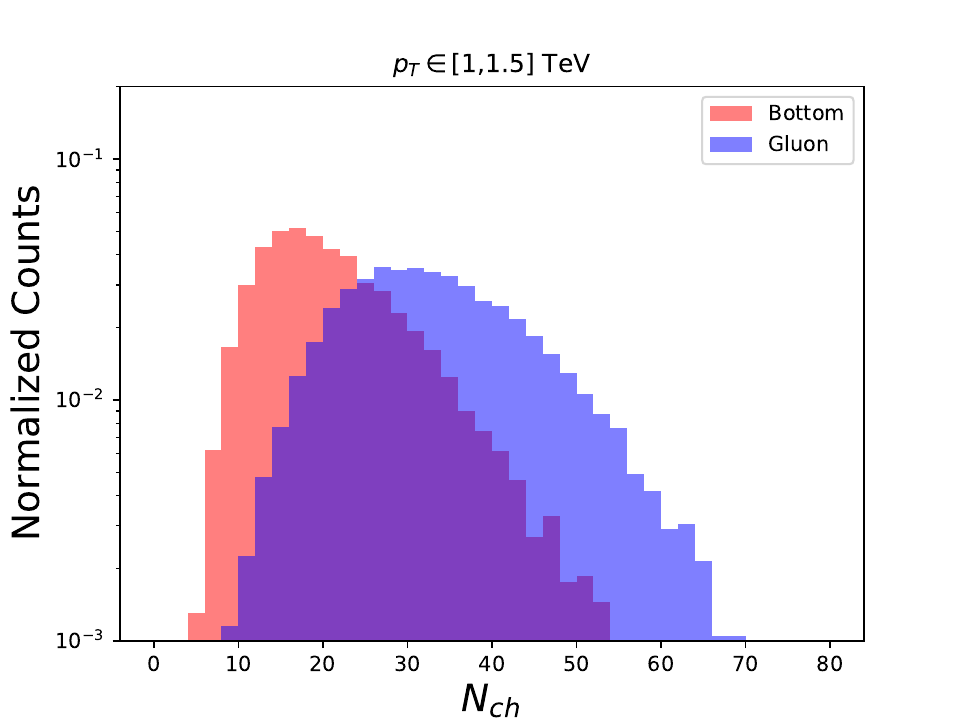}
  \includegraphics[width=0.32 \linewidth]{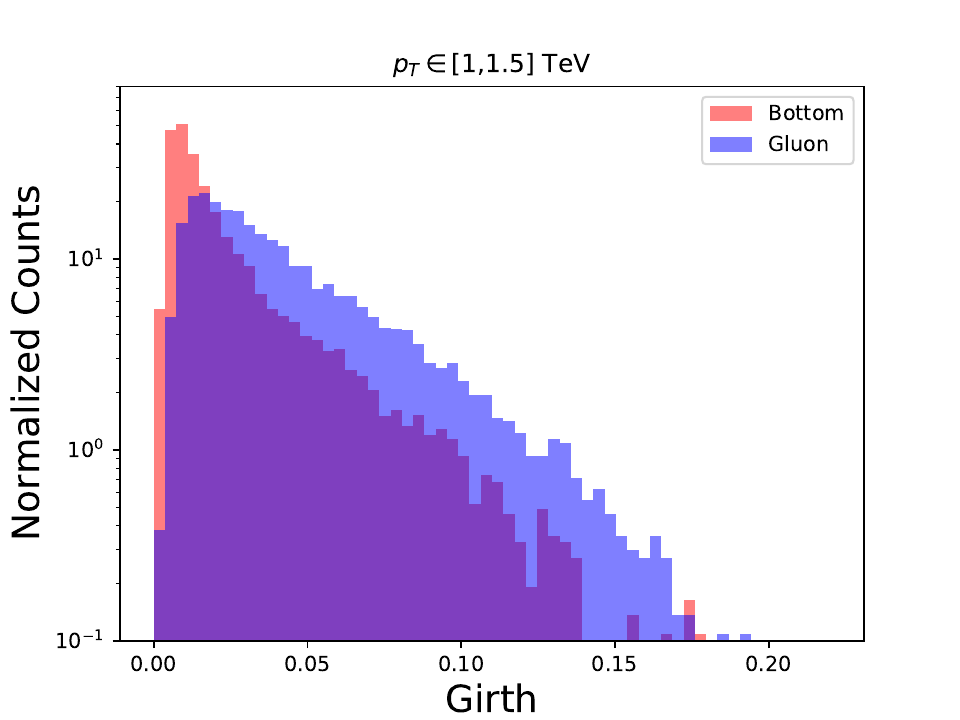}
  \includegraphics[width=0.32 \linewidth]{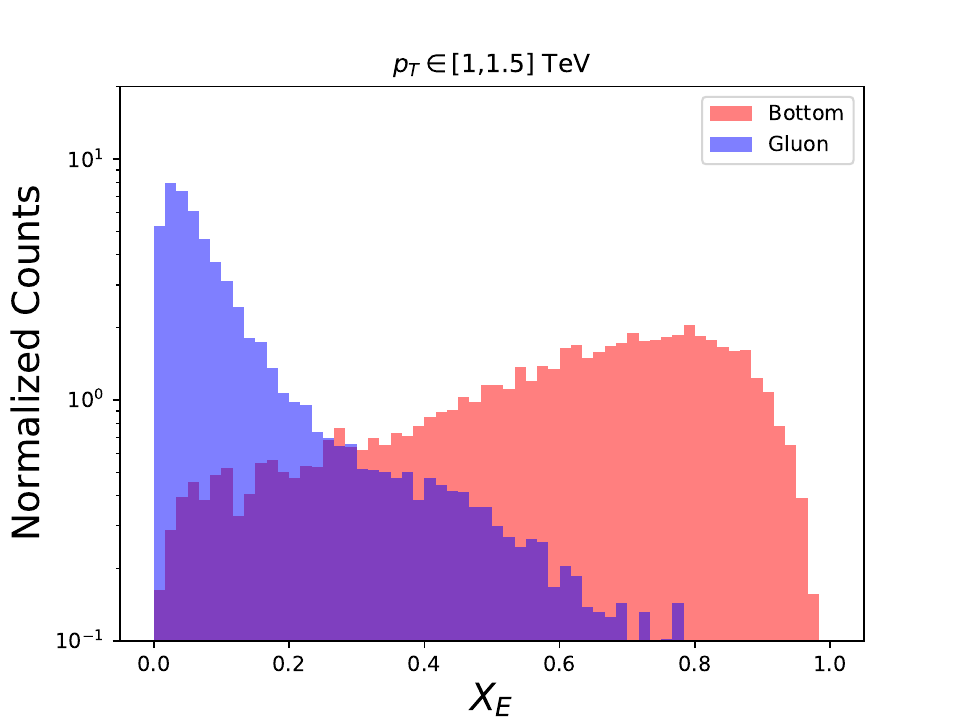}
  \includegraphics[width=0.32 \linewidth]{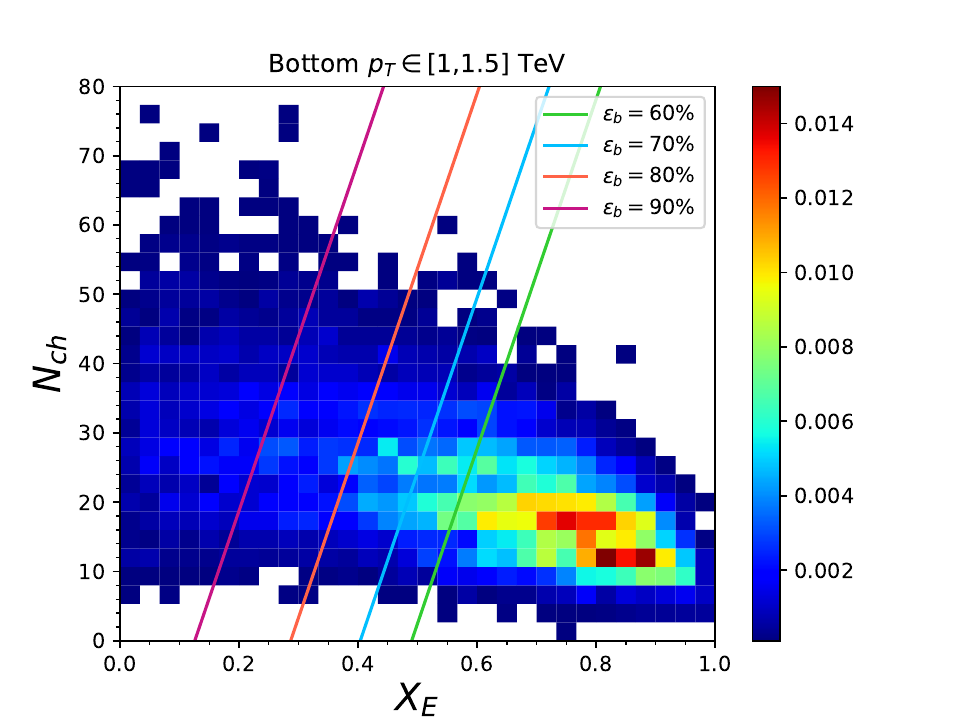}
  \includegraphics[width=0.32 \linewidth]{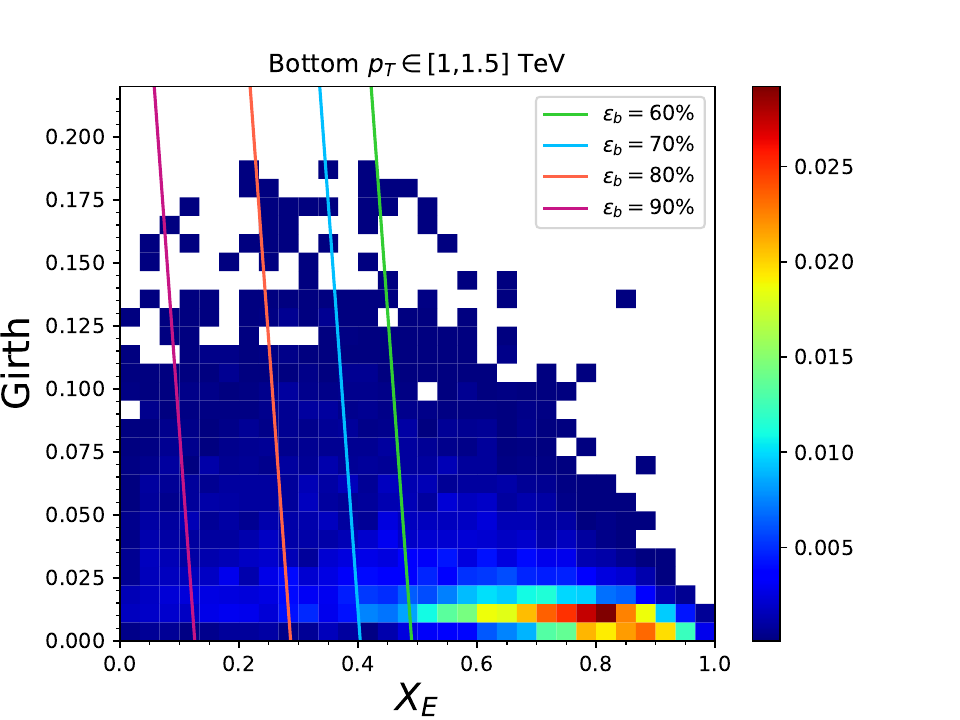}
  \includegraphics[width=0.32 \linewidth]{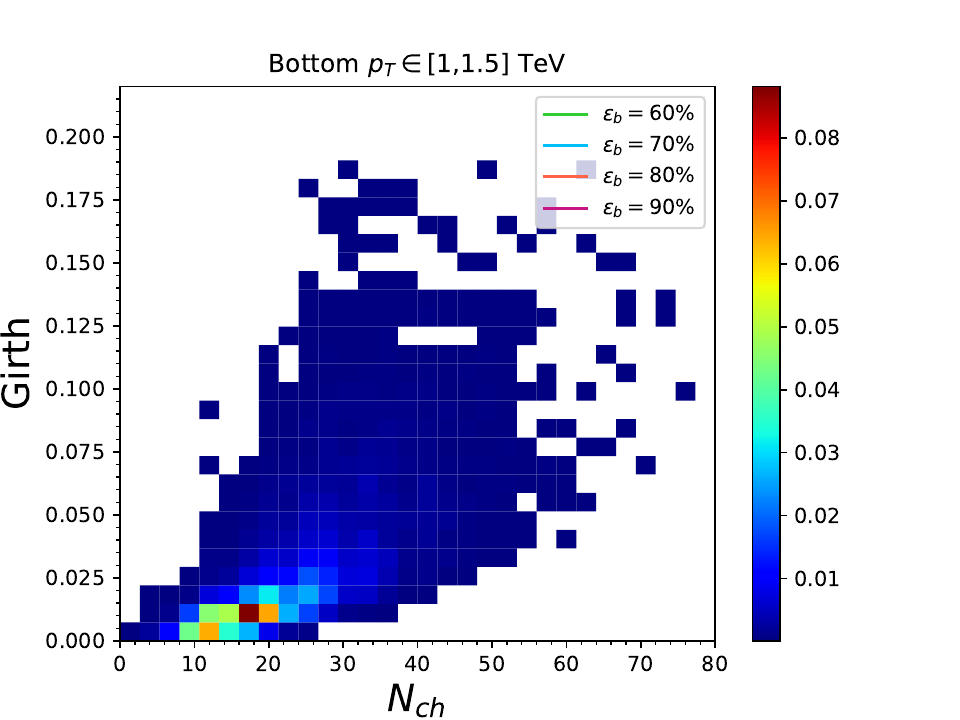}
  \includegraphics[width=0.32 \linewidth]{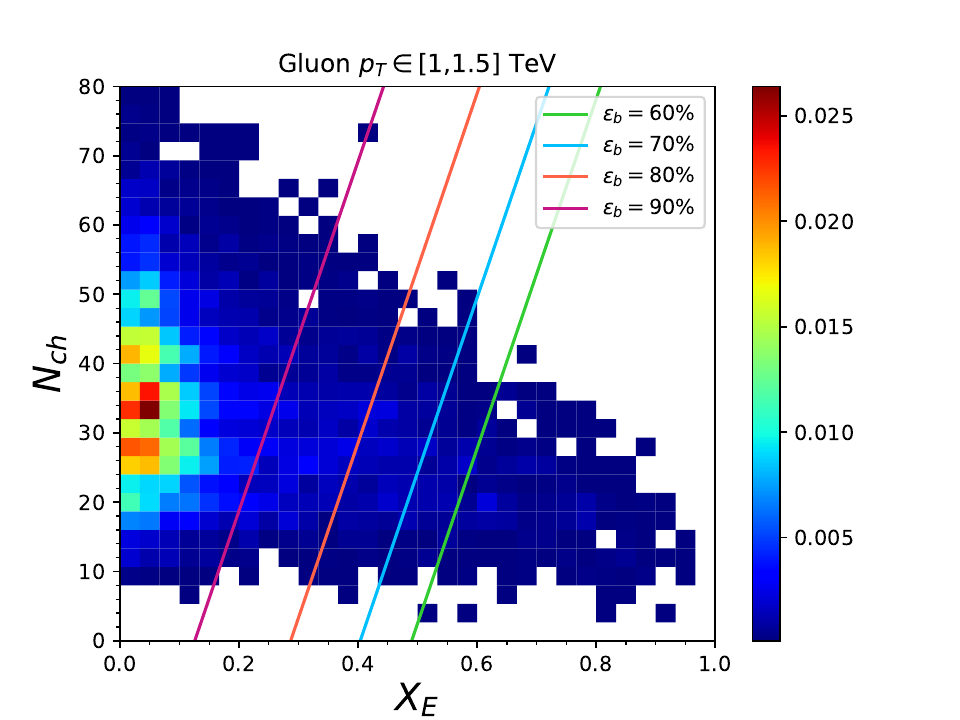}
  \includegraphics[width=0.32 \linewidth]{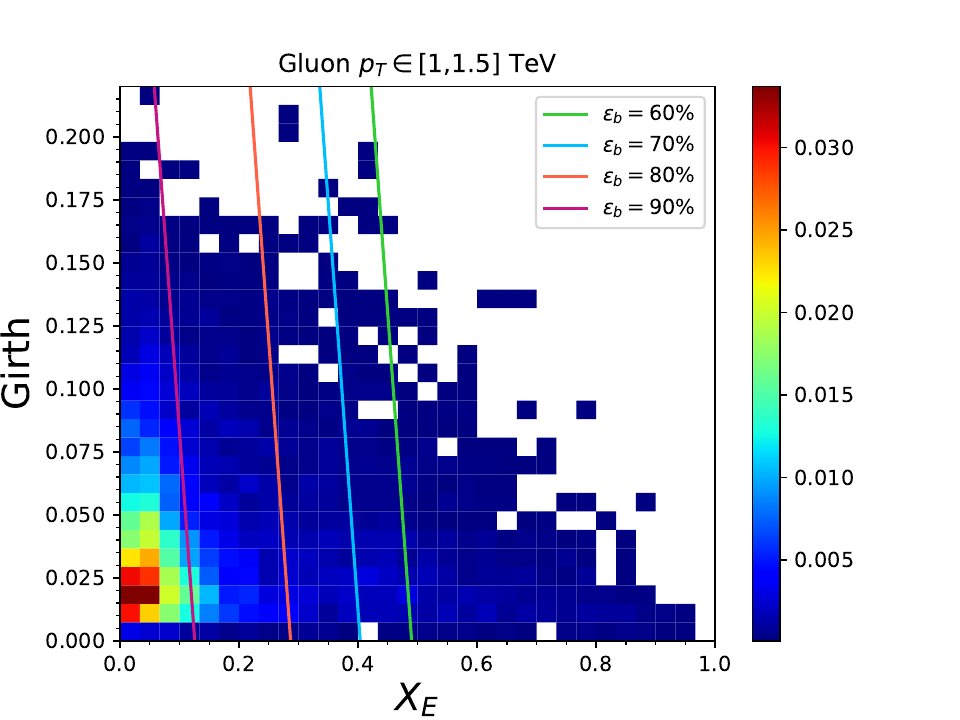}
  \includegraphics[width=0.32 \linewidth]{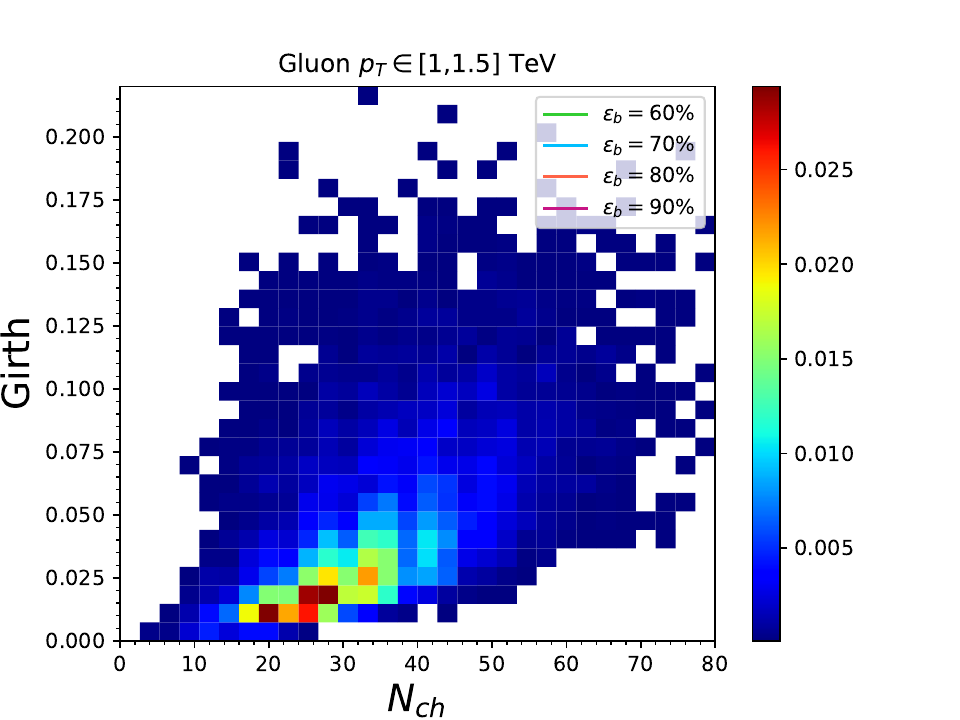}
  \newline
  \includegraphics[width=0.32 \linewidth]{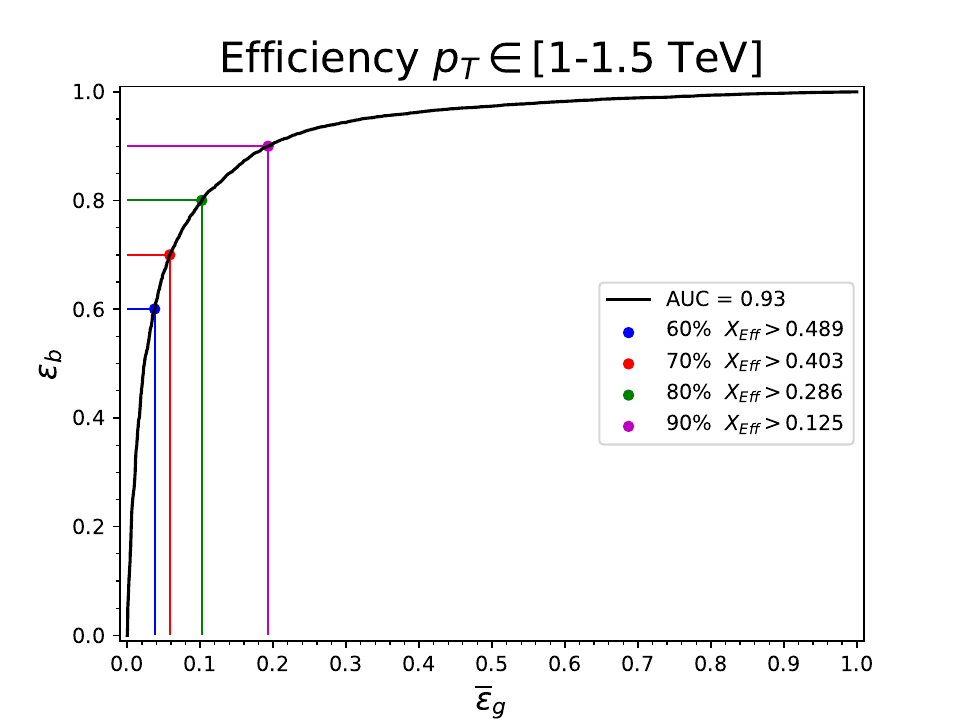}
\end{center}
\caption{Charged track multiplicity, girth and $x_E$ for jets with $p_T\in[1,1.5] \; {\rm TeV}$.}
\label{fig:1500}
\end{figure}

\begin{figure}[H]
\begin{center}
  \includegraphics[width=0.32 \linewidth]{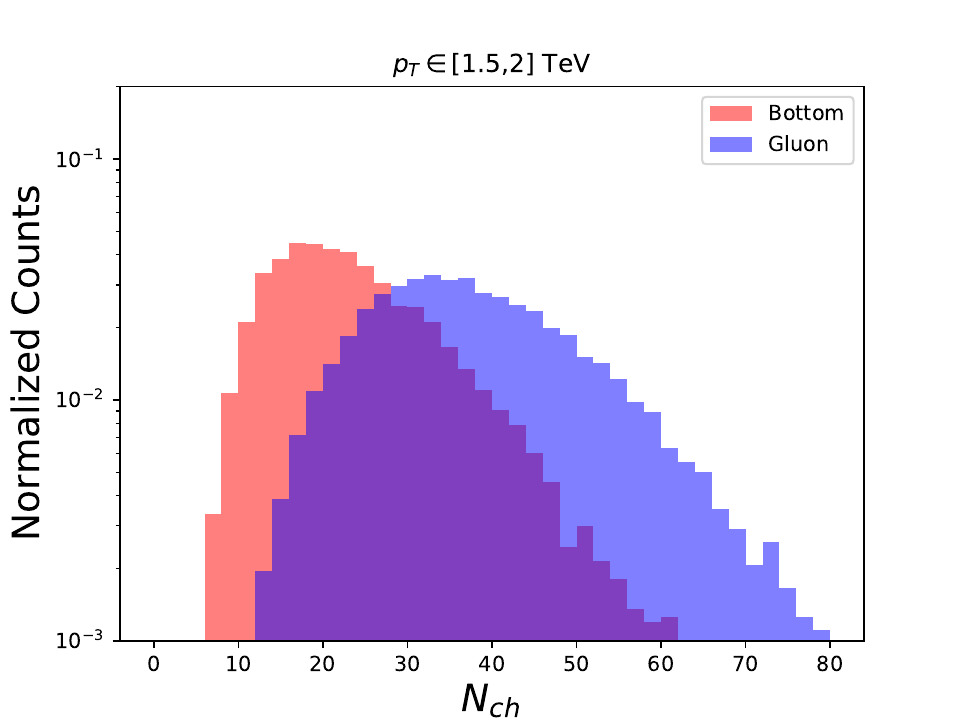}
  \includegraphics[width=0.32 \linewidth]{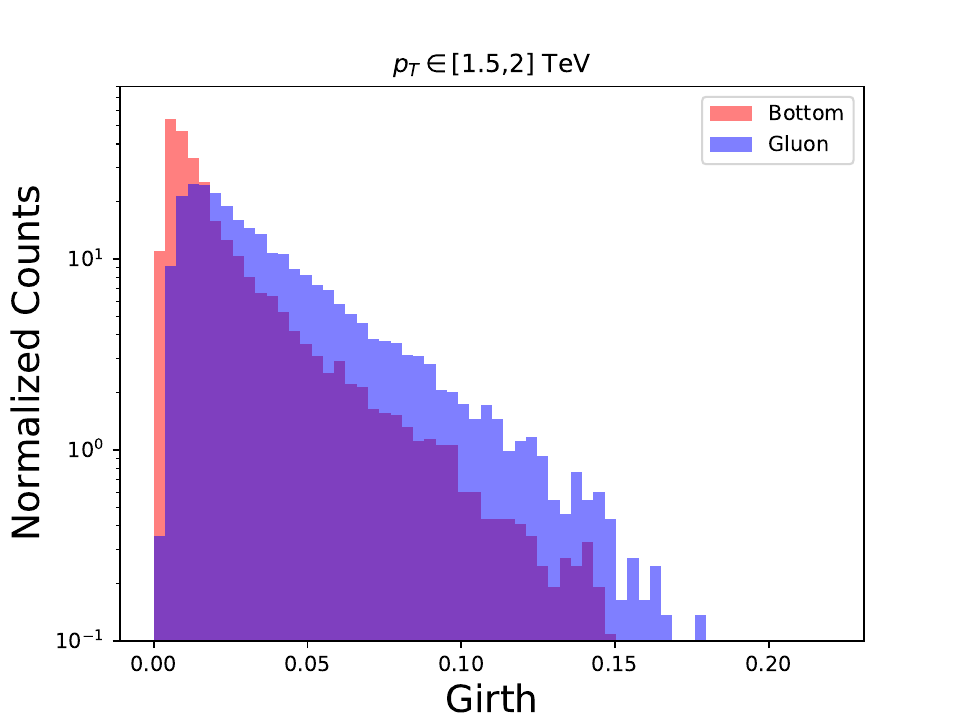}
  \includegraphics[width=0.32 \linewidth]{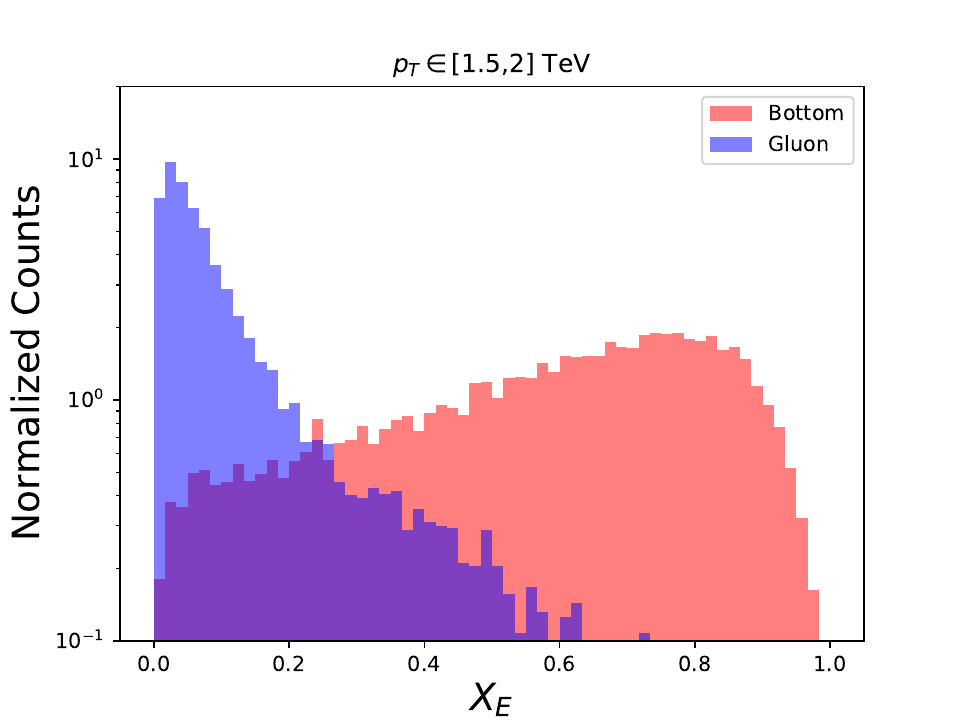}
  \includegraphics[width=0.32 \linewidth]{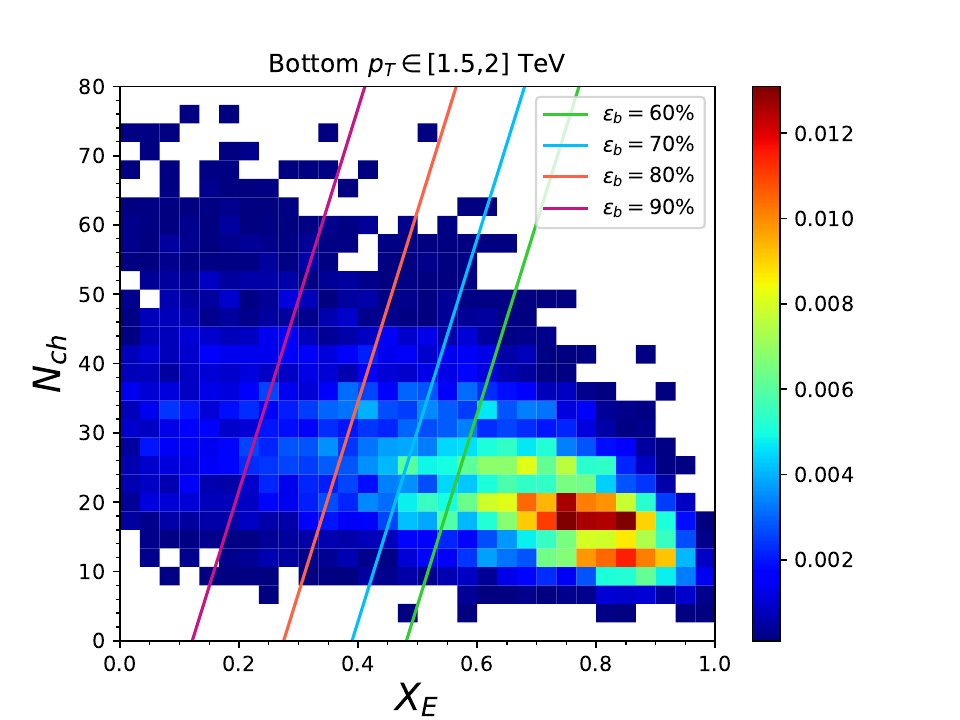}
  \includegraphics[width=0.32 \linewidth]{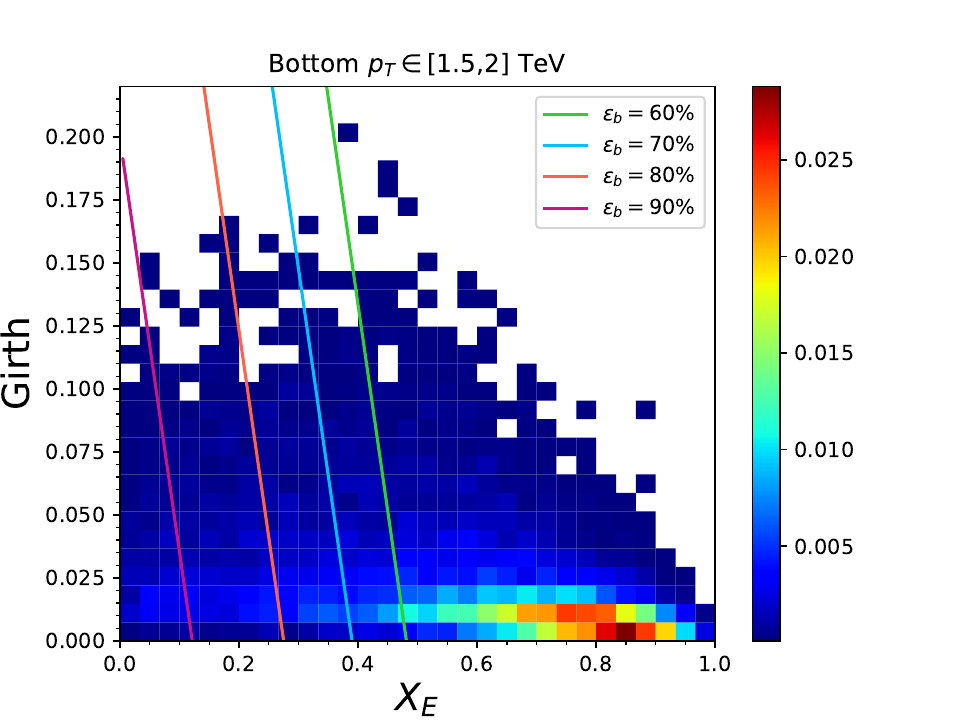}
  \includegraphics[width=0.32 \linewidth]{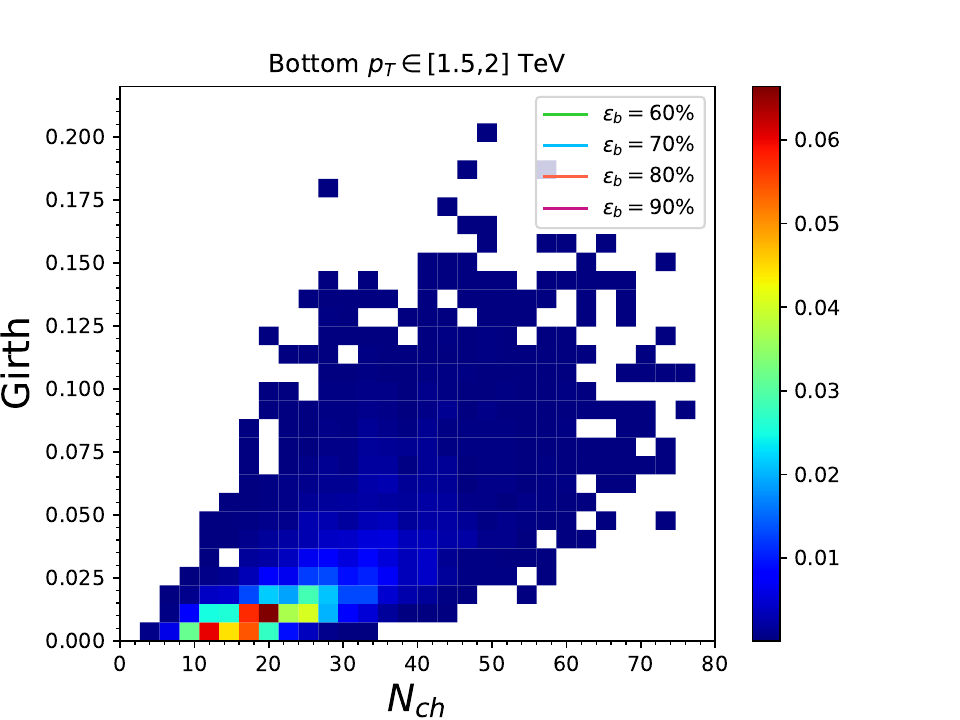}
  \includegraphics[width=0.32 \linewidth]{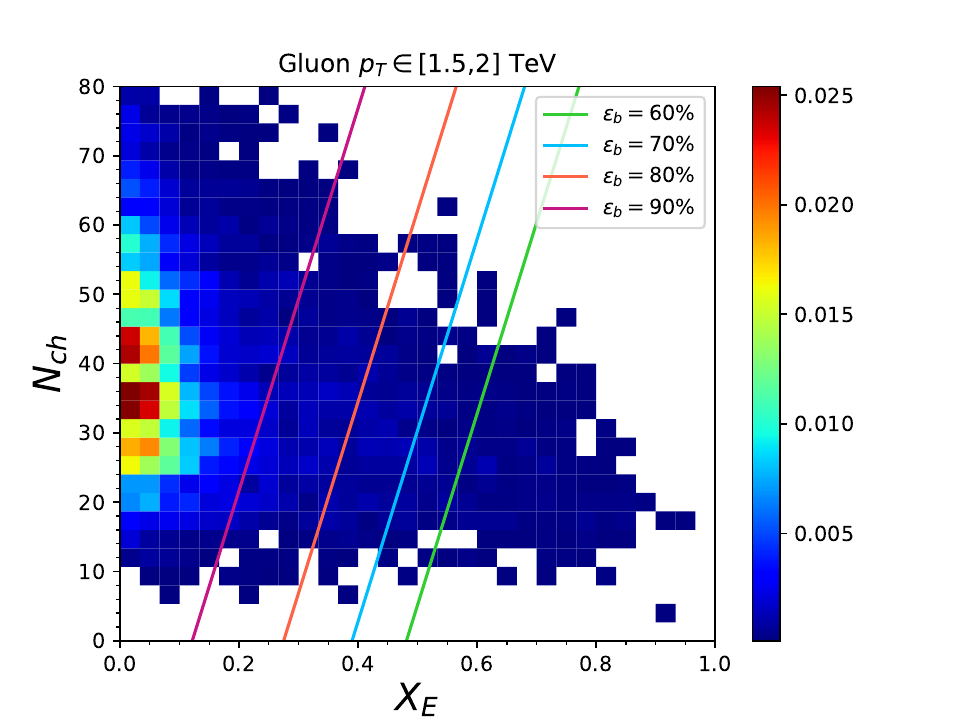}
  \includegraphics[width=0.32 \linewidth]{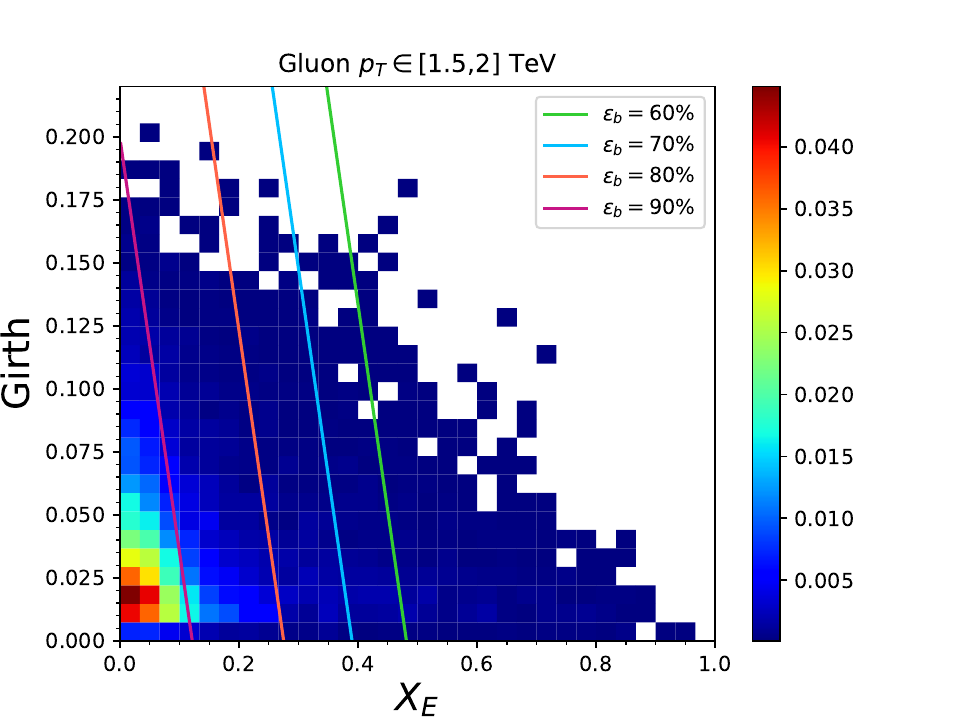}
  \includegraphics[width=0.32 \linewidth]{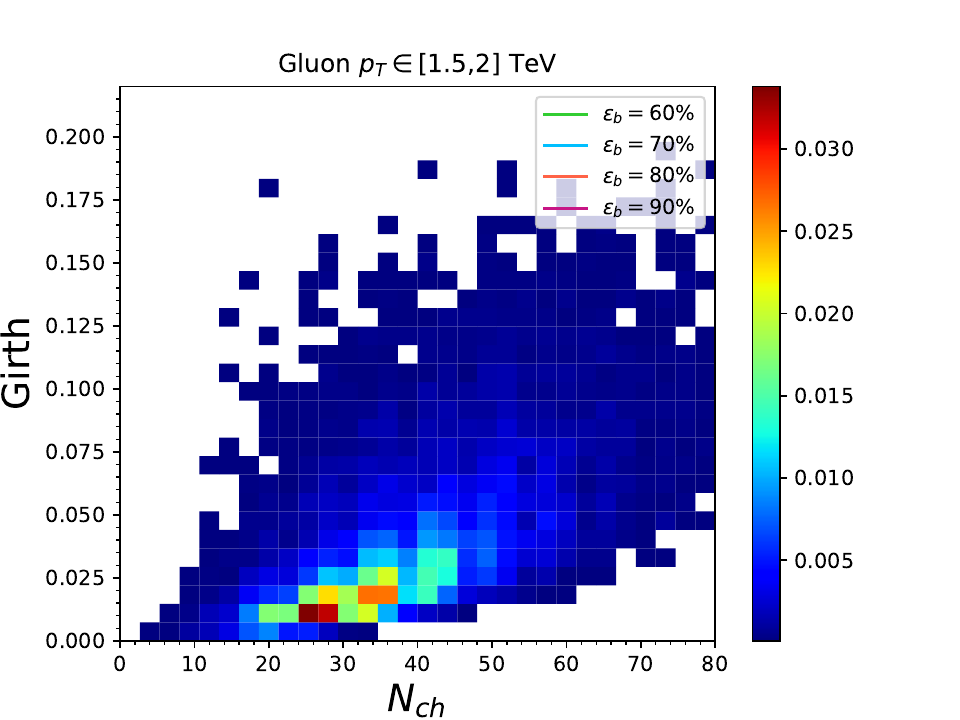}
  \newline
  \includegraphics[width=0.32 \linewidth]{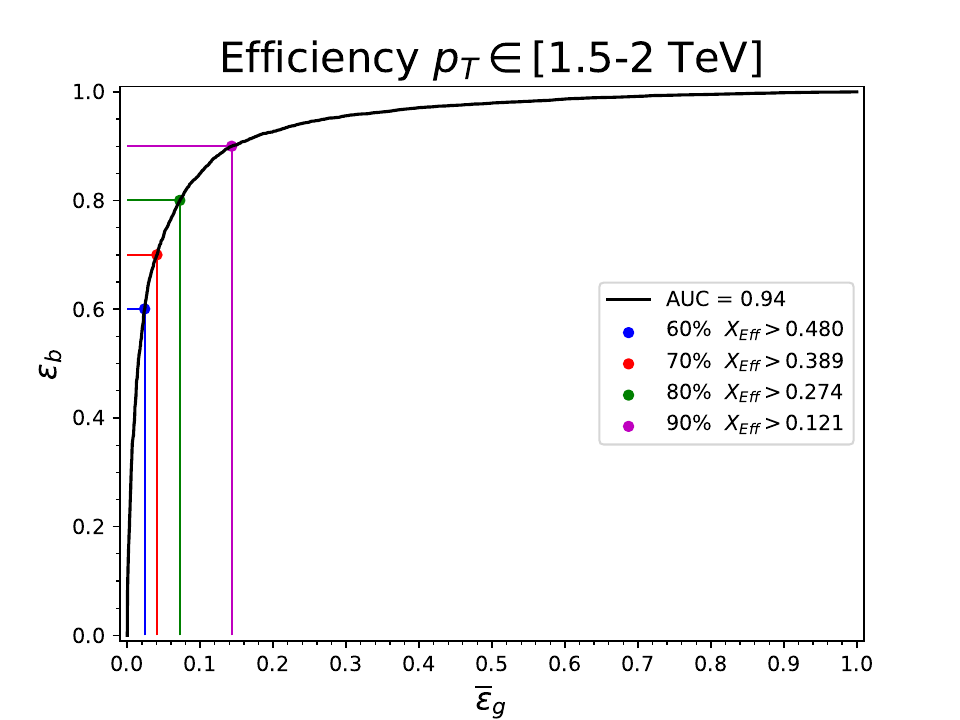}
\end{center}
\caption{Charged track multiplicity, girth and $x_E$ for jets with $p_T\in[1.5,2] \; {\rm TeV}$.}
\label{fig:2000}
\end{figure}

\bibliography{references}{}
\bibliographystyle{JHEP} 

\end{document}